\documentclass[range]{ar2e_arxiv}

\def\ts     {\thinspace} 
\def\cone {\ifmmode{{\rm C}{\rm \small I}(1-0)}\else{C\ts {\scriptsize I}(1--0)}\fi}
\def\ctwo {\ifmmode{{\rm C}{\rm \small I}(2-1)}\else{C\ts {\scriptsize I}(2--1)}\fi}
\def\conel {\ifmmode{{\rm C}{\rm \small I}\,(^3P_1\to^3P_0)}\else{C\ts{\scriptsize I}\,{\small$(^3P_1\to^3P_0)$}}\fi}
\def\ctwol {\ifmmode{{\rm C}{\rm \small I}\,(^3P_2\to^3P_1)}\else{C\ts{\scriptsize I}\,{\small$(^3P_2\to^3P_1)$}}\fi}

\def\kms    {\ifmmode{{\rm \ts km\ts s}^{-1}}\else{\ts km\ts s$^{-1}$}\fi}
\def\msol   {\ifmmode{{\rm M}_{\odot}}\else{M$_{\odot}$}\fi}
\def\aco  {\ifmmode{^{12}{\rm CO}(J=1\to0)}\else{$^{12}{\rm CO}(J=1\to0)$}\fi}
\def\cii  {\ifmmode{{\rm C}{\rm \small II}}\else{[C\ts {\scriptsize II}]}\fi}
\def\ci  {\ifmmode{{\rm C}{\rm \small I}}\else{[C\ts {\scriptsize I}]}\fi}
\def\oi  {\ifmmode{{\rm O}{\rm \small I}}\else{[O\ts {\scriptsize I}]}\fi}
\def\oii  {\ifmmode{{\rm O}{\rm \small II}}\else{[O\ts {\scriptsize II}]}\fi}
\def\oiii  {\ifmmode{{\rm O}{\rm \small III}}\else{[O\ts {\scriptsize III}]}\fi}
\def\siii  {\ifmmode{{\rm S}{\rm \small III}}\else{[S\ts {\scriptsize III}]}\fi}
\def\nii  {\ifmmode{{\rm N}{\rm \small II}}\else{[N\ts {\scriptsize II}]}\fi}
\def\hi   {\ifmmode{{\rm H}{\rm \small I}}\else{H\ts {\scriptsize I}}\fi}
\def\hii  {\ifmmode{{\rm H}{\rm \small II}}\else{H\ts {\scriptsize II}}\fi}
\def\h2     {\ifmmode{{\rm H}_2}\else{H$_2$}\fi}
\def\apjl  {{\em Ap.\,J.\,Lett.}}
\def\mnras  {{\em MNRAS}}
\def\araa  {{\em ARA\&A}}
\def\aj    {{\em Astron.\,J.}}
\def\apj   {{\em Ap.\,J.}}
\def\apjs   {{\em Ap.\,J.\,Suppl.}}
\def\aap    {{\em Astron.\,Astroph.}}
\def\nat    {{\em Nature}}
\def\nar   {{\em New\,Astron.\,Rev.}}
\def\apss   {{\em Astroph.\,\&\,Space~Science}}

\begin{document}


\input psfig.sty

\jname{..}
\jyear{2013}
\jvol{}
\ARinfo{}

\title{Cool Gas in High Redshift Galaxies}

\markboth{Carilli \& Walter}{Cool Gas in High Redshift Galaxies}

\author{C.L. Carilli$^{1}$ \& F. Walter$^{2}$
\affiliation{$^1$National Radio Astronomy Observatory, P. O. Box 0,
Socorro, NM 87801, USA \\ $^2$Max--Planck--Institut
f\"ur Astronomie, K\"onigstuhl 17, D--69117 Heidelberg, Germany
}}

\begin{keywords}

Galaxy formation; Radio lines: molecular, mm, cm; Molecular Gas; Atomic Fine
Structure lines; Galaxies

\end{keywords}

\begin{abstract}

Over the last decade, observations of the cool interstellar medium in
distant galaxies via molecular and atomic fine structure line emission
has gone from a curious look into a few extreme, rare objects, to a
mainstream tool to study galaxy formation, out to the highest
redshifts. Molecular gas has now been observed in close to 200
galaxies at z\,$>$\,1, including numerous AGN host--galaxies out to
z\,$\sim$\,7, highly starforming sub--millimeter galaxies (median
redshift z\,$\sim$\,2.5), and increasing samples of `main--sequence'
color--selected star forming galaxies at z$\sim$1.5--2.5. Studies have
moved well beyond simple detections, to dynamical imaging at
kpc--scale resolution, and multi--line, multi--species studies that
determine the physical conditions in the interstellar medium in early
galaxies.  Observations of the cool gas are the required complement to
studies of the stellar density and star formation history of the
Universe, as they reveal the phase of the interstellar medium that
immediately preceeds star formation in galaxies. Current observations
suggest that the order of magnitude increase in the cosmic star
formation rate density from $z \sim 0$ to 2 is commensurate with a
similar increase in the gas to stellar mass ratio in star forming disk
galaxies. Progress has been made on determining the CO luminosity to
H$_2$ mass conversion factor at high-$z$, and the dicotomy between
high versus low values for main sequence versus starburst galaxies,
respectively, appears to persist with increasing redshift, with a
likely dependence on metallicity and other local physical conditions.
Studies of atomic fine structure line emission are rapidly
progressing, with some tens of galaxies detected in the exceptionally
bright \cii\ 158$\mu$m line to date. The \cii\ line is proving to be a
unique tracer of galaxy dynamics in the early Universe, and, together
with other atomic fine structure lines, has the potential to be the
most direct means of obtaining spectroscopic redshifts for the first
galaxies during cosmic reionization.

\end{abstract}

\maketitle

\section{Introduction}

\subsection{Motivation}

The last few years have seen remarkable progress in the study of the
cool, molecular gas content of galaxies, using centimeter and
(sub--)millimeter telescopes. The cool gas content is a critical
parameter in galaxy evolution, serving as the immediate fuel for star
formation in galaxies. The state of the field in 2005 was reviewed by
Solomon \& vanden Bout (2005; see also Omont 2007). At that point,
only a few handful of extreme starburst galaxies and luminous active
galactic nuclei (AGN) host galaxies had been detected in molecular gas
emission at significant lookback times, hardly anything was known
about the gas excitation, and there were no detections of atomic fine
structure lines. Research in recent years has resulted in an explosion
in the number and type of galaxies detected in molecular line
emission, as well as in atomic fine structure line emission, in the
distant Universe. Detailed multi--transition, multi--species
follow--up has been performed to determine the physical conditions of
the gas in some of the brightest high--redshift systems known. The
results are proving extremely telling for our understanding of galaxy
formation and evolution and suggest that the molecular gas content
of galaxies increases significantly with look--back time.

It is a good time to review the field of molecular line observations
of high redshift galaxies for two reasons. First is the dramatic
advance that has been made over the last decade, both in the number of
galaxies detected, as well as the characterization of the molecular
properties of these galaxies through followup observations. Second is
the imminent full operation of revolutionary telescopes, the Atacama
Large Millimeter/Submillimeter Array, ALMA (Wootten \& Thompson
2009; Andreani 2010), and the Karl J.~Jansky Very
Large Array, JVLA (Perley et al.\ 2011), both of which promise to
explore this evolution of the universal molecular gas content to an
order of magnitude greater level of detail and sensitivity than
previously possible. This review of the field, at this temporal cusp
of knowledge, captures the current state of the field, and frames the
fundamental questions that will be addressed with the next generation
instruments.

\subsection{Galaxy formation and the need for cool gas observations}
\label{intro_notes}

Galaxy evolution has been the subject of many reviews in recent years
(e.g. Shapley 2011, Renzini 2006, Giavalisco 2002, Silk \& Mamon 2012) and the last decade
has seen dramatic advances in our understanding of cosmic structure
formation. Cosmic geometry, the mass-energy content of the Universe,
and the initial density fluctuation spectrum, are now known to better
than 10\% (Spergel et al.\ 2007, Komatsu et al.\ 2011). Structure
formation through gravitational instabilities has been calculated in
exquisite detail through numerical studies (e.g. Springel et al.\
2005, Klypin et al.\ 2011), and observationally verified through
studies of galaxy distributions (e.g., Peacock et al.\ 2001, Reid et
al.\ 2010). And the cosmic star formation rate density (the `star
formation history of the Universe', SFHU), and stellar mass build-up,
have been quantified back to first light and cosmic reionization
(e.g. Bouwens et al.\ 2011a, Coe et al.\ 2013), within 1\,Gyr of the
Big Bang.  Studies of galaxy formation are now turning attention to
the evolution of the cool gas, the fuel for star formation in
galaxies. In this section, we briefly summarize some of the general
conclusions on galaxy formation that are relevant to our subsequent
review of the gaseous evolution of galaxies.

Three main epochs have been identified in the SFHU , starting with a
steady rise during cosmic reionization from $z \sim 10$ to 6
(e.g. Bouwens et al.\ 2011b; Bouwens et al. 2012, Coe et al.\ 2013), corresponding to
the epoch when light from the first galaxies reionizes the neutral
intergalactic medium (IGM) that pervaded the Universe (Fan et al.\
2006, Finkelstein et al.\ 2012). The comoving cosmic star formation rate density then peaks at
$z \sim 1$ to 3. This range is known as the `epoch of galaxy
assembly,' during which about half the stars in the present day
Universe form (Shapley 2011; Marchesini et al.\ 2009; Reddy et al.\
2008). Last comes the order of magnitude decline in the comoving
cosmic star formation rate density from $z \sim 1$ to the present
(e.g. Lilly et al.\ 1996, Madau et al.\ 1996).

The study of galaxy formation takes on the challenge to explain this
observed star formation history of the universe in the context of
$\Lambda$CDM, the hierarchical dark matter halo model. To understand
galaxy formation we must investigate how stars and star formation are
distributed over dark matter halos with different masses as a function
of time. The most important feature of our current understanding of the
field is that star and galaxy formation is inefficient: only $\sim$5\%
of all baryons (i.e. atoms of all kind) are in stars and dark stellar
remnants at redshift zero (e.g., Fukugita \& Peebles 2004).

Galaxies with a baryonic mass of slightly more than that of the Milky
Way ($\sim 5\times 10^{10} M_{\odot}$) are most efficient in
converting the available baryons into stars ($\sim 15-20\%$,
e.g. Moster et al.\ 2010). One key observational result is that this
`typical' galaxy mass has not greatly changed since z$\sim$3 (e.g.,
Marchesini et al.\ 2009, Ilbert et al.\ 2010).  Dark matter halos and
their baryon content, on the other hand, have grown by two orders of
magnitude over that time span (e.g. Springel et al.\ 2005).  It
appears that dark matter halos with a total mass of $\sim10^{12}
M_{\odot}$ are, at all cosmic times, the most efficient star formation
factories. For such halos, the star formation rate at different epochs
is observed to roughly follow the cosmological accretion rate as
predicted by $\Lambda CDM$: star--formation rates are observed to
increase systematically with redshift, in a regular fashion such that
a galaxy  `main sequence' (defined below) is established, with a relatively
small scatter in star formation rate for a given galaxy stellar mass
(e.g. Noeske et al.\ 2007). At $z \sim 0$, the cosmic star formation
rate density is dominated by galaxies with star formation rates $\le
10$ M$_\odot$\,yr$^{-1}$ (FIR luminosities $\le 10^{11}$ L$_\odot$). By
$z \sim 2$, the dominant contribution shifts to galaxies forming stars
at $\sim 100$ M$_\odot$\,yr$^{-1}$ (Murphy et al.\ 2011; Magnelli et
al.\ 2011). Once the halo and the galaxy grow beyond this mass,
further growth through star formation is marginal (e.g. Peng et al.\
2010) and accretion of other galaxies (merging) becomes the dominant
evolutionary channel (van der Wel et al.\ 2009).

Massive galaxies with low star formation activity, if any, are
observed at all redshifts $z \leq 3$ (e.g. Franx et al.\ 2003, Kriek
et al.\ 2006) and their existence remains a puzzle as there is no
trivial mechanism that prevents gas from cooling and forming stars in
more massive halos. Many ideas abound: shock heating of
in--falling gas (Kere{\v s} et al.\ 2005, Dekel \& Birnboim 2006),
feedback from active galactic nuclei (AGN, Croton et al.\ 2006; Bell
et al.\ 2012), motivated by ubiquitously observed super--massive black
holes in massive galaxies and the coincidence of the peak of QSO and
star--formation activity at $z\sim 2$ (Hopkins et al.\ 2006), and
stabilization of gaseous disks as a result of bulge formation (Martig
et al.\ 2009). Feedback by AGN--driven winds appears to be required to
explain the evolution of young star forming galaxies into red,
bulge--dominated galaxies at intermediate redshift (e.g., Feruglio et
al.\ 2010), while powerful radio jets from AGN may be needed to heat
the intercluster gas around massive galaxies at late cosmic epochs,
thereby inhibiting further late--time growth of massive galaxies
(Fabian 2012; McNamara \& Nulsen 2007).

We briefly clarify our use of the terms `starburst' and `main
sequence' (MS) for star forming galaxies in this review. These classifications have
arisen in two, parallel situations. First, studies find that the
majority of starforming galaxies at both high and low redshifts define
a `main sequence', in which there is a relatively tight distribution
(dispersion $< 0.3$dex) in specific star formation rate
(sSFR\,=\,SFR/stellar mass) versus stellar mass. The sSFR is
typically a slowly decreasing function of increasing stellar mass, and, at a
given stellar mass, the sSFR increases by a factor 20 from z=0 to 2
for these MS galaxies (Sargent et al. 2012; Rodighiero et al. 2011).
However, at all redshifts the distribution function in sSFR requires a
second component at a factor 4 to 10 times higher SFR than
the nominal main sequence.  This `starburst' component constitutes a
few percent of the distribution by number, and about 10\% in terms of
the contribution to the cosmic star formation rate density.  

Second, as presented in Sec.~\ref{sf_law}, there also appears to be a
parallel dual-sequence in the Far-IR to CO luminosity ratio, with
factor few higher ratios for starburst versus main sequence galaxies.
The implied gas consumption timescales may be an order of magnitude,
or more, shorter in starburst systems than in main sequence galaxies
(modulo the conversion factor, see Sec.~\ref{alpha_highz}).  There is
some evidence to support the notion that starbursts are associated
with major gas rich mergers, although this remains an area of open
investigation. These two sequences of star forming galaxies are in
addition to early-type galaxies discussed above, which are typically
higher stellar mass, and show an order of magnitude, or more, lower
sSFR.

In order to understand why star formation efficiency peaks at a
certain halo mass and then declines for more massive systems, the
interplay between gas accretion, cooling, star formation and feedback
must be understood. Most of our current understanding of galaxy
formation, as briefly summarized above, is based on studies of the
stars, star formation, and ionized gas. There remains a major gap in
our knowledge, namely, observations of the cool gas: the fuel for star
formation in galaxies. Put simply, current studies probe the products
of the process of galaxy formation, but miss the source. If we can
trace the presence of cold gas and its distribution in different
galaxies and halos over cosmic time, the puzzle of the efficiency of
star and galaxy formation can be unraveled.  Numerous observational
and theoretical papers have pointed out this crucial need for
observations of the cool molecular gas feeding star formation in
galaxies (Dressler et al.\ 2009; Genzel et al.\ 2008, Obreschkow \&
Rawlings 2009a; Bauermeister et al.\ 2010).

We review the current status of observations of the cool gas content
of galaxies, as measured via the rotational transitions of common
interstellar molecules, and via the atomic fine structure line
transitions, predominantly \cii. Our focus is on results
since the reviews of Solomon \& vanden Bout (2005) and Omont (2007).
Our review is primarily observational. We present
tools and concepts of studying the interstellar medium in distant
galaxies (Sec.~\ref{concepts}), then summarize observational results
for different galaxy types at high redshift
(Sec.~\ref{high_z_summary}). We then discuss implications of the
recent observations (Sec.~\ref{diagnosis})
and what they tell us about conditions in early galaxies and galaxy
formation in general (Sec.~\ref{gas_history}). We end by raising some of the key
questions that can be addressed with new facilities: the JVLA and
ALMA (Sec.~\ref{summary_points} and Sec.~\ref{future}).

We only consider molecular and fine structure line emission in this
review. For a review of the few rotational molecular absorption line
systems seen at high redshift to date, see Combes (2008) and Carilli
\& Menten (2002).

{\bf SIDEBAR: Over the last two decades, the star formation history of
the universe, and the build up of stellar mass, has been well
quantified as a function of galaxy environment and luminosity, back to
cosmic reionization ($z \sim 10$).  It is clear that massive galaxies
form most of their stars early, and that the majority of star
formation occurs at $z > 1$. The dominant contribution to the cosmic
star formation rate density shifts to higher star formation rate
galaxies with redshift. The next major step in the study of galaxy
formation is the delineation of the cool gas content of galaxies, and,
in particular, the molecular interstellar medium (ISM) out of which
stars forms, as a function of cosmic time.}

\section{Concepts of observing cool gas} 
\label{concepts}

\subsection{Heating and cooling of the starforming ISM in galaxies}
\label{heating}

At high redshift, the ionized ISM can be studied through a combination
of optical emission lines (e.g. Ly$\alpha$, H$\alpha$) and far--infrared (FIR) atomic fine
structure lines (e.g. \cii, \nii, \oiii). The neutral medium can be
studied with FIR atomic fine structure lines (e.g. \oi, \ci,
\cii). Unfortunately, HI 21cm emission from galaxies cannot be studied
at redshifts z$>$0.5 due to limited sensitivity, and must await the
full square kilometer array (Carilli \& Rawlings 2003). The atomic
phase in high redshift systems can also be studied through absorption
measurements through individual lines--of--sights (Wolfe et al.\ 2005).

In the molecular medium, the high gas densities protect molecules
against UV photodissociation, because of the shielding by dust and
self--shielding of H$_2$ (Tielens 2005, Lequeux 2005, Dyson \&
Williams 1980). The molecular gas phase is thought to immediately
preceed star formation (e.g. Leroy et al.\ 2008, Schruba et al.\ 2011)
and this phase is thus most relevant to study a galaxy's potential to
form new stars. The other phases of the ISM cannot form stars directly
(but see, e.g., Glover \& Clark 2012), unless they cool sufficiently
to form cold and dense molecular gas. The processes in the ISM are
highly dynamic, with new gas being accreted (e.g. through mergers or
cold mode accretion, CMA, Sec.~\ref{accretion}) and gas being lost
through both stellar and black hole feedback processes
(Sec.~\ref{outflows}).

The temperature and density of the ISM is a result of environment, and
is determined by a balance of heating and cooling. There are several
mechanisms that lead to the heating of the molecular gas
(e.g. Goldsmith 1978). Deep within dark molecular clouds, the main
heating source is through cosmic rays (i.e. protons and electrons
accelerated to GeV energies). Cosmic rays ionize $\rm H_2$ molecules, and
the free electrons then transfer excess kinetic energy to other $\rm H_2$
molecules.

In molecular clouds associated with active star formation, UV heating
is also invoked to explain molecular gas heating. In this picture, O and B
stars that dominate the radiation in starforming regions (mostly in
the far-UV, 6.0\,eV\,$<$\,E\,$<$\,13.6\,eV, recall that
1\,eV\,$\sim$\,10$^4$\,K (1\,eV\,=11605\,K)), turn their interface with the molecular clouds
into photon dominated regions (PDRs; also historically known as
photo--dissociation regions; Hollenbach and Tielens 1997). On the very
surface, the CO is dissociated by UV radiation, and the dominant
emission comes from atomic fine structure lines and H$_2$
rovibrational lines, as well as dust continuum and PAH
emission. Further into the region, dust--shielding and self--shielding
allows for the persistence of CO, with the gas heated mainly by
electrons that are released from the surfaces of dust grains due to UV
absorption.

Related to the UV heating in PDRs is the heating that occurs through X--ray
emission (X--ray dominated region, or XDR) emerging from AGN and/or
hot plasmas heated by SNe, where the harder input spectrum of the
X--rays penetrates further into the molecular clouds than the UV
radiation (e.g. Meijerink et al.\ 2006). Both PDR and XDR regions are
discussed in Sec.~\ref{pdr_xdr_models}. Mechanical shock models have
also been invoked to explain the extreme excitation conditions in
nuclear starburst or other dense regions (e.g. Flower \& Pineau des
Forets 2003, Kristensen et al.\ 2008, Meijerink et al.\ 2013; 
Stacey et al. 2010; Nikola et al. 2010).

In terms of gas cooling, atomic fine structure lines have long been
noted as being the dominant coolant of interstellar gas in star
forming galaxies (Spitzer 1978), in particular, in cooler regions
where permitted lines of Hydrogen cannot be excited ($< 10^4$ K;
Sec.~\ref{fsl_intro}). As forbidden transitions, the lines are
typically optically thin, and hence avoid line trapping (resonant
scattering) in high column density regions. A few lines have
ionization potentials that are higher than hydrogen (13.6\,eV) and are
thus cooling lines of the ionized medium only (e.g. \nii,
\oiii). Others have lower ionization potentials, and thus also trace
the neutral ISM (e.g., \cii\, \oi, \ci). Up to a few percent of the
FUV energy from star formation in galaxies can go to gas heating via
photo--electrons, which is reradiated by fine structure lines,
principally the \cii\ 158$\mu$m line (and in some cases \oi).  The
majority of the stellar radiation goes into dust heating that is balanced through
FIR emission, or is radiated directly into the Universe, in roughly
equal proportions (depending on geometry and dust content, e.g., Elbaz
2002).

\subsection{Tracing molecular gas: observable frequencies}

The molecular gas mass in galaxies is dominated by molecular hydrogen,
H$_2$. Given the lack of a permanent dipole moment, the lowest
ro--vibrational transitions of H$_2$ are both forbidden, and perhaps
more importantly, have high excitation requirements (the first
quadropole line lies 500\,K above the ground, significantly higher
than temperatures in giant molecular clouds). As a consequence, only a very small
fraction of the cool molecular gas can be studied through H$_2$
emission in the infrared.  This is the reason that historically,
emission from tracer molecules have been used to detect molecular gas
in galaxies, and from which the total molecular gas mass is then
deduced. The molecule of choice has traditionally been carbon
monoxide, CO, since it is the most abundant molecule after H$_2$, has
low excitation requirements ($\sim$5\,K for first excited state, see
Sec.~\ref{temp_dens}), and is easily observed from the ground (3\,mm
band) in its ground transition.

In general, tracer molecules show quantized rotational states
populated based on collisions and the radiation field. For a linear
polar molecule of moment of inertia, $I$, the orbital angular momentum
is given by (Townes \& Schalow 1975) $L=n\hbar$ and the corresponding 
rotational energy is $E_{rot}=\frac{L^2}{2I}=\frac{n^2\,\hbar^2}{2I} =
\frac{J(J+1)\hbar^2}{2I} $ with $\Delta$J\,=\,$\pm$1 (conservation of
angular momentum). The energy released from level $J$ to $J$--1 is: $
\Delta E_{rot} = [J(J+1)-(J-1)J]\frac{\hbar^2}{2I}=\frac{\hbar^2 J}{I}
= h \nu_{line}$.

In reality this approximation is not strictly valid, as centrifugal
forces will increase with $J$ so that the bond (distance between
atoms) will stretch, which in turn changes $I$. This effect leads to
frequencies that are slightly lower than the first harmonic, eg. for
CO, the line--spreading via rotational stretching of higher order
transitions is of order 10\,MHz to 15\,MHz, or effectively 20 to
30\,km\,s$^{-1}$, allowing for unique redshift determinations based on
just two transitions (e.g., Wei\ss\ et al.\ 2009).  Depending on
molecule, there can be other additional fine and hyperfine structures
overlaid (i.e., magnetic and electrostatic interactions within the
molecule; e.g., Riechers et al.\ 2007a for the case of CN).

\subsection{Gas Temperatures \& Critical Density}
\label{temp_dens}

The kinetic temperature of the H$_2$ molecules, T$_{kin}$ is
determined by the velocity distribution of the molecules following the
Maxwell--Boltzmann distribution. The excitation of other tracer
molecules, such as CO, is mostly determined by the number of
collisions with H$_2$ molecules as they are very abundant, massive and
have a high cross section. It is often assumed that this kinetic gas
temperature equals the temperature of the dust T$_{dust}$ at high
densities.  However it should be noted that the heating and cooling
processes of the dust and molecular gas phases are quite different and
therefore thermal balance is not required.

Typically, rotational transitions are expressed as a function of
critical density, $n_{crit}$, ie. the density at which collisional
excitation balances spontaneous radiative deexcitation:
$n_{crit}$\,=\,A/$\gamma$ where $A$ is the Einstein coefficient for
spontaneous emission, $A\propto \mu^2 \times \nu^3$ in units
of s$^{-1}$, $\gamma$ is the collision rate coefficient in units
of cm$^3$\,s$^{-1}$, and $\mu$ is the dipole moment of the molecular
transition under consideration in the J\,=\,1 state.

The Einstein coefficient $A$ (see Tab.~\ref{tab_molecules}) is
determined entirely by the physical properties of the molecule and is
proportional to the frequency cubed (i.e. higher--J transitions have
higher deexcitation rates).  The collision rate coefficient $\gamma$
however depends on the temperature of
the gas ($\gamma=<\sigma \times v>$ where $\sigma$ is the collision
cross section and $v$ is the velocity of the particle). Collision rate
coefficients for the excitation of CO by H$_2$ are given by
Flower et al.\ (1985) and Yang et al. (2010), and are typically $\sim
3$ to $10 \times 10^{-11}$\,cm$^3$\,s$^{-1}$ for the low--J transitions and
temperatures of 40--100\,K.

Stars are created in cores of molecular clouds that have much higher
densities than the bulk of the gas traced by CO(1--0). This is the
reason why  molecules with significantly higher dipole
moments (leading to a higher $A$ coefficient as $A\propto \mu^2$ and thus
higher critical density) are typically observed. Typical dense gas tracers and
their dipole moments are: CS: 1.958\,D [Debye], HCO$^+$: 3.93\,D, HCN:
2.985\,D, for comparison CO has a dipole moment of 0.110\,D (Sch\"oier
et al.\ 2005). The downside of choosing these high density tracer
molecules is their low abundance and resulting faint line fluxes (as
discussed in Sec.\ref{dense_highz}).

As an example, Tab.~\ref{tab_molecules} summarizes the Einstein
coefficients for the different transitions of CO and HCN. The critical
density for these molecules is given in the last column of the table.
It is obvious that higher--J transitions
need increasingly high critical densities to be visible. Einstein
coefficients and collision rates for other molecules can be found in
Sch\"oier et al.\ (2005). 

\subsection{Brightness Temperature and line luminosities}

Historically, radio astronomers express the surface brightness of a
source as a Rayleigh--Jeans (RJ) brightness temperature for a
blackbody at given temperature through Planck's law. Measurements of
brightness temperature and surface brightness are thus equivalent
measurements.  Rather than expressing the temperature in terms of the
Planck temperature, radio astronomers have traditionally used the low
frequency (Rayleigh-Jeans) limit ($\frac{kT}{h\nu}\gg$1):

$$ B_\nu=\frac{2h\nu^3}{c^2}\times\frac{1}{\exp(\frac{h\nu}{kT})-1} \sim
\frac{2kT\nu^2}{c^2}$$

\noindent In this limit:  $ T_B^{obs}=\frac{c^2}{2k\nu_{obs}^2}I_\nu$,
where $I_\nu$ is the surface brightness.  The RJ
approximation is valid at centimeter wavelengths.  However in the
millimeter and particular submillimeter regime
$\frac{kT}{h\nu}$\,=\,0.7[$\lambda/1mm][T/10K]$, the RJ approximation
is no longer valid in many regions of
astrophysical interest, and RJ brightness temperatures can no longer
can be interpreted as a physical temperature. If one is interested in
true temperatures, full Planck temperatures should be derived instead.

At high redshift, there are two commonly used ways to express line
luminosities. One, $L_{line}$ is expressed as the source luminosity in
$L_\odot$ or other rational units. The second, $L'_{line}$, is
expressed via the (areal) integrated source brightness temperature, in
units of K\,km\,s$^{-1}$\,pc$^2$. The following equations can be used
to derive these two luminosities (Solomon et al.\ 1992):

$$ L'_{line}= 3.25 \times 10^7\times S_{line} \Delta v
\frac{D^2_L}{(1+z)^3\nu_{obs}^2} ~~\rm K\,km\,s^{-1}\,pc^2$$ 

$$ L_{line} = 1.04 \times 10^{-3} \times S_{line} \Delta v D^2_L
\nu_{obs} ~~ \rm L_{\odot}$$

\noindent where $S_{line} \Delta v$ is the measured flux of the line
in Jy\,km\,s$^{-1}$, $D_L$ is the luminosity distance in Mpc, and
$\nu_{obs}$ is the observed frequency. Solving for $S \Delta v$ and
equating leads to: $L_{line} = 3\times10^{-11} \nu_r^3 L'_{line}$
where $\nu_r$ is the rest frequency of the line. Note that $L'_{line}$
is directly proportional to the surface brightness T$_B$, i.e. the
$L'_{line}$ ratio of two lines give the ratio of their intrisic,
source--average surface brightness temperatures T$_B$. If the
molecular gas emission were to come from thermalized, optically thick
regions, $L'_{line}$ is constant for all J levels.

{\bf SIDEBAR: The question often arises which quantity to quote in a
  paper, $L_{line}$ or $L'_{line}$? Both have their justification:
  e.g. if one is interested in comparing the power that is being
  emitted through a given line to calculate the cooling capability
  (e.g. in relation to the FIR luminosity, $L_{line}/L_{\rm FIR}$) one
  uses the $L_{line}$ definition. The $L'_{line}$ is commonly used to
  translate measured CO luminosities to H$_2$ masses using the
  conversion factor $\alpha$. Also, for thermalized molecular gas
  emission $L'_{line}$ is approximately constant for all
  transitions. It is good advice to give both quantities in a paper,
  but in all cases also the measured integrated fluxes of the lines $S_{line}
  \Delta v$ (in units of Jy\,km\,s$^{-1}$).}

\subsection{CO luminosity to total molecular gas mass conversion factor}
\label{alpha_lowz}

The conversion factor relating CO(1--0) luminosity to total molecular
gas mass (dominated by H$_2$) in the nearby Universe has been reviewed
recently by Bolatto et al.\ (2013). We briefly summarize the main
points in this section.

Molecular gas in the Milky Way and nearby galaxies is predominantly in
giant molecular clouds with overall sizes up to 50\,pc. Hence, early work on the conversion factor
focused on GMCs. In nearby systems (including the Milky Way) the
clouds can typically be spatially resolved, and hence the conversion
factor considered, $X_{\rm CO}$, is the ratio of column density (H$_2$
molecules per cm$^2$) to CO velocity integrated surface brightness (K
km s$^{-1}$), while for distant galaxies spatially integrated
quantities are usually measured, and the conversion factor
$\alpha_{\rm CO}$ is then the ratio of total mass (in M$_\odot$) to total
CO line luminosity (K km s$^{-1}$ pc$^2$). The ratio, $X/\alpha$=$4.6\times 10^{10} \rm\, pc^2\,cm^{-2}\,M_\odot^{-1}$ = the number of
hydrogen molecules per solar mass, converted to the appropriate units
and corrected by a factor 1.36 for Helium.

Numerous techniques have been used to determine the conversion factor
in GMCs, including: (i) a comparison of CO columns derived from
isotopic measurements to H$_2$ columns derived from optical extinction
measurements using a standard dust--to--gas ratio (Dickman 1978; 1975;
Dame et al.\ 2001), (ii) a comparison of $\gamma$--ray emission to CO
surface brightness, where the $\gamma$--rays result from the
interaction between cosmic rays with H$_2$ molecules (Strong \& Mattox
1996, Bloemen 1989, Hunter et al 1997), and (iii) modeling GMCs as
self--gravitating clouds.  In the latter case, a key discovery was the
correlation between line width and cloud size, with line width
increasing as the square root of cloud size (the `Larson relations';
Larson 1981). This functional form implies a constant surface density
$\sim 100$ M$_\odot$ pc$^{-2}$ for GMCs (Solomon et al.\ 1987), and
that the CO luminosity is linearly proportional to cloud virial
mass. Indeed, this relationship provides the theoretical
under--pinning of the use of CO luminosity to derive total gas mass in
the case of optically thick emission (Solomon et al.\ 1987; Dickman et
al.\ 1986), where luminosity is dictated principally by line
width. Bolatto et al. (2013) summarize these results, and conclude
that a value of $\rm X_{\rm CO}\ = 2\times 10^{20}$ H$_2$ molecules per
cm$^{-2}$ /(K km s$^{-1}$), or $\alpha_{\rm CO} \sim 4$ M$_\odot$/(K km
s$^{-1}$ pc$^2$), is appropriate for GMCs in the MW and nearby spiral
galaxies, with a factor two (0.3 dex) scatter.

Draine et al.\ (2007) hypothesized that, if the abundance of the
elements can be assumed to be proportional to the Oxygen abundance,
and adopting a Milky--Way fraction of elements tied up in grains, then
$\rm M_{\rm dust}/M_{\rm gas} = 0.010 [(O/H)]/[(O/H)_{MW}]$. Using
observations of nearby galaxies, they showed that the gas columns
derived from the IR thermal dust emission are consistent with $\rm
X_{\rm CO} = 4 \times 10^{20}$cm$^{-2}$ /(K km s$^{-1}$ and
metallicities between 0.3 and 1 solar for their galaxy sample. Leroy
et al.\ (2011) use an empirical dust--to--gas ratio approach in nearby
galaxies and obtain similar results, showing that the lowest
metallicity galaxies clearly have much higher conversion factors.
They suggest that decreased dust shielding in low metallicity
environments leads to CO--free, but still H$_2$ rich, cloud envelopes
(see also Schruba et al.\ 2012; Sandstrom et al.\ 2013).

A different picture has emerged for nearby nuclear starburst galaxies,
including the nuclei of dwarf starbursts like M\,82 and NGC\,253,
corresponding to Luminous Infrared Galaxies, or LIRGs (L$_{\rm IR} \sim
10^{11}$ L$_\odot$), and ultraluminous infrared galaxies (ULIRGs) such
as Arp\,220 (L$_{\rm IR} \sim 10^{12}$ L$_\odot$). It was noted
early--on that a Milky Way conversion factor leads to a molecular gas
mass larger than the dynamical mass in some of these systems (Bryant
et al.\ 1999).  In their seminal analysis of CO radiative transfer and
gas dynamics in starburst nuclei of ULIRG on scales $<$1\,kpc,
Downes \& Solomon (1998) find a characteristic value of $\alpha_{\rm CO}
\sim 0.8$ M$_\odot$/(K km s$^{-1}$ pc$^2$) in these systems.  The
lower value of $\alpha$ implies more CO emission per unit molecular
gas mass. They hypothesize that much of the CO emission is not from
virialized GMCs, but from an overall warm, pervasive molecular
inter--cloud medium. In this case the linewidth, and hence line
luminosity (for optically thick emission), is determined by the total
dynamical mass (gas and stars), as well as ISM pressure. Narayanan et
al.  (2012) show that the lower $\alpha$ value in starbursts is due
to warm gas that is heated by dust (at densities $> 10^4$, such
energy exchange is efficient), as well as large, non-virial line
widths from the GMCs.  More recently, Papadopoulos et al. (2012)
suggest that the molecular gas heating processes in nuclear starbursts
may be very different than is typically assumed for PDRs, with cosmic
rays and turbulence dominating over photons.

In general, the value of $\alpha$ is likely a function of local ISM
conditions, including pressure, gas dynamics, and metallicity, and
remains an active area of research for nearby galaxies (Ostriker \&
Shetty 2011; Narayanan et al.\ 2012; Narayanan \& Hopkin 2012;
Papadopoulos et al.\ 2012, Leroy et al. 2013, Schruba et al. 2012,
Sandstrom et al. 2013). We note that the conversion factor has been
calibrated using observations of CO(1--0) in nearby galaxies. These
low order transitions redshift to centimeter wavelengths at z$>$2.

\subsection{Modeling the line excitation}
\label{model_excitation}

Many observations of galaxies at high--redshift are unresolved, and
studies of the global CO excitation play an important role in
constraining their average molecular gas properties. This CO
excitation, i.e. the relative strengths of the observed rotational
transitions, is sometimes referred to `CO spectral line energy
distributions', `CO SLEDs' or `CO excitation ladder'. We here use
the term `CO ladder', as the measured quantity is the
emission of a given rotational transition $J$. We note that some
authors also chose to plot spectral power distributions (i.e. plotting
$L$ instead of $S_\nu \times dV$ as a function of rotational number J).

The excitation temperature determines the population of the molecular
levels through the Boltzmann distribution. Under the assumption of
local thermodynamical equlibrium (LTE) this excitation temperature
will be equal to the kinetic temperature of the gas. `Thermal'
excitation means that the population of all levels is according to the
Boltzmann distribution. The excitation is `sub--thermal' if the
population of the high levels is less than that given the kinetic
temperature of the gas, as occurs in lower density environments where
collisional excitation cannot balance spontaneous emission rates. We
briefly discuss the commonly used techniques to model the excitation
of the various levels of molecular line emission.

\subsubsection{Escape propability/LVG models}
\label{lvg_models}

One basic tool commonly used to model the excitation of the molecular
gas emission is the large velocity gradient (LVG) method (e.g. Young
\& Scoville 1991). In the following we will discuss this model for the
CO molecule, but the same modeling can be done for any molecule. The
model calculates for a given temperature T$_{kin}$, H$_2$ density, CO
abundance ([CO]/[H$_2$]), and velocity gradient $dv/dr$, how the
various levels of the CO are populated through collisional excitation
with H$_2$. Some models and their dependence on temperature and
density are illustrated in Fig.~\ref{fig_co_seds} (see also Figures in
van der Tak et al.\ 2007). Since the CO emission is optically thick
(at least in the low--J transitions) the emission would have
difficulties escaping the cloud, which is why a velocity gradient is
introduced in the model that describes how many photons can eventually
leave the cloud. The justification for implementing such a gradient is
that in reality the molecular medium is turbulent which enables CO
photons to escape their parental clouds. The LVG codes also take the
redshift of the source into account which dictates the temperature of
the cosmic microwave background (CMB, Sec.~\ref{cmb}). Once the
occupation numbers of the different levels are calculated, the optical
depths $\tau$ for the transitions as well as the Rayleigh--Jeans
brightness temperatures T$_{B}$ (which would be constant in LTE) and
the resulting line intensities are derived.

In this way, for a given set of parameters, the expected line
intensities for any molecule can be derived (assuming certain
abundances of the molecule under consideration). In principle, the
measurement of the line emission ladder of a number of different molecules thus
puts tight constraints on the temperture and density of the gas. This
assumes that these lines probe the same volume, which may not hold for those
molecules tracing very high densities (e.g. HCO$^+$ and HCN). Of
particular interest is adding different isotopomers, such as $^{13}$CO
or C$^{18}$O as these are more optically thin than the 
$^{12}$C$^{16}$O emission, and are thought to originate from the same
volume as $^{12}$C$^{16}$O, at least on GMC scales.

\subsubsection{PDR and XDR models}
\label{pdr_xdr_models}

In PDR (Photo--dominated regions) and XDR (X--ray dominated regions)
models a cloud is exposed to a radiation field from which the
temperature and density distribution of the H$_2$ molecules is
derived. With the resulting values for T$_{ex}$ and density a code
similar to LVG is then run to calculate the line intensities of the
rotational transitions of the molecules under consideration. A
difference with respect to the LVG models is that abundances of
species are calculated based on the radiation field and the gas column
density. The main difference between the PDR and XDR codes is that
PDRs only exist at the surface of clouds (where they emit fine
structure lines of \ci, \cii\ and \oi, and rotational lines of CO in
the submillimeter). In the XDR phase (further inside the cloud) the
\oi, \cii, and \siii\ lines are main coolants in the sub--millimeter
regime (Meijerink \& Spaans 2005). One limitation of most PDR / XDR
models is that they are based on (one--dimensional) infinite slabs --
i.e. they do not have a confined volume. As a consequence no mass
estimates are typically given by the codes. The gas temperatures
calculated in PDR/XDR models are still quite uncertain and depend on
the code used, especially in the high--denisty, high--UV case (e.g.,
R\"ollig et al.\ 2007). Given the high excitation measured in nearby
galaxies based on Herschel data (up to J\,$\sim$\,30,
Hailey--Dunsheath et al.\ 2012), shock heating is also invoked to
explain extreme CO excitation (see also Meijerink et al.\ 2013).

One complication in all the models above is that it is now established
that the high--J rotational transitions of some molecules, e.g. HCN,
are not only excited through collisions (as the critical densities of
some of the detected high--J lines are extremely high,
$\sim$\,10$^9$\,cm$^{-3}$, i.e. they are not even being reached in the
cores of GMCs). One potential mechanism to excite the observed high--J
HCN rotational modes is through the infrared stretching and bending
modes of HCN at 3, 5 and 14 microns (this is referred to as `infrared
pumping' of the rotational levels, e.g. Wei\ss\ et al.\
2007a). Similar pumping may be in place even in the case of CO
emission and may explain the very high--J excitation seen in some
local galaxies (above references, and Harris et al.\ 1991 for
radiative trapping leading to enhanced mid--J CO emission).

\subsection{Water lines}

Water is thought to be one of the most abundant molecules in galaxies,
present predominantly in icy mantels of dust grains in cold
environments (Tielens et al.\ 1991; Hollenbach et al.\ 2008).  In
warmer environments, water in the gas phase is thought to play an
important role in cooling  (Neufeld \& Kaufmann 1993, Neufeld et
al.\ 1995). The rotational transitions of water have high Einstein A
values, and thus very high critical densities ($> 10^8$\,cm$^{-3}$),
i.e. collisional excitiation can only happen in the very centres of
dense cloud cores and other excitation mechanisms, in particular
infrared pumping, are typically invoked.

Naturally, water lines at low redshift are very difficult to observe
from the ground given the Earth's atmosphere. Herschel Space
Observatory observations enabled the first detections of submillimeter
lines of H$_2$O in nearby galaxies, other than Masers (Mrk\,231: Van
der Werf et al.\ 2010, Gonzales--Alfonso et al.\ 2010, M\,82: Wei\ss\
et al.\ 2010) These observations have yielded very different results
between the objects: Mrk\,231 revealed a rich spectrum of H$_2$O lines
but the ground--state lines remained undetected (Van der Werf et al.\
2010). In M\,82 no highly excited emission was found (Panuzzo et al.\
2010) but many low--excitation lines (Wei\ss\ et al.\ 2010). Generally
speaking, the water emission line spectrum is very complex and not
straightforward to interpret if only one or a few lines are measured
(Gonzalez--Alfonso et al.\ 2010). We discuss recent high--redshift
detections of water lines in Sec.~\ref{dense_highz}.

\subsection{Atomic Fine Structure lines}
\label{fsl_intro}

The atomic fine structure lines (FSLs) are major coolants of cooler
interstellar gas (Section Sec.~\ref{heating}).  Table~\ref{tab_molecules}
summarizes key parameters of important sub--millimeter atomic fine
structure lines.

\subsubsection{Singly Ionized Carbon:}

\cii\ is the strongest line from star forming galaxies at radio
through FIR wavelengths (eg. \cii\ line fluxes are typically $> 10^3$
times stronger than CO(1--0) in star forming galaxies), and in
particular, \cii\ 158$\mu$m is the strongest line from the cooler gas
in galaxies  ($< 10^4$\,K). 

The ratio of \cii/FIR luminosity for the Milky Way is 0.003, and this
value holds in nearby disk galaxies, with a relatively large scatter
(factor three; e.g. Malhotra et al.\ 1997, 2001). However, the ratio
appears to drop significantly at FIR luminosities above 10$^{11}$
L$_\odot$. One explanation for this drop is a reduction in the heating
efficiency by photoeletric emission from dust grains in high radiation
environments due to highly charged grains. This explanation is
supported by the fact that the \cii/FIR ratio is also a decreasing
function of increasing dust temperature (e.g., Malhotra et al.\ 2001,
Luhman et al.\ 2003). High dust opacity/absorption at 158$\mu$m may
also decrease the \cii/FIR ratio in high luminosity systems.

\cii\ has a lower ionization potential than \hi\ (11.3\,eV vs.\
13.6\,eV), hence \cii\ traces both the cold neutral medium (CNM) and
ionized gas. This makes \cii\ less easy to interpret, in particular if
the emission cannot be resolved spatially. The \cii\ luminosity is not
a simple function of star formation rate, nor is there a simple
dependence between \cii\ luminosity and the total mass of the ISM. In
an early study, Stacey et al.\ (1991) argue that roughly 70\% of the
total \cii\ emission from nearby spiral galaxies comes from PDRs, and
recent Herschel observations shows a possible correlation between the
\cii\ luminosity and the 11.3$\mu$m PAH feature, suggesting a close
correlation between PDRs and \cii\ emission (Sargsyan et al.\ 2012).
Herschel imaging has also shown significant differences in the spatial
distribution of the \cii\ emission and CO emission on scales $\le$ 300
pc in nearby galaxies (Mookerjea et al. 2011; Rodrigues-Fernandez et
al. 2006) (see also Cormier et al.\ 2010).

A clear trend for increasing \cii/FIR ratio with decreasing metallicity
has been established (Cormier et al.\ 2010; Isreal \& Mahoney 2011).
Lower metallicity leads to lower dust and higher UV mean free paths.
The decreased dust--shielding in PDR regions leads to deeper
photo--dissociation into the molecular clouds and increased
photo--electric heating efficiencies leading to higher \cii\
luminosities relative to the FIR. High--redshift observations of \cii\
are discussed in Sec.~\ref{fls_cii}.

\subsubsection{Atomic Carbon: [CI]}

Because the $^3$P fine--structure system of atomic carbon forms a
simple three--level system, detection of both optically thin carbon
lines, \conel\ and \ctwol\, enables one to derive the excitation
temperature, neutral carbon column density and mass, independent of
any other information (e.g., Ojha et al.\ 2001, Wei\ss\ et al.\ 2003,
Walter et al.\ 2011).  A combination of this method (using \ci) with
the aforementioned CO excitation ladder is particularly powerful as it
eliminates some of the degeneracy frequently found in CO radiative
transfer models under the assumption that [CI] traces the same regions
as CO (see discussion in Sec.~\ref{excitation}).

Studies of atomic carbon in the local universe have been carried out
in molecular clouds of the galactic disk, the galactic center, M82 and
other nearby galaxies (e.g., White et al.\ 1994; Stutzki et al.\ 1997;
Gerin \& Phillips 2000; Ojha et al.\ 2001; Israel \& Baas 2002; Israel
2005). These studies have shown that \ci\, is closely associated with
the CO emission independent of environment.  Since the critical
density for the \conel\, and CO(1--0), lines are both $n_{\rm cr}
\approx 10^3\,{\rm cm}^{-3}$, this suggests that the transitions arise
from the same volume and share similar excitation temperatures
(e.g. Ikeda et al.\ 2002). We discuss high--redshift detections of
\ci\ in Sec.~\ref{fsl_ci}.

\subsubsection{Other fine structure lines: \nii, \oi, \oiii}

Beyond \cii\ and \ci, there are a number of other far--infrared fine
structure lines that are potentially important physical diagnostics
for the ISM, in particular the lines from \oi  63$\mu$m, \oiii
52$\mu$m and 88$\mu$m, and \nii 122$\mu$m and 205$\mu$m. While
typically 10 times weaker than \cii, the oxygen line strengths
increase dramatically with the hardness of the radiation field, such
that in AGN environments the \oiii 88$\mu$m can be stronger than
\cii\ (Spinoglio et al. 2012; Spinoglio \& Malkan 1992;
Genzel \& Cesarsky 2000, Stacey et al.\ 2010). Most of these lines
are at even higher frequencies than \cii\ and are thus difficult to observe
even in the high--redshift universe.

The ionization potential of carbon, oxygen, and nitrogen are 11.3\,eV,
13.6 and 14.5 eV, respectively. Hence, \cii\ traces both the neutral
and ionized medium, while oxygen and nitrogen trace
the ionized medium only (except for \oi). The \nii 122$\mu$m /\oiii  88$\mu$m ratio is a
sensitive measure of hardness of the radiation field since these two
lines have similar critical densities for excitation ($\sim 300$ to
500 cm$^{-3}$), but very different ionization potentials (35\,eV for
\oii).  \cii\ 158$\mu$m and \nii\ 205$\mu$m have essentially the same
critical density for thermal excitation ($\sim 45$ cm$^{-3}$), hence
the ratio of \cii$_{158}$/\nii$_{205}$ indicates the fraction of
\cii\ emission from ionized gas, assuming only a gas phase C/N
abundance (Oberst et al.\ 2006, Walter et al.\ 2009a, Decarli et
al.\ 2012, Ferkinhoff et al.\ 2010, 2011). In this case, the
\cii$_{158}$/\nii$_{205}$ ratio for ionized gas ranges between 3 and 4
(Fig. 2 in Oberst et al). Oberst et al.\ (2006) show that the
\nii  122$\mu$m /\nii  205$\mu$m ratio is a densitometer, with a value of
$\sim 0.5$ for densities below 10$^2$ cm$^{-3}$, rising to 10 for
densities greater than 10$^3$ cm$^{-3}$.

\subsection{Relation to far--infrared emission and SFRs} 
\label{dust}

In this review, we concentrate on the molecular gas properties of high
redshift galaxies. To put the physical properties of the ISM into
perspective, we will also use the far--infrared (FIR) emission and star
formation rates in galaxies. For reference we here give the equations
that are commonly used to derive IR/FIR luminosities and star
formation rates from single--band FIR continuum measurements. We stress that the
numbers that follow are appropriate for `dusty' SEDs typically found at
high redshift, but that the relations will be a strong function of
galaxy properties, such as the star formation activity and dust optical depth. 
The following equations are to first order independent of
redshift for z$>$1, given the inverse K--correction in the
sub--millimeter regime (e.g. Blain et al.\ 2002; Kennicutt \& Evans
2012).

The IR luminosity $L_{\rm IR}$ is defined from 8--1000\,$\mu m$,
whereas the FIR luminosity $L_{\rm FIR}$ is commonly defined from
40--400\,$\mu$m (Sanders et al.\ 2003).  $L_{\rm IR}$ is typically
$\sim$30\% higher than $L_{\rm FIR}$ for `dusty' SEDs. Some authors
also define FIR from 42--122\,$\mu$m (Helou et al.\ 1985) -- this
latter definition gives $L_{\rm FIR}$ that are 20--30\% smaller that
the $L_{\rm FIR}$ definition above.

In measured flux densities at 250\,GHz and 350\,GHz (850$\mu$m), the
following relations are commonly used: $L_{\rm FIR}[L_\odot] \sim
1.2\times 10^{12} S_{350 \rm GHz}[mJy]$ (e.g. Genzel et al.\ 2010,
Pope et al.\ 2006, Magnelli et al.\ 2010), $L_{\rm FIR}[L_\odot] \sim
3 \times 10^{12} S_{250 \rm GHz}[mJy]$ (e.g. Bertoldi et al.\ 2003,
Omont et al.\ 2001), with significant dependencies on the actual shape
of the SED.  As Scoville (2012) points out, a continuum measurement
around 1\,mm wavelengths (observed frame) is not suited to give good
estimates of $L_{\rm IR}$ given the unknown intrinsic dust SED, at
least for redshifts z$<$4. However, it gives a good estimate on the
dust mass (given that the dust is optically thin, and that the dust
temperatures are not found to vary greatly).

The SFR can be calculated from the IR luminosities following: $SFR
\sim \delta_{MF} \times 1.0 \times 10^{-10} L_{\rm IR}$ where
$\delta_{MF}$ depends on the stellar population (this assumes that the
galaxy is dusty i.e. that all of the power radiated by young stars is
absorbed by dust and re--emitted in the infrared; also that there is no
contribution of the AGN to the FIR emission). For a range of
metallicities, starburst ages ($<$\,100\,Myr) and initial mass
functions (IMFs) Omont et al. (2001) find:
0.8$<\delta_{MF}<$2. Typically a $\delta_{MF}$\,=\,1.8 is appropriate
for a Salpeter IMF (e.g. Kennicutt et al.\ 1998), and
$\delta_{MF}\sim$1.0 for a Chabrier IMF.

We adopted FIR luminosities for all high--redshift sources in which
line emission as been detected (Tab.~3). These FIR luminosities have
been derived using various different methodologies and therefore
should be treated with caution. Wherever available, we adopted the FIR
luminosities given by indivdual studies but caution that not all
authors have adopted the same `FIR' definition. E.g. L$_{\rm FIR}$ is
frequently adopted to go from 8--1000\,$\mu m$, whereas strictly
speaking this is the definition of the IR luminosity (see above). For
the purpose of this review we simply set L$_{\rm FIR}\sim$L$_{\rm
IR}$. For each source in the table we have computed the FIR
luminosities as follows: 1) if a $L_{\rm FIR}$ value is already
reported in the literature (e.g., based on dust SED fitting of
multiple photometric information) we use the most up-to-date estimate
available; 2) for a number of color--selected galaxies, only an
estimate of the SFR is available. In this case, we convert it into
dust luminosity using the relation $\log L_{\rm FIR} = \log {\rm SFR}
- \log(1.3) +10$, where $L_{\rm FIR}$ is in solar units and SFR is in
solar masses per year (see Genzel et al., 2010). 3) If only a
continuum estimate is available, we use the 850\,$\mu$m flux to infer
continuum luminosity, assuming the Arp\,220 template by Silva et
al. (1998) and integrating from 8\,$\mu$m -- 1000 $\mu$m. To first
order, $\log L_{\rm FIR}\approx \log (1.2 \times F_\nu (850\,\mu{\rm
m}))+12$, where $F_\nu (850\,\mu{\rm m})$ is the flux density at 850
$\mu$m, expressed in mJy. If 850 $\mu$m measurements are not
available, we use the 1200 $\mu$m measurement instead (in this case,
$\log L_{\rm FIR}\approx \log (F_\nu (1200\,\mu{\rm m}))+12.36)$).

\subsection{Role of CMB}
\label{cmb}

The temperature of the CMB at redshift $z$ is given by: $ T_{CMB}(z)=
2.73\times(1+z)$. At a redshift of z\,=\,6, $T_{CMB}$\,=\,19.1\,K,
meaning it is warmer than the dust temperature of Milky Way at
present. As the dust will always be heated by photons of the CMB
(through absorption), the minimum dust temperature will be that of the
CMB (and the kinetic temperature of the gas will equal the thermal
dust temperature at high enough densities, da Cunha et al.\ 2013). As brightness temperature
measurements are by definition always done with respect to the
background (i.e. the CMB) it will be increasingly difficult to detect
cold dust emission at high redshift.

The molecular gas is affected in two ways by the CMB: First, the
higher temperature of the CMB will lead to an increase of the line
excitation, and thus the line luminosities (e.g., Obreschkow et al.\
2009a, da Cunha et al.\ 2013). Second, the observing background
against which the line is measured also increases. The relative effect
of the CMB will depend on the intrinsic excitation temperature of the
gas (Combes et al.\ 1999, Obreschkow et al.\ 2009a, da Cunha et al.\
2013). For example, for molecular gas at a temperature much higher
than the CMB, the relative decrease in line flux will be much lower
than in a case where the molecular gas temperature is almost equal to
the CMB. As a result of this, the shape of the observed CO ladder for
a source with given temperature and density will also change as a
function of redshift: The peak of the SED will shift to higher--J
values as the lower--J transitions are most easily affected by the
effects of the CMB (da Cunha et al.\ 2013).  A suppressed measurement of the CO lines also
implies that the conversion factor $\alpha$ will change as a function
of CMB temperature.

{\bf SIDEBAR: It should be noted that all CO detections at very high
  redshift (z$>$5), i.e. where T$_{\rm CMB}$ is significant, are to
  date from galaxies that harbor hyper--star forming environments
  (e.g. the z$\sim$6 quasars and the few known SMGs at z$>$5 have SFRs
  of $\sim$1000\,M$_\odot$\,yr$^{-1}$) with accompanying high
  temperatures of the ISM, ranging from 35--50\,K, or at least a
  factor two warmer than the CMB.  However it is clear that in lower
  temperature environments, more typical of lower luminosity star
  forming galaxies, the consequences of the CMB will be significant:
  the lowest--J lines will be faint at z$>4$, and other cool ISM
  tracers may be required, such as \cii.}

\section{Molecular gas at high redshift}
\label{high_z_summary}

\subsection{Introduction}

Historically, different groups have been focusing on different
selection techniques in their searches for molecular gas in high
redshift systems. This has lead to different source categories and
types which we will broadly follow for simplicity. But it is important
to point out that there is major overlap in the properties of the
sources discussed in the following. Many high--redshift quasars are
found to be bright in the submillimeter regime, but they are
classified as `quasars' not `submillimeter galaxies' as they have been
originally selected as being quasars. Likewise, `submillimeter
galaxies' (hereafter: SMGs) have typically been selected in blind
submm surveys of the sky, but some show signatures of a quasar.
However they are still typically referred to as SMGs as they have been
discovered as such (e.g., Ivison et al.\ 1998; Alexander et al.\
2005).

Considering telescopes, most work has been done the IRAM Plateau de
Bure interferometer (PdBI), the Very Large Array (VLA) and it
successor (EVLA/JVLA), the Owen's Valley Radio Observatory (OVRO) and
its successor CARMA (Combined Array for Research in Millimeter
Astronomy), the Australia Telescope Compact Array (ATCA), and the IRAM
30\,m, James Clerk Maxwell Telescope (JCMT), 100\,m MPIfR Effelsberg,
NRAO Green Bank (GBT) and Nobeyama Radio Observatory (NRO) single dish
telescopes.  These facilities have under gone a series of improvements
in receiver and antenna performance, as well as correlator upgrades
leading to dramatic improvements in bandwidth. The latter has been
particularly important in detecting the often broad lines seen in
high--redshift galaxies in the absence of precise redshifts.  Also,
customized receivers on single--dish telescopes have been built with
very large bandwidths to blindly detect CO emission in high--redshift
galaxies, eg. the ZSPEC on the CSO (Gromke et al.\ 2002), and the
ZSPECTROMETER on the GBT (Harris et al.\ 2010). These systems have
lead to successful blind searches, aided in part by strong lensing of
galaxies discovered in recent very wide--field Herschel surveys.

Figure~\ref{fig_co_coverage} shows a plot of the frequency coverage
for CO and other molecules for some available telescopes, with all
actual measurements superposed. This plot emphasizes the
complementarity of the centimeter and millimeter telescopes to probe a
broad range in molecular gas and atomic fine structure transitions,
providing a detailed diagnostic suite to study the ISM in early
galaxies.

\subsection{Quasars}
\label{quasars}

Quasars were the first targets for submm continuum observations at
high redshift (e.g. Omont et al.\ 1996). Extensive subsequent work has
shown that 1/3 of optically selected quasars are detected in submm
continuum observations with mJy sensitivity, and this fraction remains
roughly constant from $z \sim 1$ to 6 (e.g., Wang et al.\ 2008a,b; Beelen
et al.\ 2006; Priddey et al.\ 2003).  Searches for CO emission have
systematically detected CO in submm--detected quasar samples. More
recently, \cii\ is now being detected in quasar host galaxies,
including the most distant quasar known with a spectroscopic redshift
at $z$\,=\,7.1 (Venemans et al.\ 2012).

As we shall see below (Sec.~\ref{excitation}), of all objects detected at
high redshifts, quasar host galaxies show the highest excitation for
the molecular gas (Barvainis et al.\ 1997, Wei\ss\ et al.\ 2007b,
Riechers et al.\ 2009a) and, in the few cases where the emission could
be resolved, the CO is distributed in compact ($\le$ few kpc) gas reservoirs
(Walter et al.\ 2004, Riechers et al.\ 2008a, 2008b, 2009b, 2011a;
Carilli et al.\ 2002). The resulting star formation rate surface densities
can thus be very high ($\sim$1000 M$_\odot$\,yr$^{-1}$ kpc$^{-2}$,
Walter et al.\ 2009b).  This suggests the presence of a coeval starburst
with the SMBH growth (Walter et al.\ 2004, Wang et al.\ 2010, Coppin et
al.\ 2008). Typically the gas excitation in the quasar host galaxies
can be modeled with one gas component only, and there is no evidence
for extended molecular gas reservoirs around these objects (Weiss
et al.\ 2007a, 2007b, Riechers et al.\ 2009, Riechers 2011a).

The molecular medium in quasar host galaxies is amongst the best
studied at high redshift, and typical molecular gas masses are a few
$\times$ 10$^{10} (\alpha/0.8)$ M$_\odot$, with resulting short implied
gas consumption times $\sim$10$^7$\,yr (Sec.~\ref{sf_law}). Evidence for
molecular outflows in quasars has recently been presented in two cases
(Sec.~\ref{outflows}).

\subsection{Submillimeter galaxies}
\label{smgs}

Submillimeter galaxies were classically selected from the
(sub--)millimeter maps obtained at 850\,$\mu$m by SCUBA at the JCMT and
at 1.2\,mm with MAMBO at the IRAM 30m (e.g. Smail et al.\ 1997, Hughes
et al.\ 1998, Ivison et al.\ 2000, 2007, Dannerbauer et
al.\ 2004). Given the sensitivities of the sub--millimeter cameras of
typically $\ge$1\,mJy, detections are by definition Hyperluminous
infrared galaxies (HyLIRGs), with $L_{\rm FIR} \sim 10^{13}$ L$_\odot$,
implying star formation rates $\sim$1000 M$_\odot$\,yr$^{-1}$. Early
searches for molecular gas emission then revealed large gas reservoirs
(Frayer et al.\ 1998, Frayer et al.\ 1999 using OVRO) and extensive campaigns, in
particular done at the PdBI (Neri et al.\ 2003, Greve et al.\ 2005,
Tacconi et al.\ 2006, 2008, Engel et al.\ 2010, Bothwell et al.\ 2010,
2013) have characterized the molecular interstellar medium in
exquisite detail. In a few cases the molecular reservoirs could be
resolved (Ivison et al.\ 2010a, 2011, Riechers et al.\ 2011b, Carilli et
al.\ 2011, Hodge et al.\ 2012), leading to substantial sizes of
$\sim$10 kpc in a number of galaxies, although compact CO emission has
been seen in a few cases as well (Carilli et al.\ 2002a; Tacconi et
al.\ 2008). In all cases, a common trait for SMGs is extremely high
optical extinction in the main regions of star formation traced by
CO and thermal dust emission. This trait is accentuated by the case of
the first SMG discovered in a submm deep field, HDF\,850.1 (Hughes et
al.\ 1998), which still remains unidentified in the deepest optical and
near--IR images (Walter et al.\ 2012a).

The average excitation of the molecular gas in SMGs is less extreme
than in the quasars (e.g. Wei\ss\ et al.\ 2007b, see
Sec.~\ref{excitation}), which may be attributed to the fact that the
star formation in some SMGs proceeds on more extended scales than in
the compact quasar hosts. Typical gas reservoir masses are of order a
few $\times$ 10$^{10}(\alpha/0.8)$ M$_\odot$ which implies that the
gas consumption times are short, of order $\sim$10$^7$ yr
(similar to the quasars). Even though the redshift distribution of
SMGs is thought to peak around z\,$\sim$\,2.5 (Chapman et al.\ 2003),
there is a significant tail towards higher redshift (Schinnerer et
al.\ 2008, Daddi et al.\ 2009a, Riechers et al.\ 2010a, 2013; Coppin et
al.\ 2009, Walter et al.\ 2012a; Younger et al. 2007; Yun et al. 2012).

FIR--bright galaxies have recently also been blindly detected by large
areal mapping with Herschel (sometimes referred to as
`Herschel--selected galaxies' or HSGs) at shorter wavelengths than in
the original SCUBA/MAMBO selection (Negrello et al.\ 2010) and thus
have a different selection function (either warmer dust temperatures
or lower redshifts) than the traditional SMG selection (see also Greve
et al.  2008). The Herschel selected galaxies are also typyically a
factor few lower in FIR luminosity than classical (SCUBA-selected)
SMGs.  These galaxies have also been shown to harbor massive
reservoirs of gas (Cox et al.\ 2011, Combes et al.\ 2012a, Riechers et
al.\ 2011a, Harris et al.\ 2012). A significant fraction of the
brightest HSGs are gravitationally lensed.

The SMG surveys and related work showed a 1000--fold increase in the
space density of ULIRGs from z\,=\,0 to z$\sim$2.5 (Hughes et al.\ 1998,
Blain et al.\ 2002). Like in the case of the quasar
host galaxies, it has been proposed that the SMGs pinpoint the
formation of a massive galaxy at high redshift (Swinbank et al.\ 2006,
Daddi et al.\ 2009a,b, Hickox et al.\ 2012).

Like low--z ULIRGs, QSOs and SMGs lie significantly above the
so--called SFR--M$_\star$ `main sequence' (Sec.~\ref{sf_law},
Sec.~\ref{intro_notes})), however, they have other properties that are
often dissimilar to nearby nuclear starbursts, such as more extended
gas disks in many cases (Sec.~\ref{imaging}), cooler average dust and
gas temperatures, and disk--like fine structure line ratios
(Sec.~\ref{fsl}).  Overall, SMGs are certainly extreme starbursts, but
they are likely a heterogeneous population, including compact
starbursts in gas--rich major mergers, massively accreting disk
galaxies, and enhanced star formation likely due to gravitational
harrassment in dense proto--cluster regions (Hayward et al. 2012;
Hodge et al. 2012).

\subsection{Radio Galaxies} 
\label{radiogals}

Radio galaxies are identified in wide--field radio surveys at cm
wavelengths, and the radio emission is related to AGN jet activity,
typically on scales $>\,10$kpc.  Radio galaxies were the first very
high redshift galaxies discovered, and they remain the best beacons to
massive, clustered galaxy formation at high redshift (Miley \& de
Breuck 2008). In the standard AGN unification model, radio galaxies
are simply radio loud quasars seen with the jets closer to the sky
plane, such that the broadline region is obscured by the accreting
dusty torus (so--called type--II AGN).

Radio galaxies were also among the first high--redshift sources in
which molecular gas was detected (Scoville et al.\ 1997, Papadopoulos
et al.\ 2000, De~Breuck 2003a,b, 2005, Greve et al.\ 2004). Like some
quasars, radio galaxies are often bright in (sub)millimeter continuum
emission and have similar gas masses to the quasars and SMGs. CO
imaging of radio galaxies often reveals multiple components on tens of
kpc scales, likely indicating major gas--rich mergers (De Breuck
2003a,b, 2005, Emonts et al.\ 2011). The most recent imaging study by
Ivison et al.\ (2012) shows multiple, gas rich components over tens of
kpc, indicating a merging, starburst proto--cluster environment.
 
\subsection{Color Selected Star--Forming Galaxies (CSG)}
\label{cssfg}

Major progress was achieved in recent years in star forming galaxies
selected via their optical or near--IR colors at z$\sim$1.5 to
3. These galaxies have been selected in three different ways. The
first strategy was through the near--IR BzK color selection, which
selects galaxies by their 4000\AA\ break (Daddi et al.\ 2004).  In
2008, Daddi et al.\ demonstrated that this color selection
successfully selects gas--rich galaxies.  Daddi et al.\ (2010) then
used this selection technique to target 6 sBzK galaxies with stellar
masses $> 10^{10}$ M$_\odot$, that were also detected in the radio
(but not the sub--millimeter). All  galaxies were detected in CO
emission, implying large reservoirs of molecular gas in galaxies that
are not forming stars at the extreme rates seen in quasars and SMGs
(Daddi et al.\ 2008; 2010).

A second color--selected sample is the rest--frame UV color selection
technique, the BM/BX selection (Steidel et al.\ 2004).  Tacconi et
al.\ (2010)  used this technique to identify a sample of CSG at
z$\sim$2 for CO observations. They also detect large gas reservoirs in
the majority of their sample. In the following we use the term CSG
(color--selected star forming galaxies) for both the BzK and BM/BX,
and related, color selection techniques. 

A third color--selected sample is the Star--Forming Radio--selected
Galaxies (SFRGs). These objects are rest--frame UV--color selected to be
$z > 1$ glaxies, and then further identified as 20\,cm radio continuum
sources of $\sim 50$ to 100$\mu$Jy but are not detected in the
sub--millimeter.  Chapman et al.\ (2008) and Casey et al.\ (2011)
detected molecular gas reservoirs with masses of $\sim$10$^{10}$
M$_\odot$ in about half their sample (which includes a few BzK galaxies
studied by Daddi et al.\ 2008). Indeed, it has been found that there is
significant overlap between the various color selection techniques
(Grazian et al.\ 2007).

The term `normal star forming galaxies' at high redshift is often used
for these CSG. The intrinsic star formation rates are high, $\ge 100$
M$_\odot$\,yr$^{-1}$ (Daddi et al.\ 2010, Tacconi et al.\ 2010),
however, these galaxies follow a similar CO to FIR luminosity ratio as
low redshift spirals (Sec.~\ref{sf_law}), and they lie on the sSFR
`main sequence', and as described in Sec.~\ref{intro_notes}. Moreover,
galaxies of this FIR luminosity make a dominant contribution to the
overall cosmic SFR during the epoch of galaxy assembly
(Sec.~\ref{intro_notes}).

Further studies of the CSG galaxy population has shown that they are
extended on 10~kpc scales in gas and stars (Daddi et al.\ 2010,
Tacconi et al.\ 2010; Tacconi et al. 2012; Sec.~\ref{imaging}) and
that they have gas excitation that is lower than what is found in
quasar hosts or SMGs (Dannerbauer et al.\ 2009; Aravena et al. 2011;
see Sec.~\ref{excitation}). However, such measurements are currently
restricted to low--J lines only (up to J\,=\,3).

\subsection{MIPS/24 micron--selected Galaxies}

Spitzer 24\,$\mu$m--selected galaxies at z$\sim$2 have been studied in
molecular gas emission by Yan et al.\ (2010). At the typical redshift
of the sample, the 8\,$\mu$m PAH feature, a tracer for star formation,
is shifted in the 24 micron band. Even though these sources are bright
at 24 $\mu$m and are thus forming stars at high rates, they are mostly
undetected at 1.2\,mm using MAMBO. Still they often have CO
luminosities comparable to SMGs. These sources often show evidence for
mergers, and some host dust--obscured AGNs (Yan et al.\ 2010, see also
Iono et al.\ 2006a). This population is likely a hodgepodge of
sources, including type 2 quasars, SMGs and hot dust sources.

\subsection{Lyman--Break Galaxies, Ly$\alpha$ Emitters, and Ly$\alpha$
Blobs}

Progress has been made on detecting CO emission from strongly lensed
LBGs, corresponding to color-selected galaxies at $z \ge 3$. Baker et
al.\ (2004) and Coppin et al.\ (2007) detected CO 3-2 emission in two
LBGs at $z \sim 3$ that are lensed by a factor of 30. Riechers et
al. (2010b) detect CO 1-0 emission in these two systems, and find
relatively low excitation, and gas masses $\sim 4 \times$
10$^{9}(\alpha/4)$ M$_\odot$, after correcting for
lensing.  The first detection of a more massive, unlensed LBG at $z=
3.2$ has been reported in Magdis et al.\ (2012a).

CO emission has also been searched for in high--z Ly$\alpha$ emitting
galaxies (LAE) and Ly$\alpha$ emitting `blobs'. Observations of one
strongly lensed LAE at $z$\,=\,6.5 resulted in a stringent upper limit
to the gas mass of $4.9\times 10^9 (\alpha/0.8)$ M$_\odot$ (Wagg et
al.\ 2009). Ly$\alpha$ blobs are typically large (tens of kpc), and
their exact origin remains uncertain, possibly being the remnants of
radio--mode feedback as seen in the Ly--$\alpha$ halos around powerful
radio galaxies (Miley \& De Breuck 2008), or even representing cooling
gas in dense filaments from the IGM (Dekel et al.\ 2009). Searches for
CO emission have led to non--detections at $z \sim 3$ (Yang et
al.\ 2012), and $z$\,=\,6.6 (Wagg \& Kanekar 2012), with gas mass
limits of order $10^{10} (\alpha/0.8)$ M$_\odot$.

The advent of ALMA early science has opened the very real and exciting
prospect of detecting \cii\ 158$\mu$m emission from typical LBGs and
LAEs at very high (z$>$6) redshift. Even with limited capabilities
(1/3 of the final array), a number of typical LBGs and LAEs have
already been detected with relative ease in \cii\ at $z \sim 5$
(Riechers et al. 2013; Wagg et al. 2012; Carilli et al. 2012;
Sec.~\ref{fls_cii}).

\subsection{Table of all high--z ISM detections}

We have compiled a table of all high--redshift (z$>$1) detections of
the molecular interstellar medium. Most of the detections are of the
CO line, but there are also detections of higher--density molecular
gas tracers (observed only in quasar hosts), such as CN, HCN, HNC and
HCO$^+$ and the CO isotopomer $^{13}$CO. For many sources, multiple--J
CO detections are available. The table also includes measurements of
the atomic fine structure lines, most notably \cii, but also \nii,
\oiii\ and \ci. For convenience we have also added the FIR
luminosities of these sources, derived through heterogenous ways (see
footnotes in the table), following the equations in
Sec.~\ref{dust}. The table will be available online through the homepage of
the Annual Reviews in Astronomy and Astrophysics.

In total, the interstellar medium has now been detected in close to
200 sources at $z>1$.  In the following we summarize the basic
observational parameter space (CO line luminosities and excitation,
full width half maximum, sub--millimeter continuum emission, redshift
distribution) through some key plots to interpret this rich set of
observational data. In all plots we use the same color coding for
different groups of galaxies (as defined in
Sec.~\ref{high_z_summary}).  These are the QSOs (quasars), SMGs
(submillimeter--selected galaxies, including one Extremely Red Object
(ERO) and Herschel--selected galaxies, HSG), CSGs (color--selected
star forming galaxies, selected through BzK and BM/BX selection
techniques), RG (radio galaxies), SFRG (star forming radio galaxies),
Spitzer 24$\mu$m selected galaxies and LBG (Lyman--break
galaxies). The classification is therefore entirely based on the
selection technique, not on the actual physical properties.

\subsection{Historical note}

Since the earliest detection of molecular gas in the $z$\,=\,2.28
quasar IRAS\,F\,10214 by Brown \& Vanden Bout (1992) and Solomon et
al.\ (1992) there has been a steady increase in the number of line
detections at high redshift. This is illustrated in the left panel of
Figure~\ref{fig_history} where the cumulative number of detections is
presented as a function of discovery year. From this plot it is clear
that over the first decade or so, the detections were dominated by
quasars, with SMGs picking up in larger numbers around the turn of the
millennium. Only recently have CSG been added to the list in
significant numbers. The right panel of Figure~\ref{fig_history} shows
the redshift distribution of all galaxies with a line detection.

\section{Observational diagnosis of the cool ISM in distant galaxies}
\label{diagnosis}

\subsection{Molecular gas excitation}
\label{excitation}

As noted in Sec~\ref{model_excitation} the excitation of molecular gas
in high--redshift galaxies provides important information regarding
the average physical properties of the gas, in particular its
temperature and density. This is illustrated in Fig.~\ref{fig_co_seds}
where the expected CO excitation is shown as a function of density and
temperature. One important caveat is that typically only
galaxy--integrated line fluxes can be measured at high redshift. These
measurements are then fed into LVG (Sec.~\ref{lvg_models}) or PDR/XDR
(Sec.~\ref{pdr_xdr_models}) models that assume that the line emission
is emerging from the same cloud/volume (i.e. that they share the same
physical conditions). The validity of this assumption is
questionable, as it is known that the different rotational transitions
have different critical densities: the higher--J transitions need much
higher densities than the low--J ones to be excited. This situation
complicates further if unresolved measurements of high--density
tracers (e.g. HCN, HCO$^{+}$) are added to the `single component'
models. Another caveat is that the molecular gas excitation measurements in
high redshift galaxies are typically restricted to the mid--J levels
of CO emission. 

With these caveats in mind there are a number of important results
emerging from high--redshift CO excitations analyses. The main result
is that the different source populations at high redshift show
distinctly different excitation (see compilation of all high--redshift
CO excitation in Fig.~\ref{fig_co_sed_all}). The CO
excitation of quasar host galaxies can in essentially all cases be
modeled by a simple model with one temperature and density out to the
highest--J CO transition with a typical gas density of
log(n(H$_2$)[cm$^{-3}$])\,=\,3.6--4.3, and temperatures of
$T_{kin}=40-60$\,K. A prominent example is the highly excited quasar
host APM\,08279+5255 (Wei\ss\ et al.\ 2007a). Measurements of the CO
ground transition in a number of quasar host galaxies (Riechers et al.\ 2011a) 
have shown that the measured CO(1--0) flux is
what was expected based on a single--component extrapolation from
higher--J CO transitions, i.e. there is no excess of CO(1--0) emission
in these sources (Riechers et al.\ 2006a, 2011a). One interpretation of
this finding is that the molecular gas emission comes from a very
compact region in the centre of the quasar host, which is confirmed by
the few (barely) resolved measurements of quasar hosts (Sec.~\ref{imaging}). It
should be noted though that there is recent evidence that an additional
high--excitation component, likely related to the AGN itself, is
needed to explain the elevated high--J line fluxes in some sources,
e.g. PSS\,2322+1944 (Wei\ss\ et al., in prep.) and J\,1148+5251
(Riechers et al., in prep.).

The submillimeter galaxies, on the other hand, show (i) less excited
molecular gas, and (ii) excess emission in the CO(1--0) ground
transition. This is again illustrated in Fig.~\ref{fig_co_sed_all},
where the orange symbols of the SMGs are on average at lower fluxes
compared to the quasars (red symbols). On average, the typical density
of SMGs is log(n(H$_2$)[cm$^{-3}$])\,=\,2.7-3.5, and temperatures are
in the range of T$_{kin}=30-50$\,K. Observations in the CO(1--0) line
of a few SMGs have furthermore demonstrated that an additional cold
component is needed to explain the observed excitation (Ivison et al.\
2011, Riechers et al.\ 2011b). This implies that the total gas mass of
the SMGs is underestimated if mid--J CO transitions are used to
calculate masses assuming constant brightness temperature (see
Sec.~\ref{alpha_highz}). A few SMGs have been resolved spatially and
show more extended emission in the ground transition than in higher
transitions (Ivison et al.\ 2011; although cf. Hodge et al. 2012).

Up to the mid--J measurements available for high redshift sources the
CO excitation of the SMGs and QSOs roughly follow those of the centres
of nearby starburst galaxies (e.g., M\,82: Mao et al.\ 2000; Ward et
al.\ 2003, see dashed line in Fig.~\ref{fig_co_sed_all}; NGC\,253:
Bradford et al.\ 2003; Bayet et al.\ 2004, and Henize 2-10: Bayet et
al.\ 2004). It should be noted that the excitation in Galactic
molecular cloud cores can be equally high (e.g. Habart et al.\ 2010,
Leurini et al.\ 2012; van Kempen et al.\ 2010, Manoj et al.\
2012). However, in Galactic clouds such very high excitation is only
seen in pc scales cloud cores, while in nearby starburst nuclei high
excitation is seen on scales of order 100 pc scales, and for
high--redshift galaxies the scale can extend to a few kpc.

As stated in Sec.~\ref{high_z_summary}, the seperation between SMGs
and QSO host galaxies is largely due to historical selection effects
and there are sources that fullfill both definitions. The different
excitation conditions in the two groups however argue that broadly
speaking gas--rich quasars and SMGs represent different stages of
galaxy evolution, with the quasars being more compact and harboring
more highly excited gas than the SMGs. In a simplistic picture, the
quasars could be the result of a merger, in which the molecular gas
concentrates in the centre of the potential well, while the
SMGs would then constitute merging systems that have not fully
coalesced. The latter is supported by the fact that high--resolution
observations of many SMGs show multiple emission components
(e.g. Engel et al.\ 2010, Tacconi et al.\ 2010). SMGs and quasars 
also show similar clustering
properties (Hickox et al. 2012).  

To date, only a few excitation measurements exist for the CSGs
(Dannerbauer et al.\ 2009; Aravena et al.\ 2010), and only up to CO
3-2.  In the case of BzK--21000 Dannerbauer et al.\ (2009) showed that
the J\,=\,3 emission must be significantly subthermally excited,
resembling more that of the Milky Way than the higher--excitation
systems discussed above. These CSGs have lower star formation rates
than found in the quasars and SMGs, and have large ($\sim$10\,kpc)
sizes -- the combination of these apparently leads to a less extreme
gas excitation than found in the hyper-starburst galaxies at high
redshift. Observations of higher order CO transitions are required in
these systems.

Bothwell et al. (2012) suggest that there is a correlation between CO
luminosity and line width in high redshift starbursts, likely relating
to baryon-dominated gas dynamics within the regions probed.  In
Fig.~\ref{fig_l_fwhm}, we plot the CO line widths (FWHM) vs.\ CO
luminosity for a broader range of CO detected distant galaxy
types. Considering only the hyperstarbursts (SMGs and quasar hosts),
both occupy similar parameters space in this diagram, and there is no
significant correlation between CO luminosity and linewidth.  However,
the CSG clearly segregate to smaller line widths relative to SMGs, by
about a factor $\sim$two, and, while the scatter is large, the median CO
luminosity for CSGs is smaller by a similar factor. 

\subsection{CO luminosity to total molecular gas mass conversion}
\label{alpha_highz}

Until a few years ago, the standard practice for high redshift
galaxies was to use the starburst conversion factor to derive gas
masses.  This practice was justified in part on the extreme
luminosities, high inferred gas densities based on CO excitation, and
in some cases on the direct observation of compact star forming
regions (scales $\sim 2-4$ kpc; Tacconi et al.\ 2008, Momjian et al.\
2007, Ivison et al.\ 2010a, 2012, Carilli et al.\ 2002,
Walter et al.\ 2009). Also, in some cases it appeared that, like
nearby nuclear starbursts, the gas mass exceeded the dynamical mass
when using a Milky Way conversion factor (Solomon et al.\ 1997,
Tacconi et al.\ 2008, Carilli et al.\ 2010, Walter et al.\ 2004,
Riechers et al.\ 2008, 2009).

As other populations of galaxies are now being detected in CO emission
at high redshift, such as color--selected star forming galaxies
(Sec.~\ref{cssfg}), the question of the conversion factor has become
paramount.  The last few years have seen a few attempts at the first
direct measurements of the CO luminosity to gas mass conversion
factor.

A minimum $\alpha$ can be derived assuming optically thin emission,
and adopting a Boltzmann distribution for the population of states. In
this case, the CO(1--0) luminosity, L$_{\rm CO(1-0)}$, simply counts the
number of molecules in the J\,=\,1 state via: $\rm N_{J=1} =
L_{\rm CO(1-0)}/(A \times h\nu)$, where A is the Einstein A coefficient
($\rm 6\times 10^{-8}~ s^{-1}$), and $\nu = 115$ GHz. The total number
of CO molecules is then determined by the partition function, ie. the
relative fraction in the J\,=\,1 state, which we designate f. For
example at T\,=\,10, 20, and 40 K, f\,=\,0.47, 0.31, and 0.19,
respectively.  One can then convert to total number of H$_2$ molecules
assuming a CO/H$_2$ abundance ratio, which for Galactic GMCs is
10$^{-4}$.  Multiplying by the mass of an H$_2$ molecule, and
including a factor 1.36 for the Helium abundance, leads to the total
gas mass. Converting to the relevant units, leads to: $\rm \alpha =
(0.09/f)$ M$_\odot$ [K\,km\,s$^{-1}$ pc$^{2}$]$^{-1}$, at
$z$\,=\,0. \footnote{Including a correction for the relative
contribution of the CMB increases these values by about 35\% at $z =
2.5$ (Ivison et al.\ 2011).}
  
Tacconi et al.\ (2008) use spatially resolved spectral imaging to
derive either virial or rotational masses for a sample of SMGs and
color selected galaxies. Galaxy sizes are estimated from the CO
imaging and/or optical or near--IR imaging (see also Neri et al.\ 2003;
Thomson et al.\ 2012). Tacconi et al.  compare the dynamical masses
with the sum of the gas mass (modulo $\alpha$) and stellar mass
(derived from near--IR photometry), including a small correction (10\%
to 20\%) for dark matter. They conclude that for the SMGs, the value
of $\alpha$ is most likely close to the starburst value ($\alpha \sim
1$), while for the CSG a value $\alpha \sim 4.8$ is favored. 

Daddi et al.\ (2010) have performed a similar analysis on CSGs, based
on (marginally) spatially resolved imaging with the PdBI of CO 2--1
emission of a sample of three $z \sim 1.5$ sBzK galaxies. Their
dynamical analysis is guided by numerical simulations of clumpy,
turbulent disks, to allow for significant non--circular motions
(Bournaud et al.\ 2009). They find that the measurements are
consistent with $\alpha_{\rm CO} = 3.6 \pm 0.8$ M$_\odot$/(K km s$^{-1}$
pc$^2$).

The best rotational dynamical analysis of an SMG to date is the JVLA
observation of GN\,20 at $z$\,=\,4.0 (Fig.~\ref{fig_CO_im}) by Hodge
et al.\ (2012; see Sec.~\ref{imaging}). They show that the CO(2--1)
emission is well characterized by a rotating disk of 7\,kpc radius
with a large rotation velocity of $575 \pm 100$ km s$^{-1}$ and
internal velocity dispersion of 100 km s$^{-1}$.  They derive a
dynamical mass of $5.4\pm 2.4 \times 10^{11}$ M$_\odot$. Subtracting
the stellar mass, including a 15\% dark matter contribution leads to
$\alpha \sim 1$ M$_\odot$/(K km s$^{-1}$ pc$^2$).

The main uncertainties in the dynamical analysis are the current crude
estimates of the dynamical masses based on marginally resolved imaging
data, and the standard pit--falls in estimating the stellar masses
based on SED fitting, in particular for starburst--type systems (Ivison
et al.\ 2011). 

Magdis et al.\ (2011) use a metallicity--dependent dust--to--gas ratio
approach to estimate $\alpha_{\rm CO}$. They analyze two galaxies in the
GOODS--N deep field with excellent rest--frame IR SEDs, one SMG
(GN\,20), and one CSG (BzK 21000).  From these, and using the Draine
\& Li (2007) dust models, they derive a dust mass for each
source. Metallicities are derived from the stellar mass -- metallicity
relation for the CSG (Erb et al.\ 2006), and a starburst model for
GN\,20.  They then use the Solomon et al.\ (1997) metallicity
dependent dust to gas relation to obtain an estimate of the gas masses
of $9\times 10^{10}$ M$_\odot$ for the CSG and $1.5\times 10^{11}$
M$_\odot$ for the SMG.  Comparing these masses to the CO(1--0)
luminosities leads to $\alpha \le 1.0$ for the SMG, while for the
CSSGF they find $\alpha \sim 4$, consistent with earlier
estimates. These results have been generalized in the more
comprehensive study by Magdis et al.\ (2012b).

Ivison et al.\ (2011) take two different approaches to derive
molecular gas masses, and hence infer $\alpha$ in SMGs. First, they
propose a model in which some fraction of the CO emission occurs at
the `Eddington limit' in dense star forming regions (a `maximal
starburst'), ie. a self--gravitating disk supported by
starburst--driven radiation pressure on the dust grains (Thompson et
al.\ 2005; Thompson 2009; Krumholz \& Thompson 2012). Thompson (2009)
derive a maximum star formation efficiency in this case of: $\rm
SFE_{max} \equiv L_{\rm IR}/M_{gas} < 500$ L$_\odot$ M$_\odot^{-1}$,
and an areal star formation rate density of 1000 M$_\odot$\,yr$^{-1}$
kpc$^{-2}$.  This maximum surface density is seen to hold on scales
from star forming cloud cores in GMCs ($\sim 1$pc) to the nuclear
starburst regions of low--z ULIRGs.

Ivison et al. then make a correction to the total gas for the fraction
of cold, quiescent gas not involved in active star formation based on
the CO excitation.  Under these assumptions, the molecular gas mass
can be estimated from the IR luminosity.  For 4 of their 5 SMGs, the
values are between $\alpha$\,=\,0.4 and 0.7, while for the last source
the value is 3.7. This latter source also has by far the lowest CO
excitation.

Second, Ivison et al.\ (2011) use radiative transfer modeling to
derive physical conditions within the clouds, and infer gas masses in
a manner analogous to the low redshift ULIRG analysis of Downes \&
Solomon (1998).  Unfortunately, LVG models have a fundamental
degeneracy between density, temperature, and non--virial gas
kinematics, when using just CO excitation
(Sec.~\ref{lvg_models}). They find the CO SEDs are consistent with
either $\alpha \sim$ 4.5 in cold, dense gas with large non--virial
motions, or $\alpha \sim 1$ in warm, more diffuse gas closer to virial
dynamics.

Genzel et al.\ (2012) consider the effect of metallicity
on the conversion factor $\alpha$ in a sample of CSG at $z \sim 1$
to 2.  They have obtained estimates for the metallicities from optical
\nii/H$\alpha$ measurements. They find that galaxies with similar
sizes, SFRs, and dynamical masses, but different metallicities (and
stellar masses; Erb et al.), show very different star formation
efficiencies, with the CO to FIR luminosity ratio decreasing
dramatically with decreasing metallicity, at fixed FIR luminosity. If
$\alpha$ is the same in each system, then this would imply a much
shorter gas consumption timescale (gas mass/SFR) in low metallicity
galaxies vs.\ high metallicity galaxies.

Alternatively, Genzel et al.\ (2012) propose adopting a standard star
formation efficiency relating gas mass to star formation rate
(Sec.~\ref{sf_law}).  In this case, the decrease in CO luminosity is
not due to a lower gas mass but an increase in $\alpha$ at low
metallicity.  Such a trend is expected and observed at low metallicity
in nearby galaxies (Sec.~\ref{alpha_lowz}), since dust shielding
becomes less efficient, and CO in molecular cloud envelopes are
photo--dissociated much more deeply into the cloud (Leroy et al.\
2011; Bolatto et al.\ 2013, Schruba et al.\ 2012). Genzel et al.\
2012; Stacey et al. 1991) derive the empirical relationship: $\rm
log(\alpha_{\rm CO}) = 12 - 1.3\times[12 + log(O/H)]$. The implication is
that stars can form in CO--poor molecular gas, and that the expected
CO luminosities of low metallicity galaxies at high redshift are lower
than previously predicted based on standard $\alpha$ values (Glover \&
Clark 2012; Papadopoulos \& Pelupessy 2010).

Magnelli et al. (2012) perform a similar analysis of the conversion
factor based on dust-to-gas ratios, for a mixed sample of SMGs and CSG
at $z = 1$ to 4. They adopt a similar dust mass calculation based on
the FIR luminosity, and a metallicity-dependent dust-to-gas ratio as in
Genzel et al. (2012). Their results are consistent with a low value of
$\alpha$ for starbursts, and a factor five higher value for CSG. They
also find a trend of decreasing $\alpha$ with increasing dust
temperature, and while the trend may be continuous, they recommend
that for a dust temperatures $< 30$K, a Milky Way conversion factor is
appropriate, while for higher dust temperatures the starburst value
should be used, at least for solar metallicity systems. 

{\bf SIDEBAR: Many conclusions in astronomy are not based on a single,
absolutely compelling direct measurement, but on a series of
measurements, modeling, and consistency arguments from which the
weight of evidence leads to a `concordance model'. This approach
applies to the the CO--to--gas mass conversion factor in distant
galaxies. The current measurements of $\alpha_{\rm CO}$ are principally
based on dynamical imaging and modeling, dust--to--gas measurements,
radiative transfer modeling, and most recently, on star formation
efficiencies. These measurements suggest a nuclear starburst value
$\sim 0.8$ M$_\odot$/(K km s$^{-1}$ pc$^2$)$^{-1}$ for SMGs and quasar
hosts, and a Milky Way GMC value of $\alpha_{\rm CO}\sim 4$ M$_\odot$/(K
km s$^{-1}$ pc$^2$)$^{-1}$ in CSG, with a likely further increase in
low metallicity galaxies. These values are consistent with the
generally extended, clumpy disk--like CO distribution, lower CO
excitation, and Milky Way CO/FIR (FIR derived from SFR) ratios in CSG,
compared to the often (although not exclusively) more compact,
merger--like morphologies, higher excitation, and starburst CO/FIR
ratios in SMGs and quasars.}

\subsection{Atomic Fine Structure lines}
\label{fsl}

\subsubsection{Atomic Carbon: \ci}
\label{fsl_ci}

In a recent study, Walter et al.\ (2011) summarize the \ci\
observations of 13 high--redshift galaxies (this compilation included
work by Barvainis et al.\ 1997, Wei\ss\ et al.\ 2003, 2005a, Pety et
al.\ 2004, Wagg et al.\ 2006, Ao et al.\ 2005, Riechers et al.\ 2009a, Casey
et al.\ 2009, Lestrade et al.\ 2010, Danielson et al.\ 2011).  These
systems studied in \ci\ are amongst the brightest emitters both in the
submillimeter regime and in CO line emission, and many of them are
lensed. The main finding was that the \ci\ properties of
high--redshift systems do not differ significantly to what is found in
low--redshift systems, including the Milky Way. In addition, there are
no major differences in \ci\ properties between the QSO-- and
SMG--selected samples.  The $L'_\cone$/$L'_{\rm CO}$ ratios
(0.29$\pm$0.12) are similar to low--z galaxies (e.g., $0.2\pm0.2$,
Gerin \& Phillips 2000).

As argued in Sec.~\ref{fsl_intro}, measurements of both \ci\ lines can
constrain gas excitation temperatures of the molecular gas,
independent of radiative transfer modeling. For the available sample,
a carbon excitation temperature of 29.1$\pm$6.3\,K was derived (Walter
et al.\ 2009). This temperature is lower than what is typically found
in starforming regions in the local universe despite the fact that the
sample galaxies have star formation rate surface densities on kpc
scales of 100's of M$_\odot$\,yr$^{-1}$\,kpc$^{-2}$. However, the
temperatures are roughly consistent with published dust temperatures
of high--redshift starforming galaxies (Beelen et al.\ 2006,
Kov{\'a}cs et al., 2006, 2010, Harris et al.\ 2012). Low carbon excitation as well as low
dust temperatures could indicate that the measurements include a
significant amount of gas/dust unaffected by star formation. 

The \ci\ abundances in the current high--z galaxy sample of
(X[\ci]/X[H$_2$]\,=\,(8.4$\pm3.5)\times$10$^{-5}$) are comparable,
within the uncertainties, to what is found in local starforming
environments (Walter et al.\ 2009).  The \ci\ lines are a negligible
coolant (average $L_\ci$/$L_{\rm FIR}=(7.7\pm4.6)\times10^{-6}$).
There is tentative evidence that this ratio may be elevated in the
SMGs by a factor of a few compared to the QSOs (Walter et al.\ 2009).

\subsubsection{Ionized Carbon: \cii}
\label{fls_cii}

The \cii\ 158$\mu$m line is rapidly becoming a work--horse line for
the study of the cool atomic gas in distant galaxies. The number of
\cii\ detections at high redshift has increased substantially in the
last few years using the CSO, PdBI, APEX and the SMA (Maiolino et al.\
2009; Stacey et al.\ 2010; Iono et al.\ 2006b, Hailey--Dunsheath et
al.\ 2010, Wagg et al.\ 2010; Walter et al.\ 2012; Cox et al.\
2011; De Breuck et al.\ 2011; Gallerani et al.\ 2012), as well as a
few detections in strongly lensed galaxies with Herschel (Ivison et
al.\ 2010b, Valtchanov et al.\ 2011).  To date, there have been about two
dozen \cii\ detections from galaxies at $z > 1$, including a number at
$z > 6$ (see Sec.~\ref{first_gal}).  ALMA during early science has
already demonstrated the ability to detect [CII] emission from
relatively normal star forming galaxies at high redshift (LBGs, LAEs)
in \cii\ emission (Wagg et al. 2012; Carilli et al. 2012; Riechers et
al. 2013).

Fig.~\ref{fig_cii_fir} shows a broad scatter in the \cii/FIR ratio at
z$>$1.  Note that the high--z galaxy sample all have FIR luminosities $\ge
10^{12}$ L$_\odot$. However, there is a trend for luminous AGN to have
the lowest ratios, and star--formation dominated galaxies (eg. SMGs) to
have ratios closer to that of the Milky Way (e.g., Stacey et
al.\ 2010).

Stacey et al.\ (2010) use the CO, FIR, and \cii\ emission to argue
that the \cii\ emission is dominated by PDRs in the star forming
galaxies. From these observations, they derive the physical conditions
in the PDRs in high redshift galaxies, finding gas densities of order
a few $\times 10^4$ cm$^{-3}$, and the strength of the FUV radiation
field: $G \sim 10^3 \times 1.6\times 10^{-3}$ erg cm$^{-2}$ s$^{-1}$,
where $1.6\times 10^{-3}$ erg cm$^{-2}$ s$^{-1}$ is the local Galactic
interstellar radiation field (ISRF, `Habing field'). They find a
further factor 10 higher $G$ for the AGN dominated systems.  Their
modeling also implies that the scale of the PDRs is $\ge$ few kpc,
much larger than for GMCs or nuclear starburst galaxies at low
redshift. The fraction of molecular gas mass associated with PDRs is
between 20\% and 100\% in their sample.

Swinbank et al. (2012) detect likely [CII] 158$\mu$m emission in two
SMGs using ALMA, from which they derive redshifts of $z \sim 4.4$ and
4.2. They suggest that the bright-end of the [CII] luminosity function
increases dramatically with redshift, with close to a thousand-fold
increase in the number density of galaxies with $\rm L[[CII] > 10^9$
L$_\odot$ from $z = 0$ to 4. We discuss [CII] detections in a number
of z$>$6 galaxies in Sec.~\ref{first_gal}.

\subsubsection{Other fine structure lines}

There have been only a few detections of fine structure lines other
than \cii\ at high redshift to date, mostly in strongly lensed systems
using principally the PdBI and Herschel. We summarize some of the
physical diagnostics that have been achieved with such observations.

In their Herschel study of SMGs at $z \sim 1.4$, Coppin et al.\ (2012)
find that the \oi\ 63$\mu$m/FIR ratio is comparable to spiral galaxies
nearby, and much higher than what is seen in low--z nuclear
starbursts. Similar results were found by Ferkinhoff et al.\ 2010 and
Sturm et al.\ (2010) in their \oi\ 63$\mu$m, \oiii\ 52$\mu$m, and
\cii\ 158$\mu$m study of a strongly lensed ULIRG at $z$\,=\,1.3 (see
also Hailey--Dunsheath et al.\ 2010).  The interesting conclusion is
that, although these systems are intrinsically ULIRG or HyLIRGs, they
do not show a deficit in the major PDR cooling lines as seen in nearby
ULIRGs. A similar conclusion has been reached based on \cii\ and \nii\
observations of lensed SMGs (Nagao et al.\ 2012; De Breuck et al.\
2011). In particular, the compilation of Decarli et al. (2012) shows a
mean value for \nii 122$\mu$m/FIR $\sim 3\times 10^{-4}$, close to
that seen in low $z$ disk galaxies, although admittedly the scatter is
large.

Conversely, Valtchanov et al.\ (2011) and Ferkinhoff et al.\ (2010;
2011) perform PDR analyses of lensed SMGs at $z \sim 3$, using FIR,
CO, \oiii\, \nii\, and/or \cii\ measurements. They find mean densities
of order 2000 cm$^{-3}$, FUV radiation field of $G$\,=\,200, and sizes
for the emitting region $\sim 0.6$ kpc. The relatively strong \oiii\
implies that the radiation field must be very hot (35,000K), either
dominated by O9.5 stars or an AGN NLR--like radiation field.

{\bf SIDEBAR: The \cii\ 158$\mu$m line is realizing its potential in
the study of the cooler atomic gas in distant galaxies. This strong
line will be a workhorse--line for redshift determinations in the
first galaxies, and in the study of galaxy dynamics at the highest
redshifts. However, the interpretation of \cii\ emission is not
straighforward, since \cii\ traces both the neutral as well as the
ionized medium, and it appears to be suppressed in high--density
regions.  Likewise, multi--fine structure line diagnostics at high
redshift have tremendous potential to probe the physical conditions in
the ISM in high z galaxies, but studies are in their infancy. The
current observations reflect the heterogeneous nature of SMGs, with
many showing spiral galaxy--like FSL properties, and a few showing
evidence for an AGN--like component.}

\subsection{Dense gas tracers and other molecules at high redshift}
\label{dense_highz}

Emission from high dipole molecules, such as HCN and HCO$^+$, comes
only from highest density regions in molecular clouds ($n_{CR} > 10^4$
cm$^{-3}$), corresponding to regions directly associated with active
star formation.  These lines are typically an
order of magnitude weaker than the integrated CO emission from star
forming galaxies, although the higher dense gas fraction in nuclear
starbursts can lead to line strengths up to 25\% of CO (Riechers et
al.\ 2011c).  To date, there remain just a handful of detections at
high redshift, mostly from strongly lensed hyper--starbursts (Riechers
et al.\ 2006b, 2010c; 2011c;  Garc{\'{\i}}a-Burillo et al. 2006, Danielson et al.\ 2011).  The two strongly lensed
hyper-starburst/AGN, APM\,0827+5255 at $z$\,=\,3.91 and the
Cloverleaf quasar (H1413+117; Barvainis et al.\ 1994) at $z$=\,2.56
play an analogous role to the Galactic molecular cloud SgrB2, as
high--z `molecule factories'. These are the only sources with multiple
transitions detected from molecules and isotopes other than $^{12}$CO,
including HCN, HNC, HCO$^+$, H$_2$O, CN, $^{13}$CO. We briefly review
the recent results on these sources.

For APM\,0827+5255, strong emission from very high order transitions
is seen (Wei\ss\ et al.\ 2007a; Riechers et al.\ 2010c).  Formally,
the critical densities for excitation for the higher order HCN and
HCO$^+$ transitions are $> 10^8$ cm$^{-3}$. The excitation suggests
emission from a region with a radius $\sim 200$pc around the AGN,
where IR pumping (Sec.~\ref{dense_highz}) plays a dominant role in
molecular excitation.  The J\,=\,6 isomer ratio of HNC/HCN\,=\,0.5 is
consistent with the IR pumping model (Riechers et al.\ 2010c).

For the Cloverleaf quasar, low and high order emission from HCN and
HCO$^+$, and CN have been detected (Solomon et al.\ 2003; Riechers
2006b, 2007a, 2011c; Barvainis et al.\ 1997; Wilner et al.\
1995). Modeling including the multiple CO transitions implies
collisional excitation by gas with a mean kinetic temperature of 50\,K
and density of 10$^{4.8}$ cm$^{-3}$, in a molecular region with a
radius $\sim 0.8$ kpc (Riechers et al.\ 2011c). Hence, in contrast to
APM\,0827--5255, the Cloverleaf is consistent with collisional
excitation in a hyperstarburst region with a radius of order 1 kpc.

The only detection of rare isotopic emission at high redshift is
$^{13}$CO 3--2 emission from the Cloverleaf (Henkel et al.\ 2010).  The
isotopic luminosity ratio for the J\,=\,3 emission is 40, eight times larger
than that seen in Milky Way GMCs. The only nearby galaxy with a ratio
approaching that of the Cloverleaf is the merging ULIRG NGC 6240,
where Greve et al.\ (2009) obtain a lower limit of 30. Henkel et
al.\ (2010) conclude that the large isotope ratio in the Cloverleaf
likely reflects a real $^{13}$C abundance deficit, by a factor of four or
so relative to Milky Way GMC value of 40 to 90 (Henkel et al.\ 1993).  

Progress has recently been made on detecting thermal emission from
(rest frame) FIR ro--vibrational transitions of water at high redshift
(van der Werf et al.\ (2011), Omont et al.\ 2011; Combes et al. 2012a).
In APM\,0827--5255, van der Werf et al.\ show that the lower level
transitions (rest frame frequencies below 1 THz) likely arise in
collisionally excited gas with kinetic temperatures of 100 K, and
clump densities of order $3\times 10^6$ cm$^{-3}$. The higher order
transitions require radiative excitation by IR radiation on a scale of
a few hundred parsecs around the AGN, with conditions similar to the
nuclear regions of the low redshift starburst/AGN galaxy, Mrk 231 (van
der Werf et al.\ 2010). They argue for distributed gas heating, as
expected for a star--formation heated PDR, and not an AGN dominated
XDR.

We note that a linear relation between L$'_{\rm HCN(1-0)}$ and
L$_{\rm FIR}$ has been found for local galaxies , including spirals
and ULIRGs (Gao \& Solomon 2004a, 2004b), unlike the non-linear
relationship between L$'_{\rm CO(1-0)}$  and L$_{\rm FIR}$. It has been
shown that this relation extends even to the dense cores of galactic
clouds (Wu et al.\ 2005). One interpretation of this finding is that
the dense molecular phase traced by HCN is immediately preceeding the
onset of star formation. However, Riechers et al.\ (2007b) have shown
that this linear relationship appears to break down in the
high--redshift systems studied to date, i.e. that they have lower HCN
luminosities than expected based on a linear extrapolation from the
low--redshift (and Galactic) measurements. This finding may indicate
even higher average gas densities in the highest--redshift systems
compared to dense environments found locally. It may also hint at an
increased star formation efficiency (or both, Riechers et al.\ 2007b).

{\bf SIDEBAR: Progress on dense gas tracers at high redshift has been
limited to strongly lensed, extremely luminous systems, due to the
limited sensitivity of existing telescopes. The results thus far
indicate a mixture of PDR/XDR heating and collisional excitation in
compact extreme starburst regions. Given the very high average
densities in dense starbursts at high redshift, cosmic ray heating and
related modeling will play an increasingly important role. In the
future, much progress is expected from much broader bandwidths, as
they will in almost all cases yield `involuntary line surveys' of
dense gas tracers. }

\subsection{Star formation laws and gas consumption}
\label{sf_law}

Quantifying the relationship between star formation rate (SFR) and gas
density (the so--called `star formation law', `Schmidt law',
`Schmidt--Kennicutt law' or `K--S law') has been a key goal in
observational astrophysics over the last 50 years, starting with
Schmidt (1959).  Any relationship between SFR and gas surface
densities has important implications for our understanding of galaxy
formation and evolution as it describes how efficiently galaxies turn
their gas into stars. Such a `law' would also serve as essential input
to hydrodynamical simulations (and other models) of galaxy evolution
that start with dark matter halos that are beeing fed by gas
infall. It should be noted that the term `law' (in a sense of a
physical law) is not appropriate to describe an empirical
relationship. However, we stick to this nomenclature as it is now
common practice to refer to the SFR--gas relationship. This
relationship has been reviewed recently for nearby galaxies by
Kennicutt \& Evans (2012). Herein we focus on $z>1$ galaxies.

The relation between star formation rate and gas density is typically
expressed in terms of surface densities
($\Sigma_{SFR}\sim\Sigma_{gas}^n$).  One complication is that the
measurement of surface densities requires resolved measurements of
galaxies. In the compilation by Kennicutt (1998a, 1998b) the
integrated star formation rates and gas masses were averaged over
entire galaxy disks to derive an average surface density for a given
galaxy. Observational capabilities have improved since, and spatially
resolved measurements of the star formation rate and gas surface
densities are now available for a number of nearby galaxies (e.g.,
Kennicutt 2007, Bigiel et al.\ 2008, 2011, Leroy et al.\ 2008, 2013,
Rahman et al.\ 2011, 2012, Liu et al.\ 2011). These spatially resolved
measurements typically have resolutions of one kiloparsec or slightly
better, i.e. they still average over many individual giant molecular
clouds that are typically $\sim$50\,pc in size. It is clear from
Galactic studies that the relationship
$\Sigma_{SFR}\sim\Sigma_{gas}^n$ must break down on very small scales:
e.g. a star forming region will ionize its immediate surrounding, thus
destroying any relationship that may be present on larger scale. This
has been quantified by Schruba et al.\ (2010) and Onodera et
al.\ (2010) who have shown that in the case of M\,33 the star
formation law on large/galactic scales breaks down on scales below
$\sim$300\,pc. There is now largely consensus that the star formation
law is `molecular', i.e. that the equation above becomes:
$\Sigma_{SFR}\sim\Sigma_{H_2}^n$. \hi\ does not appear to be intimitly
linked to the star formation process through such a simple
description, even though it is clear that \hi\ is needed to form H$_2$
to begin with (Kennicutt \& Evans 2012).

Measuring the star formation law at high redshift is significantly
complicated by the fact that few resolved CO measurements exist, and
dust and/or AGN emission confuses determination of optical sizes.  The
recent comprehensive study of Tacconi et al. (2013) shows that, where
both the size of the SF disk as well as the size of the CO emission
could be measured, reasonable agreement between the quantities was
found (see also Daddi et al.\ 2010a).  However, typically only global
measurements of SFR and molecular gas are plotted in high redshift
research.

Unlike in the local universe, where L$'_{\rm CO}$ luminosities and H$_2$
masses can be calculated from the CO(1--0) or CO(2--1) transitions
(Sec.~\ref{alpha_lowz}), the majority of high--redshift measurements
are done using higher J transitions. To put all high--redshift systems
on the same plot, the L$'_{\rm CO}$ luminosity of the CO(1--0) transition
needs to be estimated, assuming the typical excitiation of the galaxy
under consideration. Narayanan et al.\ (2010a) discusses how the slope
of the star formation law could change if different transitions of CO are
used and if the different line ratios are not accounted for
properly. We consider the excitation ladder in Sec.~\ref{excitation},
and in the analysis below, we adopt a set of canonical values for
different galaxy types, based on the (admittedly limited) available
data (Tab.~\ref{tab_lum_ratios}).

Two further complications are involved when comparing derived
quantities, such as gas mass and star formation rate. First is CO
luminosity to gas mass conversion factor, $\alpha$. As discussed in
detail in Sec.~\ref{alpha_highz}, different source populations likely
have different values of $\alpha$. And second is derivation of the SFR
from observed SEDs. Methods include: SED fitting of UV/optical/IR data
for CSGs, FIR luminosities for SMGs and AGN after correction for hot
dust heated by the AGN (Jiang et al.\ 2010, Leipski et al.\ 2010,
2013, Riechers 2011, Genzel et al.\ 2010), and radio luminosities
assuming the radio--FIR correlation (Condon 1992).

To avoid the above complications, we focus on the empirical
relationship between the observables, L$_{\rm FIR}$ and
L$'_{\rm CO(1-0)}$, for all high--redshift galaxies detected to date in
Figure~\ref{fig_sf_law}. Most of the galaxies detected at high
redshift have luminosities (both FIR and CO) that are much higher than
typical systems seen in the local universe (grey/black data
points). At the highest luminosities, the SMGs and quasar host
galaxies dominate. The entire distribution can be fit with a power law
of the form

$$ \log(L_{\rm IR})=1.37(\pm0.04)\times(\log L'_{\rm CO})-{\rm 1.74(\pm0.40)} $$

\noindent with a slope that is consistent with the power law found
when looking at the intergrated properties of nearby galaxies only
(including ULIRGs, Kennicutt 1998a). However, it is also apparent that
the CSGs have higher CO luminisities for a given L$_{\rm FIR}$ (or
SFR) compared to the QSO/SMG population (Daddi et al.\ 2008; 2010;
Tacconi et al.\ 2010; 2013).  Genzel et al.\ (2010) and Daddi et al.\
(2010b) argue that there are in fact 2 sequences visible in this plot,
one `starburst' sequence (red dashed line) and one `main sequence'/CSG
relation (grey dashed line). For the latter, Daddi et al.\ (2010b) and
Genzel (2010) derived:

$$ \log(L_{\rm IR})=1.13\times(\log L'_{\rm CO})+0.53 $$

\noindent with L$'_{\rm CO}$ in units of K\,km\,s$^{-1}$\,pc$^2$, L$_{\rm
IR}$ in units of L$_\odot$. The values given here are from Daddi
(2010b), but the relation derived by Genzel et al.\ (2010) is very
similar.

Both Genzel et al. (2010) and Daddi et al. (2010b) proceed to
calculate molecular gas masses from the CO luminosities using a
molecular gas conversion factors $\alpha$ that is about five times
higher in the case of the CSGs relative to SMGs, based on the
arguments in Sec.~{\ref{alpha_highz}. This then leads to an increased
gap between the CSG population and the SMGs/QSOs with an offset of
roughly 1 dex in L$_{\rm FIR}$. This strengthens the presence of two
different star formation regimes, one highly--efficient star formation
mode for starburst (presumably triggered by mergers and/or other
interactions) and one less--efficient and thus longer--lasting mode
for the CSGs (`main--sequence' galaxies).

The different star formation efficiencies, defined as SFE=L$_{\rm
FIR}$/L$'_{\rm CO}$ (in units of L$_\odot$/(K\,km\,s$^{-1})pc^2)^{-1}$,
are further highlighted in the right panel of Fig.~\ref{fig_sf_law}.
It is immediately obvious from this plot that the SFE of the CSG is
lower compared to the high--redshift quasars and SMGs by a factor of
few, and that the CSGs reach values that are similar to what is found
in local galaxies. The offset between the two sequences becomes 
even more pronounced when different $\alpha$ conversion factors are
used for the two high--redshift galaxy populations. 

The gas consumption timescale is defined simply as: $\tau_c =
M_{gas}/SFR$, i.e. the time it would take to use up the available
molecular gas reservoir given the current star formation rate. The
right hand of axis of Fig.~\ref{fig_sf_law} shows these timescales,
assuming the two different conversion factors relevant to SMGs and
CSGs (Sec.~\ref{alpha_highz}). The gas consumption timescale in CSGs
is of order 10$^8$ to 10$^9$ years (Daddi et al.\ 2008, 2010a,b,
Genzel et al.\ 2010, Tacconi et al.\ 2010, 2013), comparable to that in low
redshift disk galaxies (Bolatto et al.\ 2012). For comparison, the
SMGs and other HyLIRGs galaxies have extremely short consumption
timescales $\le 10^7$ yr.

In the local universe it has been shown by Kennicutt (1998a,b) that if
the dynamical timescale is taken into account (i.e. $\Sigma_{SFR}$ is
plotted as a function of $\Sigma_{gas}/\tau_{dyn}$ and not only as a
function of $\Sigma_{gas}$), all galaxies fit on one relation with
slope $\sim$\, 1.  Both Genzel et al.\ 2010 and Daddi et al.\ 2010
showed that this trend also continues at high--redshift, i.e. that a
`universal' star formation law is obtained, with no separation between
the two star forming sequences. The dynamical timescale $\tau_{dyn}$
is here defined to be the rotational period at the last measured point
in a galaxy (typically taken to be the half--light radius). We note
that the latter is not easily determined given current observations at
high redshift, and that a physical interpretation of this finding is
not straightforward. In a sense it is puzzling that the global
properties of a galaxy ($\tau_{dyn}$) appear to be related to local
star formation process. Krumholz et al.\ (2011) perform a similar
analysis, only using an estimate of the local gas free--fall time, and
conclude that all systems, from local molecular clouds to distant
galaxies (starburst and main--sequence), fall on a single star formation
law in which the star formation rate is simply $\sim 1\%$ of the
molecular gas mass per local free--fall time.

As a general comment, it should be kept in mind that there are
significant selection effects that may contribute to the apparent
functional behavior in the `SF law' plot. For example, as essentially
all galaxies have been pre--selected based on their star formation
activity, it is conceivable that large molecular gas reservoirs exist
at low L$_{\rm FIR}$; they have simply not yet been looked at.

{\bf SIDEBAR: Constraining the `star formation law', i.e. relating the
star formation rate to the available gas reservoir, has been the focus
of many studies at low and high redshift in the past decade. Even
though it is not a `law' in a physical sense, this empirical relation
has the potential power to predict one quantity from the other, sheds
light on the star formation process, and serves as vital input in
numerical simulations and galaxy formation models. At high redshift
there is a clear relation between the main observables, $\log(L_{\rm IR}$)
and $\log (L'_{\rm CO})$. Once translated into physical quantities (SFR and
H$_2$ mass) there is now good evidence for two sequences of star
formation, one `starburst' sequence, with very short consumption times,
10$^7$--10$^8$\,yr (local ULIRGs, SMGs and QSOs) and a `normal,
quiescent' sequence, $\sim$10$^9$\,yr (local spirals, CSG, `main
sequence' galaxies). }

\subsection{Imaging of the molecular gas in early galaxies}
\label{imaging}

The last few years has seen tremendous progress in high resolution
imaging of the molecular gas in high redshift galaxies, mainly by the
VLA and the IRAM PdBI. Imaging of CO emission has been performed on
the brighter sources with a spatial resolution comparable to that of
the HST ($\sim 0.15"$), or rougly 1\,kpc at z$>1$.  Following we
highlight a few of the best examples to date (see Fig.~\ref{fig_CO_im}).

\subsubsection{A typical CSG}

In their ground breaking studies of CSG, Tacconi et al.\ (2010) and
Daddi et al (2010a) find gas rich disks without extreme starbursts
(see also Tacconi et al. 2012; Sec.~\ref{cssfg}). The best imaging
study to date of a CSG is that of the CO(3--2) emission from
EGS1305123 using the PdBI at $0.6"$ resolution by Tacconi et al.\
(2010) (Fig.~\ref{fig_CO_im}).  The H$_2$ mass is $1.3\times 10^{11}
(\alpha/3.2)$ M$_\odot$, distributed in a rotating disk with a radius
of 8kpc, and a terminal rotation velocity of 200 km s$^{-1}$. While
clearly rotating, the disk also has a high internal velocity
dispersion $\ge 20$ km s$^{-1}$, implying substantial disk turbulence.

The disk is punctuated by high brightness temperature clumps with
radii $\le 2$ kpc and masses $\sim 5\times 10^9$ M$_\odot$. The gas
surface densities are $\sim 500$ M$_\odot$ pc$^{-2}$, a factor few
larger than for Galactic GMCs, and they hypothesize that these clumps
are conglomerates of a number of Galactic--type GMCs rather than a
single giant GMC, due to the relatively low internal velocity
dispersions ($\sim 20$ km s$^{-1}$) Daddi et al.\ (2010a) obtain
similar imaging results for a few CSG at $z \sim 1.5$, although at
somewhat lower spatial resolution. 

These imaging observations of CO in $z \sim 1$ to 3 CSG have been
critical to the interpretation that the CSG are gas--dominated,
rotating disk galaxies undergoing steady, high levels of star
formation (see Sec.~\ref{cssfg}). They also helped to constrain the
conversion factor $\alpha$ for these systems through dynamical
arguments (Sec.~\ref{alpha_highz}).

\subsubsection{BRI 1335--0417: tidally disturbed gas surrounding a
luminous quasar}

BRI\,1335--0417 at $z$\,=\,4.4 was among the first optically selected,
very high--z quasars to be identified with a hyper--luminous FIR host
galaxy, implying an accreting SMBH with mass $\sim 10^9$ M$_\odot$
coeval with an extreme starburst (SFR $\sim 1000$
M$_\odot$\,yr$^{-1}$) (Guilloteau et al. 1997).  The host galaxy has
been detected in CO line emission, with an implied $\rm M(H_2) \sim
8\times 10^{10}(\alpha/0.8)$ M$_\odot$ (Carilli et al.\ 2002a), as well
as strong \cii\ emission (Wagg et al.\ 2010).  BRI\,1335--0417 is a
broad absorption line quasar, indicating AGN
outflow. Fig.~\ref{fig_CO_im} (middle) shows the CO images from the of
BRI\,1335--0417 from the VLA (Riechers et al.\ 2008b).  These are the
highest quality CO images of any non--lensed high--z quasar to
date. The system shows complex morphology in the gas on a scale of
$\sim 1''$.  The molecular gas shows multiple components distributed
over $\sim 7$kpc, with a pronounced tail extending to the north,
peaking in a major CO clump $\sim 0.7''$ from the quasar, and with a
few other streams connecting to the quasar from other directions.
While there is an overall north--south velocity gradient, the general
velocity field appears chaotic.

Riechers et al.\ interpret the complex gas structure in
BRI\,1335--0417 as tidal remnants from a late--stage, gas--rich
(`wet') merger. The merger drives gas accretion onto the main galaxy,
fueling the hyper--starburst and the luminous AGN, generally
consistent with the high molecular excitation seen in quasar hosts
(Sec.~\ref{excitation}). For comparison, the SMG--quasar pair
BRI\,1202--0725 at $z\sim$\,4.7 shows a distinct SMG and quasar,
possibly corresponding to an early--stage merger, before galaxy
coalesence (Salome et al.\ 2012, Carilli et al.\ 2012).

\subsubsection{GN\,20: an SMG with a gas rich disk}

The galaxy GN\,20 is the brightest SMG in the GOODS--North field (Pope
et al.\ 2006), and the host galaxy is heavily obscured at optical
wavelengths. Daddi et al.\ (2009a) made a serendipitous redshift
determination of $z$=\,4.05 from CO emission using the PdBI (by
targeting a nearby CSG). Two other SMGs at this redshift have been
detected in CO and dust continuum emission about 20$"$ to the west
(Carilli et al.\ 2010; Hodge et al.\ 2013), and there is a clear
over--density of galaxies in this field, with 15 LBGs with $z_{phot}
\sim 4$ within a radius or $25"$ of GN\,20 (Daddi et al.\ 2009).

A long observation using the JVLA in early science of the CO (2--1)
emission (Hodge et al.\ 2012) shows that the CO is distributed in a
disk with a diameter $\sim 14$kpc (Fig.~\ref{fig_CO_im}, right). The
regions emitting in CO and dust continuum are mostly obscured in the
HST I--band image.  The only tracable optical emission is seen at the
edges of the source (Hodge et al.\ 2012). Such heavy obscuration in
the optical is characteristic of SMGs.

The GN\,20 disk dynamics are consistent with a standard `tilted--ring'
gas rotation model, with a dynamical mass of $5.4\pm 2.4$ 10$^{11}$
M$_\odot$. Observations at 1kpc resolution reveal that 30\% to 50\% of
the gas is in giant clumps with gas masses of a few $\times 10^9
(\alpha/0.8)$ M$_\odot$, brightness temperatures between 16 K and 31
K, and line widths of order 100 km s$^{-1}$ (Hodge et al.\ 2012). A
dynamical analysis suggests the clumps could be self--gravitating. The
gas surface densities of the clumps are $\sim 4000 (\alpha/0.8)$
M$_\odot$ pc$^{-2}$, more than an order of magnitude larger than
typical GMCs. An analysis of the overall galaxy dynamics has been used
to determine the value of $\alpha$ in GN\,20 (see
Sec.~\ref{alpha_highz}).

The apparent disk in GN\,20 suggests that not all HyLIRGs at very high
redshift result from an ongoing major merger. This
conclusion has also been reached for a few other SMGs at lower
redshift, where low order CO observations show large, disk--like gas
reservoirs similar to GN\,20 (Ivison et al.\ 2011; Riechers et al.\ 2011;
Greve et al.\ 2003, Ivison et al.\ 2010a,b).

In GN\,20, the clumpy disk is consistent with that expected in the
cold mode accretion model (e.g. Keres et al.\ 2005, Dekel et al.\
2006, 2009), only now scaled up by almost an order of magnitude in FIR
luminosity relative to typical CSG (Sec.~\ref{accretion}). It is
possible that the star formation in this gas rich disk has been
enhanced due to gravitational harrasment by the other SMGs and smaller
galaxies in the proto--cluster.

{\bf SIDEBAR: Spatially resolving the molecular gas emission in high
redshift galaxies is to date restricted to very few, bright, sources.
Imaging the molecular and cool atomic gas a few selected high redshift
galaxies has revealed 10\,kpc--scale, clumpy and turbulent, but
apparently rotating disks in CSG and some SMGs, as well as strongly
tidally disturbed gas distributions in some SMGs and quasar hosts. }

\subsection{Outflows}
\label{outflows}

Theoretical models without negative feedback (negative feedback =
ejection of material due to either star formation or AGN activity)
predict both a higher gas content in massive galaxies in the nearby
Universe, and a larger population of star forming massive galaxies
today, than observed. At high galactic masses, including AGN feedback
mitigate these problems in simulations, both driving gas out of the
immediate ISM of the host galaxy via AGN winds, and suppressing
further gas accretion from the IGM via large--scale radio jets (Fabian
2012).  Direct evidence for feedback has been seen in nearby galaxies,
including outflows seen in OH far--IR lines, molecular emission lines,
optical lines, and atomic fine structure lines (e.g., in the case of
M\,82: Strickland \& Heckman 2009, Walter et al.\ 2002, Martin et al.\
1998). The fact that old, gas--poor massive galaxies have now been
seen at redshifts of 2 and beyond suggest that feedback must be an
important process even earlier. 

Recent observations have detected evidence for feedback on kpc--scales
in very early galaxies.  One of the best examples of AGN feed--back
are the broad line wings of the \cii\ emission from the $z$\,=\,6.4
quasar J1148--5251 by Maiolino et al.\ (2012). They derive an outflow
velocities of 1300 km s$^{-1}$, an outflow rate of ${\dot
M}_{outfl} \sim 3500$ M$_\odot$\,yr$^{-1}$, and kinetic power of: $P_K
\sim 1.9\times 10^{45}$ erg s$^{-1}$.  This is roughly 0.6\% of the
quasar bolometric luminosity, well below the theoretical upper limit
to a radiatively driven quasar outflow of 5\% of the bolometric
luminosity (Lapi et al.\ 2005), but it is barely consistent with the
maximum kinetic power that can be driven by the associated starburst
in the quasar host (Veilleux et al.\ 2005). Likewise, theoretical
models indicate that star formation driven winds reach a maximum
velocity of $\sim 600$ km s$^{-1}$ (Thacker et al.\ 2006). Maiolino et
al. (2012) argue that the outflow is most likely AGN--driven.  The gas
consumption timescale for the outflow is comparable to that due to
star formation, of order $10^7$\,yr.

Other systems show varying degrees of feedback in the cooler gas. The
BRI 1202--0725 quasar--SMG galaxy pair does show evidence for a broad
wing in the quasar host galaxy \cii\ 158$\mu$m spectrum, but the
outflow kinetic energy is well below that of J1148+5251, and star
formation likely dominates gas depletion in the galaxy
(Wagg et al.\ 2012; Carilli et al.\ 2012). Wei\ss\ et al.\ (2012)
detect a 250 km s$^{-1}$ outflow in a $z$\,=\,2.8 quasar in both CO
and \ci. They derive a lower limit to the mass outflow rate of 180
M$_\odot$\,yr$^{-1}$, which is slightly larger than the star formation
rate in the host galaxy. 

{\bf SIDEBAR: Molecular Outflows have only very recently been detected in a
few high--redshift systems. Given that they feature (by
definition) broad and faint line wings they remained undetected by
past observations, both due to missing sensitivity and insufficient
bandwidth. Quantifying the kinetic energy and masses associated with
such outflows will provide important input in galaxy simulations in
which feedback by stellar (or AGN) activity is a key driver for galaxy
evolution.}

\section{Dense Gas History of the Universe}
\label{gas_history}

\subsection{Gas dominated disks during the epoch of galaxy assembly}

\subsubsection{Gas fraction}
\label{fgas}

One of the most striking results from the study of CSG is the
remarkably high detection rate (between 50--100\%) in CO emission
(Daddi et al.\ 2010a; Tacconi et al.\ 2010; Taconni et al. 2012). The
line strengths are comparable to those seen in the hyperstarbusts, but
the star formation rates are almost an order of magnitude smaller
(Sec.~\ref{sf_law}). Moreover, these galaxies have a space density
more than an order of magnitude larger than SMGs ($10^{-4}$ Mpc$^{-3}$
down to $M_{stars}$= 10$^{10}$ M$_\odot$), and they represent the galaxies
that dominate the integrated cosmic star formation rate during the
`epoch of galaxy assembly' (Sec.~\ref{intro_notes}).

A series of measurements and consistency arguments lead to a value of
$\alpha_{\rm CO} \sim 4$ in CSG (Sec.~\ref{cssfg}). The implied H$_2$
masses range from $3\times 10^{10}$ to 10$^{11}$
M$_\odot$. Interestingly, these gas masses are comparable to, or even
larger than, the stellar masses in these galaxies, as first pointed
out by Daddi et al.\ (2010a) and Tacconi et al.\ (2010). This ratio is
very different with respect to large disk galaxies at low redshift,
where the stellar masses are close to an order of magnitude larger
than the cool gas masses.

In Fig.~\ref{fig_gas_stars} we plot a compilation of the latest
measurements of the gas fraction, defined as $\rm f_{gas} \equiv
M_{gas}/M_{stars}$, out to $z \sim 4$, where $\rm M_{gas}$ corresponds
to the molecular component, including a 1.36 factor for Helium.  At
all redshifts, the galaxies were selected to be star forming disk
galaxies with stellar masses $>$10$^{10}$ M$_\odot$.  

While the scatter in Fig.~\ref{fig_gas_stars} at any given redshift is
large, there is a clear trend for increasing gas fraction with
redshift in massive disk galaxies. The mean value at $z \sim 0$ is:
$\rm M_{gas}/M_{stars} \sim 0.1$, which increases to $\rm
M_{gas}/M_{stars} \sim 1$ at $z > 1.5$. A functional form of $f_{gas}
\sim 0.1\times (1+z)^2$ fits the data reasonable (see also Magdis et
al.\ 2012a; Geach et al.\ 2011).  Note that the high-$z$ CSG CO samples
tend to be at the upper end of the mass range for CSGs, but are
typical of main sequence galaxies in all other respects (eg. Daddi et
al. 2010). Moveover, the recent large study of Tacconi et al. (2012)
suggests that the gas fraction in CSGs may increase with decreasing
galaxy mass, thereby excentuating the results in
Fig.~\ref{fig_gas_stars}.

We note that all points in Fig.~\ref{fig_gas_stars} are for large disk
galaxies (stellar masses $> 10^{10}$ M$_\odot$), and that the same
value of $\alpha\!\sim\!4$\,M$_\odot$/(K km s$^{-1}$ pc$^2$)$^{-1}$
was used for all sources. Moreover, while stellar masses are nominally
based on full SED--fitting, the outcome is predominantly dictated by
the observed near--IR data, corresponding to roughly rest frame R band
(Daddi priv. comm.). Hence, the ratio on the ordinate of
Fig.~\ref{fig_gas_stars} can be considered approximately empirical,
and $\propto L'_{\rm CO1-0}/R_{mag}^{rest}$.

Narayanan et al.\ (2012) consider the question of the increasing gas
fraction in galaxies with redshift, including SMGs and CSG, and
conclude that some of the effect might be attributed to a changing
value of $\alpha$, with $\alpha$ decreasing with redshift due to
higher velocity dispersions and gas temperatures in high redshift
galaxies.  

Overall, the CO measurements of CSG suggest that the peak epoch of
cosmic star formation also corresponds to an an epoch when molecular
gas masses dominates over stellar masses in common star forming
galaxies (see also Swinbank et al. 2013; Geach et al. 2011;
Magdis et al. 2012a). This fundamental change in
the baryon content of disk galaxies with redshift likely has
definitive consequences on the nature of star formation in early
galaxies (Sec.~\ref{first_gal}).

Of course, we must keep in mind that the CSG at high redshift may
not evolve into low redshift disk galaxies, and demographics suggests
that subsequent mergers can lead to substantial
morphological evolution. For instance, the space density and
clustering of the CSG at $z \sim 2$ is consistent with their
evolution into early--type galaxies at $z \sim 0$ (Shapley 2011; 
Tacconi et al.\ 2008).

{\bf SIDEBAR: The gas fraction of molecular gas versus stars in massive
disk galaxies increases by an order of magnitude from $z = 0$ to
$z \sim 2$. Hence, the peak epoch of cosmic star formation ($z \sim
2$) corresponds to the epoch when typical star forming disk galaxies
were dominated by cool gas, not stars.}

\subsubsection{Gas Accretion}
\label{accretion}

Gas resupply for star formation in galaxies has become an important
issue at both low and high redshift. The most extreme situations are
the gas consumption times in QSO host galaxies and SMGs, which are always
extremely short, in some cases as short as 10 Million years
(Sec.~\ref{sf_law}).  However, even CSGs have gas consumption
timescales substantially less than the Hubble time (Tacconi et al.
2012). This point is emphasized by Bauermeister et al.\ (2010), where
they conclude based on the short molecular gas consumption timescales
in high $z$ galaxies and the lack of evolution of the cosmic HI mass
density (from study of damped Ly$\alpha$ absorption systems), that
ultimately the gas must be accreted from the IGM.

The clear need for substantial gas resupply over cosmic time has led
to a change in thinking on the gas supply to early galaxies. As
opposed to either cooling of virialized, hot halo gas (White \& Rees
1978), or major, gas--rich mergers (Robertson et al.\ 2006), the
dominant mode of star formation during the epoch of galaxy assembly
may be driven by cold mode accretion (CMA), aka stream--fed galaxy
formation. A convergence of observations, simulations, and analytic
studies suggest that gas accretion in early galaxies occurs along cold
streams ($\rm T \sim 10^4$K) from the filamentary IGM that never
shock--heat, but stream onto the galaxy at close to the free--fall
time (Elmegreen \& Burkert 2010; Dekel et al.\ 2009a,b; Bournaud et
al.  2009; Keres et al.\ 2005, 2009).  This cool gas may form a thick,
turbulent, rotating disk (Genzel et al.\ 2011, 2008, 2006; Daddi et
al.\ 2010).  The disks are marginally Toomre--unstable (Swinbank et
al.\ 2011; Tacconi et al.\ 2010), leading to rapid fragmentation into
a few kpc--scale clumps which very efficiently form stars. These
clumps can be 100 times larger than Galactic GMCs, and 10$^7$ times
more luminous (Swinbank et al.\ 2010).  The star forming regions could
then migrate to the galaxy center via dynamical friction and
viscosity, forming compact stellar bulges (Genzel et al.\ 2008; Dekel
et al.\ 2009a; Keres et al.\ 2009).  The process leads to relatively
steady, active ($\sim 100$ M$_\odot$ yr$^{-1}$) star formation in
galaxies over timescales of order 1 Gyr, and has been termed `rapid
secular galaxy evolution' (Genzel et al.\ 2008).

In these models, the process slows down dramatically as gas supply
decreases, and the halo mass increases, generating a virial shock in
the accreting gas.  Subsequent dry (gas--poor) mergers at $z < 2$ lead
to continued total mass build up, and morphological evolution, but
little star formation in such models (Hopkins et al.\ 2010; Naab et
al.\ 2009, Kere{\v s} et al. 2009).  In this picture the majority of
stars in spheroidal galaxies are thought to form via CMA at $z \sim 2$
to 3.

Dave et al.\ (2010) suggest that the CMA phenomenon may even scale up to
HyLIRGs, and that a substantial fraction of high--z SMGs could be fed
primarily by CMA, and not major mergers.  Hydrodynamical simulations
show that quite high gas accretion rates can be achieved in large
halos at early epochs, and SFR can be elevated over the average
accretion rate by a factor 2 to 3 via (common) minor mergers and
general gravitational harassment in the dense environments of SMGs (see
also Finlator et al.\ 2006, Narayanan et al.\ 2010b). 

Physically, there may be a continuum of accretion processes, from
relatively continuous CMA, through `clumpy' accretion of
small satellites, to the rare major gas rich merger. However, current
modeling and observations favor a model in which most of the accretion 
occurs relatively continuously over timescales $\ge 10^9$ years
(Dekel et al.\ 2009; Guo \& White 2008). 

We emphasize that clumpy, turbulent but rotating gas disks are simply
a consistency argument for CMA (Shapiro et al.\ 2008), and they do not
conclusively rule out gas--rich mergers.  Robertson \& Bullock (2008)
have shown that ordered rotation of an extended gas disk can be
reestablished shortly after a major merger.  Some recent observations
detected low--metallicity gas near high--redshift galaxies; the
properties of this gas have been argued to be consistent with a CMA
picture (e.g. Ribaudo et al.\ 2011, Kacprzak et al.\ 2012). The few single
detections to date are however not sufficient to conclusively
proof the existence of CMA from an observational perspective.

{\bf SIDEBAR: The current evidence for gas accretion through cold mode
accretion at $z > 1$ is circumstantial, based on the similarity
between the predicted and observed morphologies of gas--dominated disks
in CSG. Low ionization quasar absorption lines may provide the best
means by which to prove the existence of these flows.}

\subsection{First Galaxies}
\label{first_gal}

Deep fields at near--IR wavelengths, and narrow band near--IR surveys,
are now systematically detecting galaxies at $z \sim 6$ to~10,
corresponding to the tail--end of cosmic reionization when the
Universe was less than 1Gyr old (Bouwens et al.\ 2012a,b; Pentericci
et al.\ 2011, Bradac et al. 2012; Coe et al.\ 2013). Reionization
represents the epoch when light from the first galaxies acted to
reionize the neutral intergalactic medium that pervaded the Universe
after recombination (e.g., Fan et al.\ 2006). Current studies suggest
that the typical star forming galaxies at these redshifts have lower
dust content than similar luminosity galaxies at lower redshift
(Bouwens et al.\ 2010).

The search for dust and cool gas into cosmic reionization has focused
predominantly on the host galaxies of luminous quasars with good
spectroscopic redshifts.  One of the most distant, and best studied,
of the molecular gas detections remains the quasar SDSS J1148+5251 at
$z$\,=\,6.42 (Walter et al.\ 2003; 2004, 2009; Riechers et al.\ 2009a
Wang et al.\ 2013). This (and similar) system represent the coeval
formation of massive galaxies and supermassive black holes within 1
Gyr of the Big Bang.  Large scale cosmological simulations show that
massive galaxies and SMBH can form at $z \sim 6$ via gas rich mergers,
driving extreme starbursts, and rapid accretion onto the black holes,
with subsequent black hole mergers (Li et al.\ 2007). Such systems are
thought to evolve into large galaxies in rich clusters at low--z. More
recent simulations suggest that cold accretion from the IGM may also
play a role in, and possibly even dominate, the gas resupply (Khandai
et al 2012). As the SMBH builds, feedback from the AGN expels gas from
the galaxy, and hinders further accretion, thereby terminating star
formation in the galaxy (Maiolino et al. 2012).

Exciting results have been obtained via imaging of the \cii\ 158$\mu$m
line in $z > 6$ galaxies, as demonstrated in Fig.~\ref{fig_cii_im}.
Imaging of [CII] emission from the highest redshift SDSS quasars ($z
\sim 6$) show velocity gradients indicative of rotation, with disk
scales of order a few kpc (Wang et al. 2013). In once case, a `maximal
starburst disk' is seen on a scale $\sim 1$ kpc (Walter et al.\ 2009;
Sec.~\ref{alpha_highz}). \cii\ has also been detected in the most
distant quasar with a spectroscopic redshift known, the quasar
J\,1120+0641 at $z$\,=\,7.08 (Venemans et al.\ 2012). 

The gas dynamics in distant quasar host galaxies allows for a study of
the evolution of the black hole -- bulge mass relation.  Wang et al.\
(2010) present the most detailed analysis to date out to $z \sim
6$. In this study, the black hole masses are based on standard
line--width relations of ionized gas, and are consistent with simple
Eddington arguments. They find that the median black hole--bulge mass
ratio is 15 times higher at $z \sim 6$ than today, although the
scatter is close to an order of magnitude (see also Shields et al.\
2006; Coppin et al.\ 2008). These results suggest that SMBH may
assemble before the mass in their host galaxies. Such a deviation has
been predicted in hydrodynamical simulations of very early SMBH and
massive galaxy formation (Khandai et al.\ 2012).

As pointed out in Sec.~\ref{fls_cii}, the \cii/FIR increases with
decreasing metallicity. Hence, even if the dust content of star forming
galaxies at the highest redshifts were indeed to decrease (Bouwens et
al 2010cb, Walter et al.\ 2012b), the \cii\ line may remain strong, and
will be a key redshift determinant, and a primary means to image gas
dynamics in the first galaxies, since it traces both PDRs and the
CNM. 

{\bf SIDEBAR: The presence of detectable CO, \cii\ (and dust) emission
at redshifts out to z\,=\,7 was almost inconcievable a little more
than 10 years ago. At these redshifts, the age of the universe was
$< 1$ Gyr, and there has been little time to enrich the
interstellar medium with carbon and oxygen, and then cool to form dust
and molecules. The detection of molecular emission and fine structure
line emission at $z > 6$ currently remains limited to hyperluminous
infrared galaxies.  These results reveal the coeval formation of
massive galaxies and supermassive black holes in extreme starburst
events within 1 Gyr of the Big Bang. Given the difficulties of
detecting CO at the highest redshifts, \cii\ and other fine structure
lines will likely play the dominant role in the study of the starforming ISM
at the earliest epochs.  }

\subsection{Spectral Deep Fields and the Dense Gas History of the
Universe}

Most observations of the molecular gas phase in the universe have been
restricted to galaxy populations that were pre--selected in the
UV/optical/IR/FIR, i.e. based on their star formation properties.
However, in order to obtain an unbiased census of the molecular gas
content in primeval galaxies, there is a clear need for a blind search
of molecular gas down to mass limits characteristic of the normal star
forming galaxy population, ie. a CO deep field. Such a CO deep field
has been out of reach given past instrumentation, both in terms of
sensitivity and instantaneous bandwidth. However, this situation is
now dramatically changing with the advent of new observational
facilities. Clearly, fields with the best photometric supporting data
are prefered, for rapid follow--up identification, as well as dense
spectroscopic coverage, for possible `3D--stacking'. 

As an aside, we note that a number of authors have successfully used
frequency scans to obtain redshifts from CO emission for sources that
were very bright in the FIR, but that had no easily identifiable
counterpart in the optical, and therefore no redshift determination
(Weiss et al.\ 2009; Lestrade et al.\ 2010; Walter et al.\ 2012; Cox
et al.\ 2011; Combes et al.\ 2012a, Riechers 2011, Scott et
al.\ 2011). A good example of such a redshift search using CO is the
recent GBT spectroscopic survey of Herschel--discovered SMGs, for
which redshifts were determined for 11 of 24 galaxies observed (Harris
et al.\ 2012).  Recent ALMA observations by Swinbank et al. (2012)
have shown the power of the [CII] 158$\mu$m line as a redshift
determinant at $z \ge 4$.

Cosmological simulations have been used to predict the cosmic
evolution of the dense gas history of the universe. Most of these
models and simulations start with dark--matter simulations, such as
the Millenium simulations (Springel et al.\ 2005). The dark matter
halos are then populated with model galaxies and
these evolve according to simple rules (`semi--analytical
modeling'). In this modeling, each galaxy today has a well--defined
history of growing and merging galaxies. Obreschkow et al.\ (2009b)
applied post--processing to these galaxies to subdivide their cold
gas masses into the atomic and molecular gas phases, by assuming that
the ISM pressure (Leroy et al.\ 2008, 2013) sets the phase balance between
these two phases. As discussed in Sec.~\ref{alpha_highz}, an important
complication is the CO luminosity to gas mass conversion factor. In
a follow--up paper, Obreschkow et al.\ (2009a) predicted the CO
luminosity functions for the various transitions of CO, including a
number of effects, such as heating by starbursts and CMB, and
metallicity dependence. Power et al.\ (2010) compared a number of
currently favored semi--analytical galaxy models (again applying them
to the Millenium simulation). Lagos et al.\ (2011) used a similar
methodology, but a different galaxy formation model to separate
the atomic and molecular gas phases of the interstellar medium. Lagos
et al.\ (2012) then refined their model through modeling the various
rotational transitions of CO using a PDR code. All these models are
scaled such that the CO(1--0) luminosity function at $z$\,=\,0 is
matched (Keres et al.\ 2003).

An independent approach to predicting CO and H$_2$ luminosity
functions as well as the cosmic evolution of the H$_2$/CO density is
presented in Sargent et al.\ (2013). They separate the contributions
of main--sequence and starburst galaxies to the IR luminosity
functions of galaxies at various redshifts, and use the relation
between CO luminosities and IR luminosities for these two populations to
predict CO luminosities (Sec.~\ref{sf_law}). They also use a
metallicity--dependent conversion factor to go from predicted CO
luminosities to H$_2$ masses (Sec.~\ref{alpha_highz}). Other
calculations of this type include those by Geach \& Papadopoulos
(2012) and Carilli \& Blain (2002).

Figure~\ref{fig_lum_functions} summarizes the various predictions put
forward to date. The top panel shows the predicted CO luminosities in
5 redshift bins, including $z$\,=\,0. The CO luminosity function has to date
only been measured at $z$\,=\,0 (see datapoints in top left panel by Keres
et al.\ 2003). The luminosity functions for all other redshift bins are
essentially unconstrained.

The integral under each of the luminosity curves gives the total CO
luminosity which can then, under the assumption of a given conversion
factor $\alpha$, be translated to total molecular gas for each galaxy
population. The summed density of this quantity is plotted in 
Fig.~\ref{omega_h2}, for the models discussed above. 
We note that the cosmic density in \hi\,
determined through studies of damped Ly$\alpha$ absorption line
systems, appears to show little evolution at least through the epoch
of galaxy assembly (Wolfe et al.\ 2005).

The number of high--z detections of molecular gas in CSG remains
limited, and to date no truly deep, wide field spectral search has
been performed. However, we are reaching a point where it may be
possible to set limits on the evolution of the cosmic density of
molecular gas.  We here take a very simple approach, based on the
measured stellar to gas mass ratios in existing (admittedly limited,
e.g. Aravena et al.\ 2012) samples, and the cosmic stellar mass
densities of these populations.

For five sBzK galaxies at $z \sim 1.5$ to 2, Daddi et al.\ (2010) find
a mean value of $\rm M_{gas}/M_{stars} = 1.15$. For ten BX/BM galaxies
at $z \sim 2$ to 2.5, Tacconi et al.\ (2010) find $\rm
M_{gas}/M_{stars} = 0.79$. Tacconi et al. (2012) reach similar
conclusions for CSGs at $z = 1.2$ and 2.2.  Riechers et al.\ (2010b)
measure a mean of $\rm M_{gas}/M_{stars} = 1.4$ for two lensed LBGs at
$z \sim 3$. In all cases, a value of $\alpha \sim 4+/-0.4$ was assumed
(see also Magdis et al. 2012a).  In all cases the scatter in the ratio
is close to a factor 2, and so for simplicity, we adopt $\rm
M_{gas}/M_{stars} \sim 1$ in all cases.  Note that we focus on CSG,
since their space density is more than an order of magnitude higher
than for HyLIRG, such as quasar hosts and SMGs, and their gas masses
appear comparable (although there is a question of duty cycle in the
latter samples).  In this sense, these calculations represent lower
limits.

Grazian et al.\ (2007) tabulate the total cosmic stellar mass density
in different types of galaxies at different redshifts. They also
quantify the substantial overlap between populations selected with the
different techniques.  At $z \sim 1.8$, they find the sBzK galaxies
dominate the stellar mass density of the star forming galaxy
population (we assume only star forming galaxies contribute to the
cosmic gas density), with comoving stellar mass density of
$\rho_{stars} = 5.9\times 10^7$ M$_\odot$\,Mpc$^{-3}$.  Likewise for
BX/BM/LBG at $z \sim 2.5$, for which they find $\rho_{stars} =
2.8\times 10^7$ M$_\odot$\,Mpc$^{-3}$, and at $z \sim 3.3$, with
$\rho_{stars} = 1.2\times 10^7$ M$_\odot$\,Mpc$^{-3}$.

Assuming unity ratio with the gas mass we can then establish lower
limits to the cosmic gas mass density at the respective epochs,
plotted in Fig.~\ref{omega_h2}. We also include the $z$\,=\,0
measurement of Keres et al.\ (2003). Interestingly, the sBzK limit is
already pushing the cosmic density into the modeled regime based on
e.g., the star formation history of the universe and star formation
laws.

We emphasize that this analysis is mainly illustrative, with very
substantial uncertainties.  First, we assume the unity
gas--to--stellar mass ratios hold for galaxies well below the masses
of those current observed in CO. In particular, the galaxies in the
Daddi et al.\ (2010) and Tacconi et al.\ (2010) samples were typically
at the high--mass end of the CSG distribution, although having
`main--sequence' properties otherwise (eg. sSFR, gas consumption
timescales). Interestingly, the larger sample of Tacconi et al. (2012) 
shows a trend for increasing gas fraction with decreasing stellar mass,
which would increase the cosmic gas densities in Fig.~\ref{omega_h2}.
Second, we adopt a standard GMC value of $\alpha$, when
in fact this value could increase dramatically with e.g. decreasing
metallicity, such that the GMC value radically underestimates the total
gas mass (e.g., Genzel et al.\ 2012). And third, we currently cannot
rule--out a population of lower--mass, gas rich galaxies that do not
appear in any optical survey.

{\bf SIDEBAR: Studies of the molecular medium at high redshift have
been restricted to galaxies that have been pre--selected in the
optical or infrared wavebands through their star formation
activity. This could potentially lead to a biased view of the
molecular gas properties of high--redshift galaxies. A promising way
forward is through observations of molecular deep fields,
i.e. complete frequency scans towards regions in the sky that have
superb multi--wavelength observations available. Such observations
were prohibitive given the sensitivity and bandwidth of past
facilities; a situation that will be changing with ALMA and the
JVLA. The principal remaining uncertainty in determining the cosmic
space density of molecular gas ($\Omega_{H2}(z)$) will be the
calibration of the conversion factor, $\alpha$.}

\section{Summary Points / Concluding Remarks}
\label{summary_points}

Over the last decade, observations of the cool ISM in distant galaxies
via molecular line and atomic fine structure line emission has gone
from a curious look into a few extreme, rare objects, to a mainstream
tool in the study of galaxy formation, out to the highest redshifts
(z$\sim$7). Molecular gas has now been observed in close to 200
galaxies, including numerous AGN host--galaxies, extreme starburst
SMGs, and increasing samples of `main--sequence' CSG. Studies have
moved well beyond simple detection, to dynamical imaging at kpc--scale
resolution, and multi--line, multi--species studies of the ISM in early
galaxies. Study of atomic fine structure line emission is also rapidly
accelerating, with some tens of galaxies detected in \cii\ 158$\mu$m,
and other species, at $z > 1$, including detection of the most distant
quasar with a spectroscopic redshift ($z$\,=\,7.08).

The results of these studies are extremely telling for models of
galaxy formation, providing the required complement to studies of the
stars and star formation in early galaxies. One of the most exciting
empirical result is the discovery that CSG have CO luminosities
approaching those of SMGs and quasar hosts, but FIR luminosities close
to an order of magnitude less. The higher space density of the CSG
galaxies provides a rich hunting ground for molecular line studies of
distant galaxies.  Observations suggest that the gas fraction ($\rm
M_{gas}/M_{stars}$) in massive disk galaxies increases by an order of
magnitude from $z\sim 0$ to $z \ge 1.5$. Hence, the epoch of peak
cosmic star formation density corresponds to an epoch of gas--dominated
disks.  

For the rarer, hyper--starburst galaxies, the quasar hosts and powerful
radio galaxies show the most extreme gas properties, in terms of gas
excitation, star formation `efficiency', and compact although complex,
gas morphologies. These results indicate compact, hyper--starbursts
coeval with Eddington--limited AGN accretion.  Submm galaxies are a
mixed bag of gas rich mergers and extended, gas rich disks in 
dense cosmic environments. 

Current measurements suggest that the hyper--starbursts have a low CO
luminosity to gas mass conversion factor, $\alpha \sim 0.8$,
consistent with the extreme dense ISM conditions seen in nearby
nuclear starbursts. The CSG are consistent with a Milky Way GMC value
of $\alpha \sim 4$. There is increasing evidence that $\alpha$
increases with decreasing metallicity in galaxies, and in general,
there may be a continuum of values of $\alpha$, depending on ISM
pressure, dynamics, and metallicity. The correlation between CO and
FIR luminosity suggests two populations: starburst galaxies with rapid
gas consumption timescales of a few $\times 10^7$ years, and main
sequence galaxies with gas consumption timescales an order of
magnitude longer.

The strong ISM gas cooling line from \cii\ is proving to be a key tool
in the study of the dynamics of the earliest galaxies.  \cii\ imaging
of $z > 4$ galaxies has already revealed a `maximal starburst disk' on
sub--kpc scales, likely rotating disks on few to 10 kpc--scales, and
possible tidal structures on even larger scales. ALMA has already
demonstrated the ability to detect \cii\ emission from LAEs and LBGs
at high redshift.

We have made a first attempt at quantifying the dense gas history of
the Universe, based on current observations.  While admittedly gross,
these measurements are consistent with modeling based on large scale 
cosmological simulations, and on empirical models based on assumed star
formation laws. 

\section{Future directions}
\label{future}

We re--emphasize that we stand at a cusp in knowledge, with the
breath--taking promise of the ALMA and JVLA poised to revolutionize the
study of the cool gas in early galaxies, and we have presented some
of the early science results from these telescopes herein. Later in the
decade, telescopes such as CCAT and NOEMA will also make important
contributions. 

We are no longer limited by numbers of high--z sources to study --
there are thousands of CSG, quasars, and SMGs from cosmological deep
fields. These populations have been delineated in terms of their space
density, cosmic environment, and redshift distribution in remarkable
detail.  At this stage, we feel it is the high resolution imaging
capabilities of ALMA and the JVLA that will be most incisive for
unraveling the complex processes involved in early galaxy formation.
Some of the key questions that need to be addressed include:

$\bullet$ For interpreting CO observations, the key uncertainty
remains the conversion factor, especially in low metallicity systems
where lack of dust shielding may dramatically reduce the CO content.
We expect calibrating this relationship will follow the current path
using multiple methods, including dynamics, radiative transfer modeling,
and dust--to--gas modeling, leading to a concordance of estimates, and
likely multi--parameter models involving eg. metallicity.

$\bullet$ Spatially resolved imaging (sub--kpc) of multiple CO transitions,
as well as of the thermal dust continuum emission, is needed to study
the relative distribution and excitation of the fuel for star
formation with respect to regions of active star formation.  Such
imaging also allows for a study of star formation laws as a function of
surface brightness, and not integrated quantities.

$\bullet$ Observations of high dipole moment molecules, and other complex
molecules, will allow for detailed astrochemical modeling of the dense
gas immediately involved in active star formation in distant galaxies.

$\bullet$ An inventory of fine structure lines is required to set the 
ISM gas cooling budget, AGN versus star formation indicators,  
and for metallicity determinations.  

$\bullet$ We expect that studies of the \cii\ 158$\mu$m line will play an
important role in the determination of redshifts and dynamics 
of the first galaxies, well into cosmic reionization.

The new telescopes coming on--line open up the very real possibility
of performing blind, deep field surveys for molecular gas, and atomic
fine structure line emission, from distant galaxies. Over the next few
years, we expect that the dense gas history of the Universe diagram
will be fully populated, as the CO luminosity functions and conversion
factors are quantified out to the highest redshifts. Coupled with the
near--IR through X--ray studies of stars, star formation, and AGN,
such pan--chromatic deep fields will provide a complete picture of the
conversion of gas into stars over the history of the Cosmos.

\newpage

{\bf Acknowledgements:} The authors are indebted to their many
long--term collaborators who have all greatly contributed to the field
of high--redshift molecular gas emission. They include D. Riechers,
R. Wang, A. Weiss, E. Daddi, P. Cox, R. Neri, K. Menten, F. Bertoldi,
J. Wagg, M. Aravena, J. Hodge, R. Decarli. We thank D. Riechers,
D. Narayanan, L.  Tacconi, E. van Dishoeck, G. Stacey, R. Decarli,
A. Weiss, E. da Cunha, A. van der Wel, M. Sargent for their valuable
detailed input regarding this manuscript. We thank R.\ Decarli in
particular for his help with preparing the figures. FW thanks the
Aspen Center for Physics, where part of this work was conducted. CC
thanks the Astrophysics Group, Cavendish Laboratory, Cambridge for
support while much of this maunscript was written.

\section*{Acronyms}

\begin{itemize}
\item AGN: Active galactic nucleus
\item CMA: Cold mode accretion
\item CMB: Cosmic microwave background
\item CSG: Color--selected starforming galaxy (BzK--, BM/BX--selected)
\item DGHU: Dense gas history of the universe
\item FIR: Far--infrared
\item FSL: (Atomic) fine structure line
\item GMC: Giant Molecular Clouds
\item ISM: Interstellar Medium
\item LTE: Local thermodynamic equlibrium
\item LVG: Large velocity gradient
\item PDR: Photon dominated region (aka: Photodissociation region)
\item SMBH: Supermassive black hole
\item SMG: Submillimeter Galaxy
\item QSO: Quasi--stellar object, or quasar
\item XDR: X--ray dominated region
\end{itemize}

\newpage

\begin{table}
\label{tab_molecules}
{
\caption{Fundamental parameters for frequently observed molecules and fine structure lines. Numbers for the atomic fine structure constants are taken from Stacey et al.\ (2011). Einstein $A$ coefficients, rest frequencies and collision rates $\gamma$ are taken from the Leiden Atomic and Molecular Database (Sch\"oier et al.\ 2005). Column E.P. is the excitation potential of the upper level above ground. The ciritcal density is the density at which the rate of the collisional depopulation of a quantum level equals the spontaneous radiative decay rate. We note that definition of the critical density used here is n$_{crit}$\,=\,A$\gamma$ (Sec.~\ref{temp_dens}). This is not the proper definition which includes the summation of all collisional transitions to the lower level -- such a treatment will lower the critical densities presented here. The critical densities also decrease if the lines are optically thick. For species occuring in neutral gas clouds, the collision partneers are H and H$_2$ (assumed T$_{\rm gas}$=100\,K). For species occuring in ionized gas regions, the collision partners are electrons (marked with a [$\star$] in the last column). For details see Stacey et al.\ 2011.
}
\begin{tabular}{l|cccccc}
species& trans.           &E.P.  &$\lambda$& $\nu$    &  Einstein A            & n$_{crit}$           \\
       &                  & K    & $\mu$m  & GHz      & s$^{-1}$               & cm$^{-3}$            \\
\hline
\oi    & $^3P_1\to^3P_2$  & 228  & 63.18   & 4744.8   & 9.0$\times$10$^{-5}$   &  4.7$\times$10$^{5}$ \\ 
       & $^3P_0\to^3P_1$  & 329  & 145.53  & 2060.1   & 1.7$\times$10$^{-5}$   &  9.4$\times$10$^{4}$ \\
\hline
\oiii  & $^3P_2\to^3P_1$  & 440  & 51.82   & 5785.9   & 9.8$\times$10$^{-5}$   &  3.6$\times$10$^{3}$ [$\star$] \\
       & $^3P_1\to^3P_0$  & 163  & 88.36   & 3393.0   & 2.6$\times$10$^{-5}$   &  510 [$\star$]\\
\hline
\cii   &$^3P_{3/2}\to^3\!\!P_{1/2}$& 91 & 157.74 & 1900.5 & 2.1$\times$10$^{-6}$   &  2.8$\times$10$^{3}$ \\
       &                       &    &        &        &                        &  50  [$\star$] \\
\hline
\nii   & $^3P_1\to^3P_2$  & 188  &  121.90 & 2459.4   & 7.5$\times$10$^{-6}$   &  310 [$\star$] \\
       & $^3P_1\to^3P_0$  & 70   &  205.18 & 1461.1   & 2.1$\times$10$^{-6}$   &  48  [$\star$] \\
\hline
\ci    & $^3P_2\to^3P_1$  & 63   &  370.42 & 809.34   & 2.7$\times$10$^{-7}$   &  1.2$\times$10$^{3}$ \\
       & $^3P_1\to^3P_0$  & 24   &  609.14 & 492.16   & 7.9$\times$10$^{-8}$   &  470 \\
\hline
CO     & J=1--0           &  5.5 &    2601 &  115.27  & 7.2$\times$10$^{-8}$  & 2.1$\times$10$^{3}$    \\   
       & J=2--1  	  & 16.6 &    1300 &  230.54  & 6.9$\times$10$^{-7}$  & 1.1$\times$10$^{4}$    \\  
       & J=3--2  	  & 33.2 &     867 &  345.80  & 2.5$\times$10$^{-6}$  & 3.6$\times$10$^{4}$    \\
       & J=4--3  	  & 55.3 &   650.3 &  461.04  & 6.1$\times$10$^{-6}$  & 8.7$\times$10$^{4}$    \\
       & J=5--4  	  & 83.0 &   520.2 &  576.27  & 1.2$\times$10$^{-5}$  & 1.7$\times$10$^{5}$    \\  
       & J=6--5  	  &116.2 &   433.6 &  691.47  & 2.1$\times$10$^{-5}$  & 2.9$\times$10$^{5}$    \\  
       & J=7--6  	  &154.9 &   371.7 &  806.65  & 3.4$\times$10$^{-5}$  & 4.5$\times$10$^{5}$    \\  
       & J=8--7  	  &199.1 &   325.2 &  921.80  & 5.1$\times$10$^{-5}$  & 6.4$\times$10$^{5}$    \\  
       & J=9--8  	  &248.9 &   289.1 & 1036.9   & 7.3$\times$10$^{-5}$  & 8.7$\times$10$^{5}$    \\  
       & J=10--9 	  &304.2 &   260.2 & 1152.0   & 1.0$\times$10$^{-4}$  & 1.1$\times$10$^{6}$    \\  
\hline
HCN    & J=1--0           &  4.25&    3383 &  88.63   & 2.4$\times$10$^{-05}$  & 2.6$\times$10$^{6}$    \\         
       & J=2--1 	  & 12.76&    1691 & 177.26   & 2.3$\times$10$^{-04}$  &	1.8$\times$10$^{7}$    \\        
       & J=3--2 	  & 25.52&    1128 & 265.89   & 8.4$\times$10$^{-04}$  &	6.8$\times$10$^{7}$    \\        
       & J=4--3 	  & 42.53&   845.7 & 354.51   & 2.1$\times$10$^{-03}$  &	1.8$\times$10$^{8}$    \\        
       & J=5--4 	  & 63.80&   676.5 & 443.12   & 4.1$\times$10$^{-03}$  &	3.8$\times$10$^{8}$    \\        
       & J=6--5 	  & 89.32&   563.8 & 531.72   & 7.2$\times$10$^{-03}$  &	7.1$\times$10$^{8}$    \\        
       & J=7--6 	  &119.09&   483.3 & 620.30   & 1.2$\times$10$^{-02}$  &	1.2$\times$10$^{9}$    \\        
       & J=8--7 	  &153.11&   422.9 & 708.88   & 1.7$\times$10$^{-02}$  &	1.8$\times$10$^{9}$    \\        
       & J=9--8 	  &191.38&   375.9 & 797.43   & 2.5$\times$10$^{-02}$  &	2.5$\times$10$^{9}$    \\        
       & J=10--9	  &233.90&   338.4 & 885.97   & 3.4$\times$10$^{-02}$  &	3.3$\times$10$^{9}$    \\        
\hline
\end{tabular}
}
\end{table}

\begin{table}
\caption{Factors to calculate L$'_{\rm CO(1-0)}$ from higher--J transitions up to J\,=\,5. For the SMGs and QSOs average values are quoted based on all available literature estimates. For the CSG the ratio is taken from Dannerbauer et al.\ (2009) and Aravena et al.\ (2010); their LVG model 1. The values for the Milky Way and the centre of M82 are from Wei\ss\ et al. (2005b) `--' indicates that no well--constrained value is available. We assume the same excitation as for the CSG for the LBGs, SFRGs. We adopt QSO excitation for the RGs and SMG excitation for the 24$\mu$m and ERO sources.}
\label{tab_lum_ratios}
\begin{tabular}{lccccc}
\toprule
source & SMG  & QSO & CSG & MW & M82 \\
\colrule
L$'_{\rm CO(2-1)}$/L$'_{\rm CO(1-0)}$ & 0.85  &  0.99  & 0.97 & 0.5    & 0.98   \\
L$'_{\rm CO(3-2)}$/L$'_{\rm CO(1-0)}$ & 0.66  &  0.97  & 0.56 & 0.27   & 0.93   \\
L$'_{\rm CO(4-3)}$/L$'_{\rm CO(1-0)}$ & 0.46  &  0.87  & 0.2  & 0.17   & 0.85   \\
L$'_{\rm CO(5-4)}$/L$'_{\rm CO(1-0)}$ & 0.39  &  0.69  & --   & 0.08   & 0.75   \\
\botrule
\end{tabular}
\end{table}


\newpage
\begin{figure}
\centerline{\psfig{figure=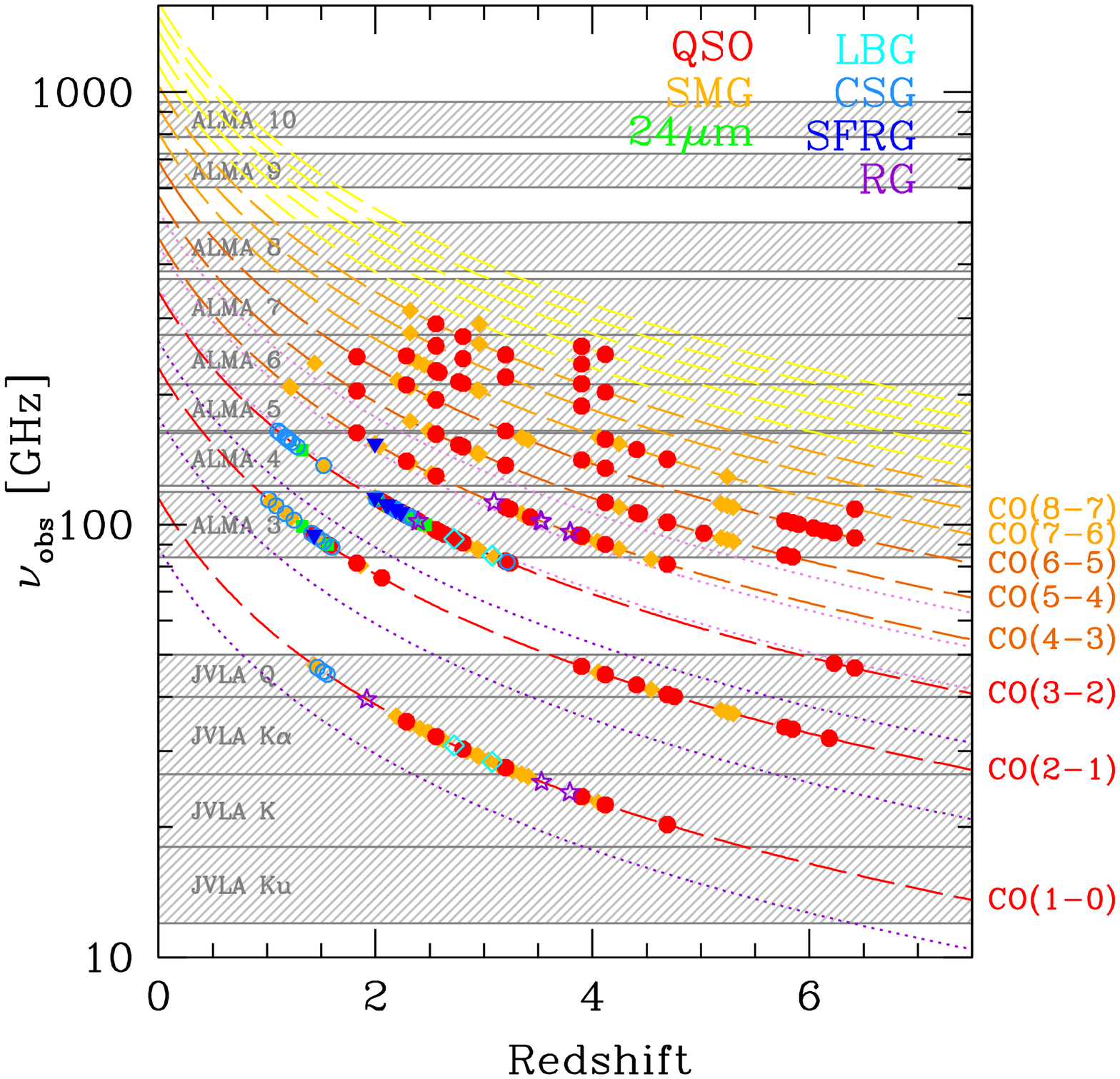,height=7cm}\psfig{figure=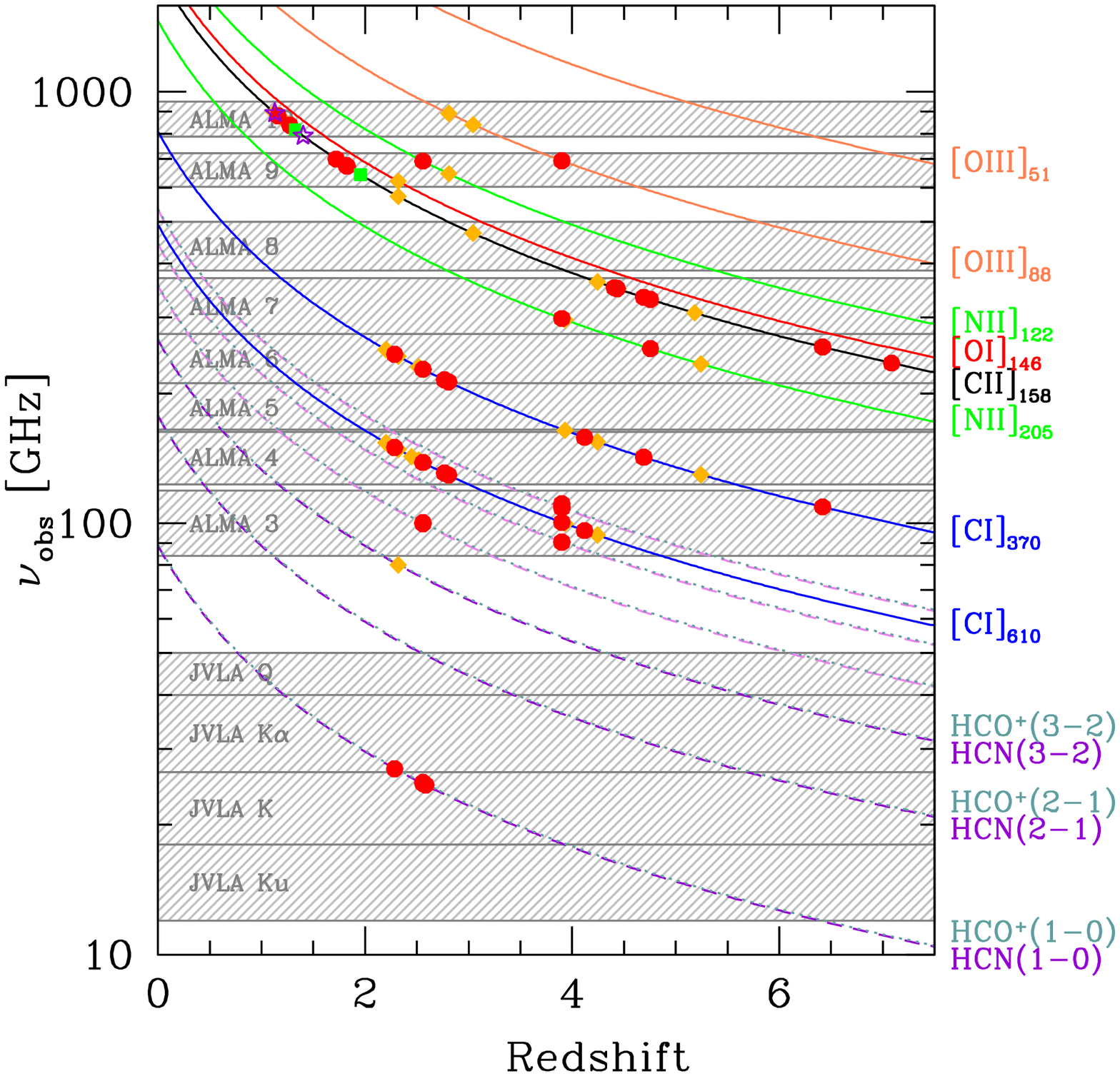,height=7cm}}

\caption{Redshifted frequencies $\nu_{obs}$ of CO transitions (left)
and other key tracers of the starforming ISM (right) as a function of
redshift $z$, following $\nu_{obs}=\nu_{rest}/(1+z)$.  The shaded
areas indicate the frequency bands covered by various
telescopes. Highlighted are the ALMA frequency bands as well as the
`high--frequency' bands of the JVLA. The colored points indicate
detection of all high redshift (z$>1$) lines. The color of the points
refer to the different source types, as explained in the left panel.
}

\label{fig_co_coverage}
\end{figure}

\newpage
\begin{figure}
\centerline{\psfig{figure=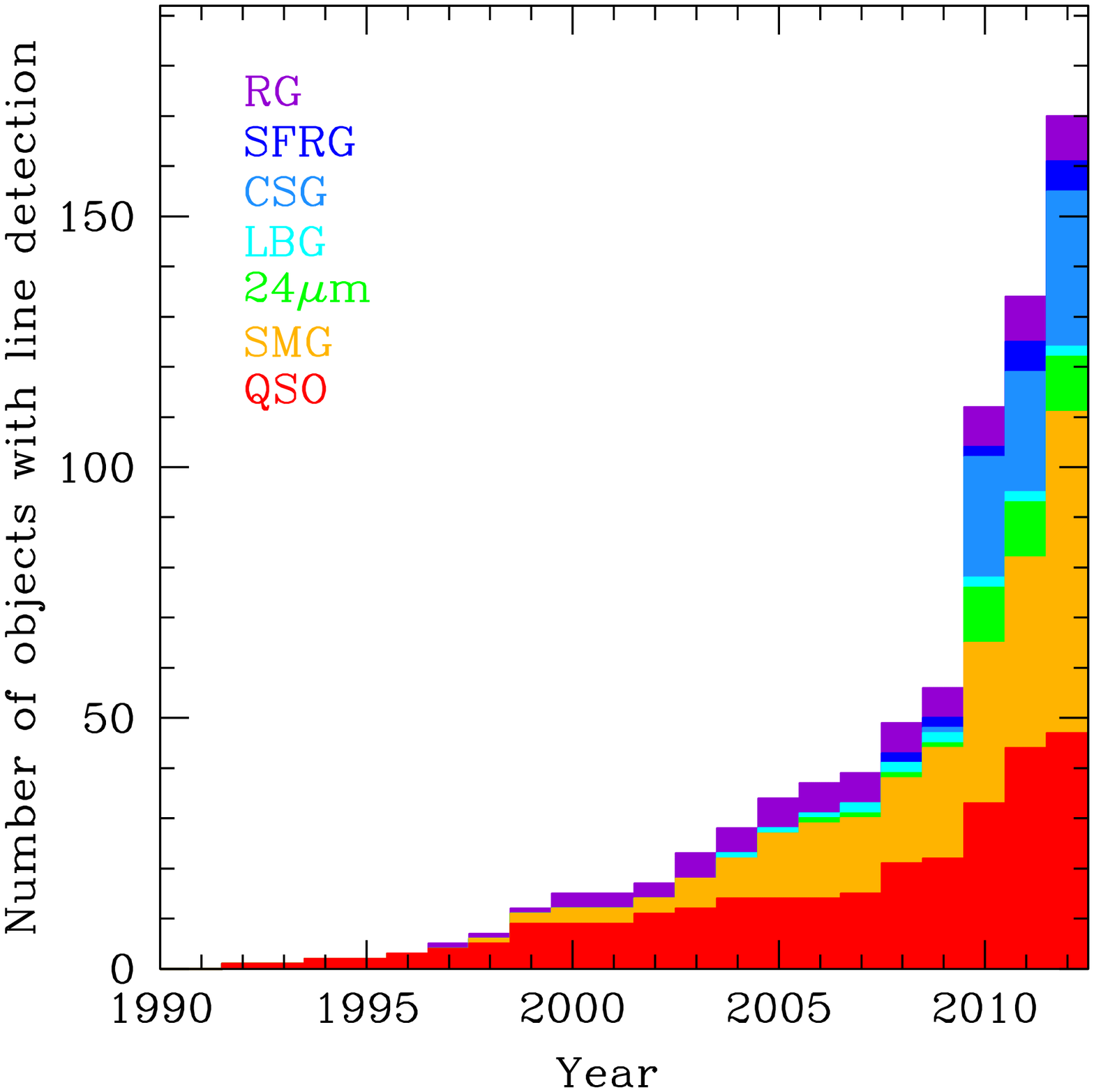,height=7cm}\psfig{figure=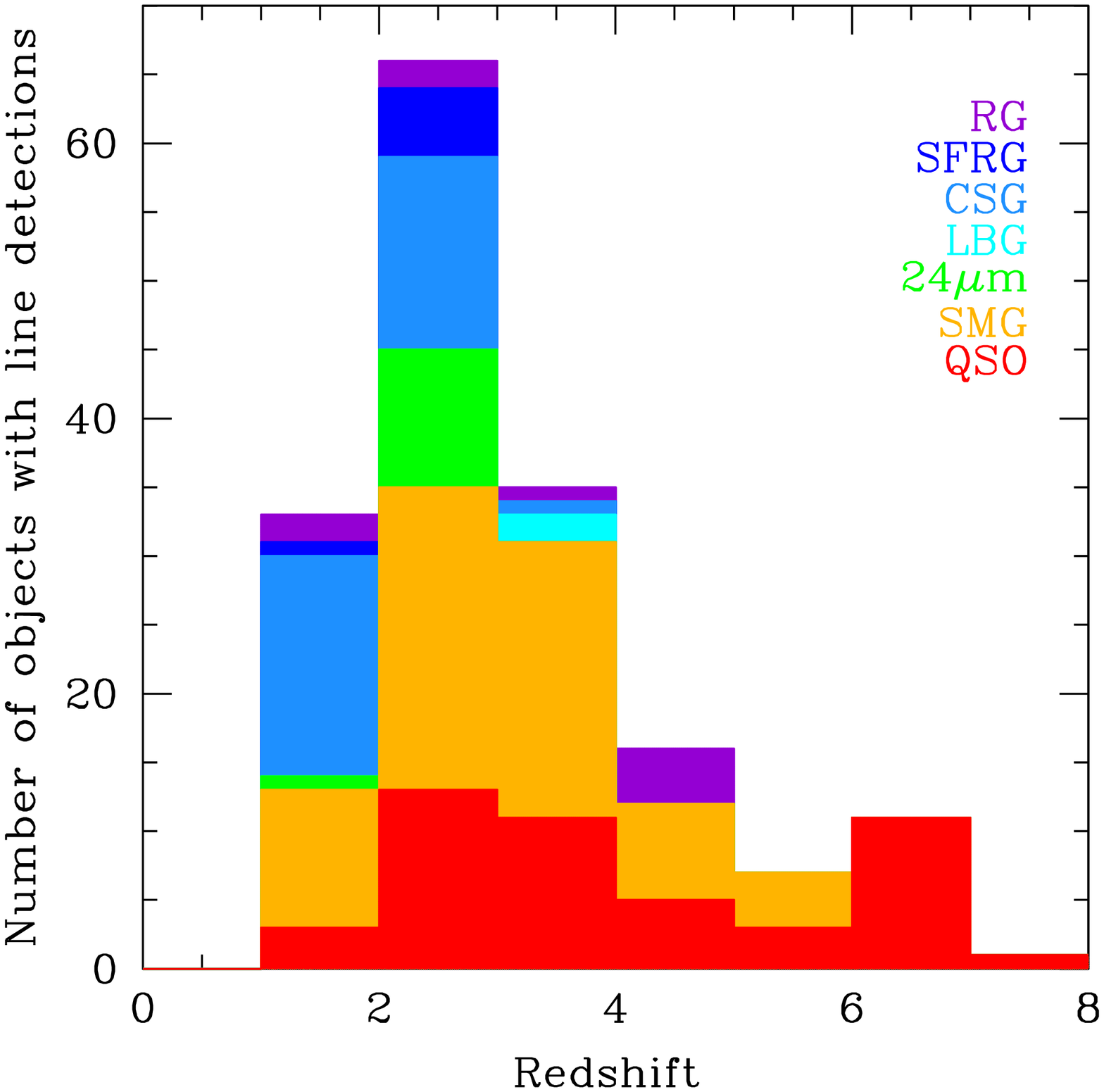,height=7cm}}

\caption{{\em Left:} Discovery history of high--redshift (z$>$1) line
detections. The cummulative number of detections is shown, and the
different colors indicate the different galaxies
populations. Historically, QSOs (Sec.~\ref{quasars}), SMGs
(Sec.~\ref{smgs}) and radio galaxies (Sec.~\ref{radiogals}) have been
the focus of most studies. In recent years, these have been
complemented by observations of `main sequence' starforming galaxies
(CSG, Sec.~\ref{cssfg}). To date close to 200 galaxies have been
detected in line emission at z$>$1. {\em Right:} Redshift distribution of all
sources for all z$>$1 line detections. The highest redshift sources
$z>5$ detected are the QSOs, with a growing contribution from
SMGs.  }

\label{fig_history}
\end{figure}

\newpage
\begin{figure}
\centerline{\psfig{figure=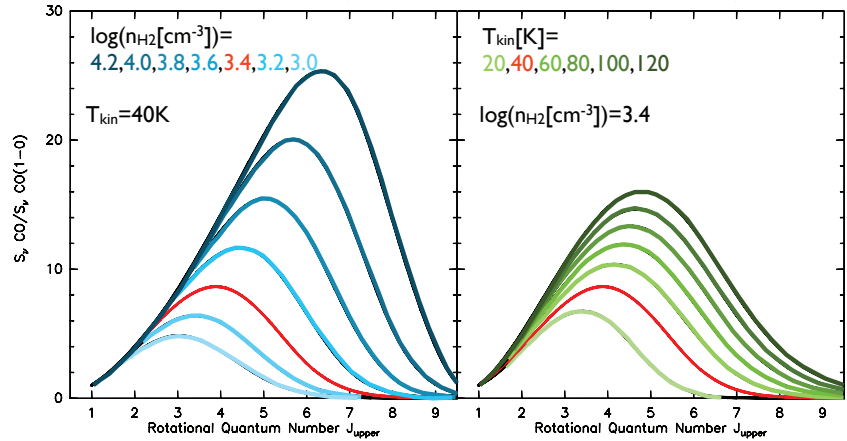,height=6cm}}

\caption {This figure illustrates how the measured CO emission ladder
changes as a function of temperature and density (adopted from Wei\ss\
et al.\ (2007). The left panel shows the effect of changing density at
fixed temperature (T$_{kin}$\,=\,40\,K. The right panel shows the
effect of varying kinetic temperatures for a fixed density
(log(n(H$_2$))\,=\,3.4).  Both panels have been normalized to the
CO(1--0) transition.  High CO excitation is achieved through a
combination of high kinetic temperature and high density. Given the
typically sparsely sampled CO excitation and large error bars in high
redshift observations, this degeneracy can not be easily broken
observationally. Additional information, such as independent estimates
of the kinetic temperature through \ci\ or dust measurements can help
to break this degeneracy. Note that increased temperatures lead to a
broader CO emission ladder, as more and more high--J levels are
populated following the Boltzmann distribution.  }

\label{fig_co_seds}
\end{figure}

\newpage
\begin{figure}

\centerline{\psfig{figure=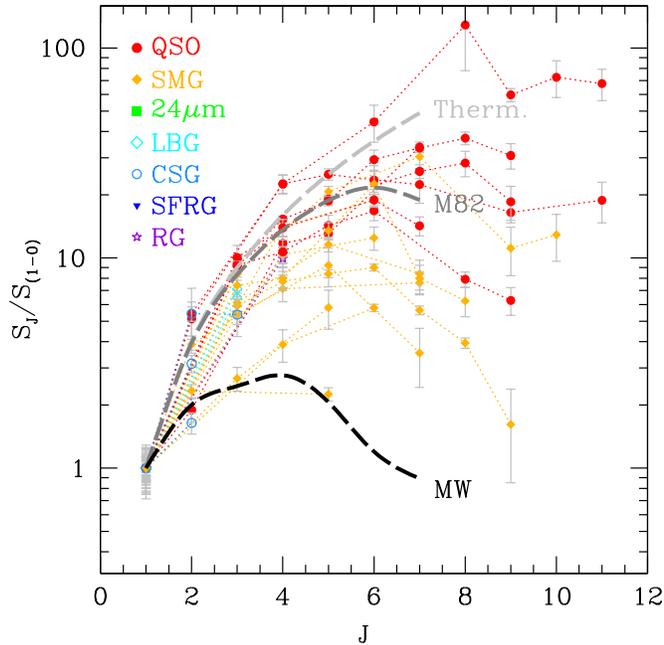,height=9cm}}

\caption {CO emission ladder of all sources where the CO(1--0) line
has been measured. The CO line flux is shown as a function of
rotational quantum number and the colors indicate the different source
types. Measurements for individual sources are connected by a dashed
line. The line fluxes have been normalized to the CO(1--0) line. The
QSOs are the most excited systems found, with an average peak of the
CO ladder at around J$\sim$6. This is consistent with the high star
formation rate and compact emission regions in their host
galaxies. The SMGs are slightly less excited on average, and their CO
emission ladder peak around J$\sim$5. The dashed thick lines indicate
template CO emission ladders for the Milky Way (black) and M\,82
(grey). The dashed dark grey line shows constant brightness
temperature on the Rayleigh--Jeans scale, i.e. $S\sim\nu^2$ (note that
this approximation is not valid for high J).  } 

\label{fig_co_sed_all}
\end{figure}

\newpage
\begin{figure}
\centerline{\psfig{figure=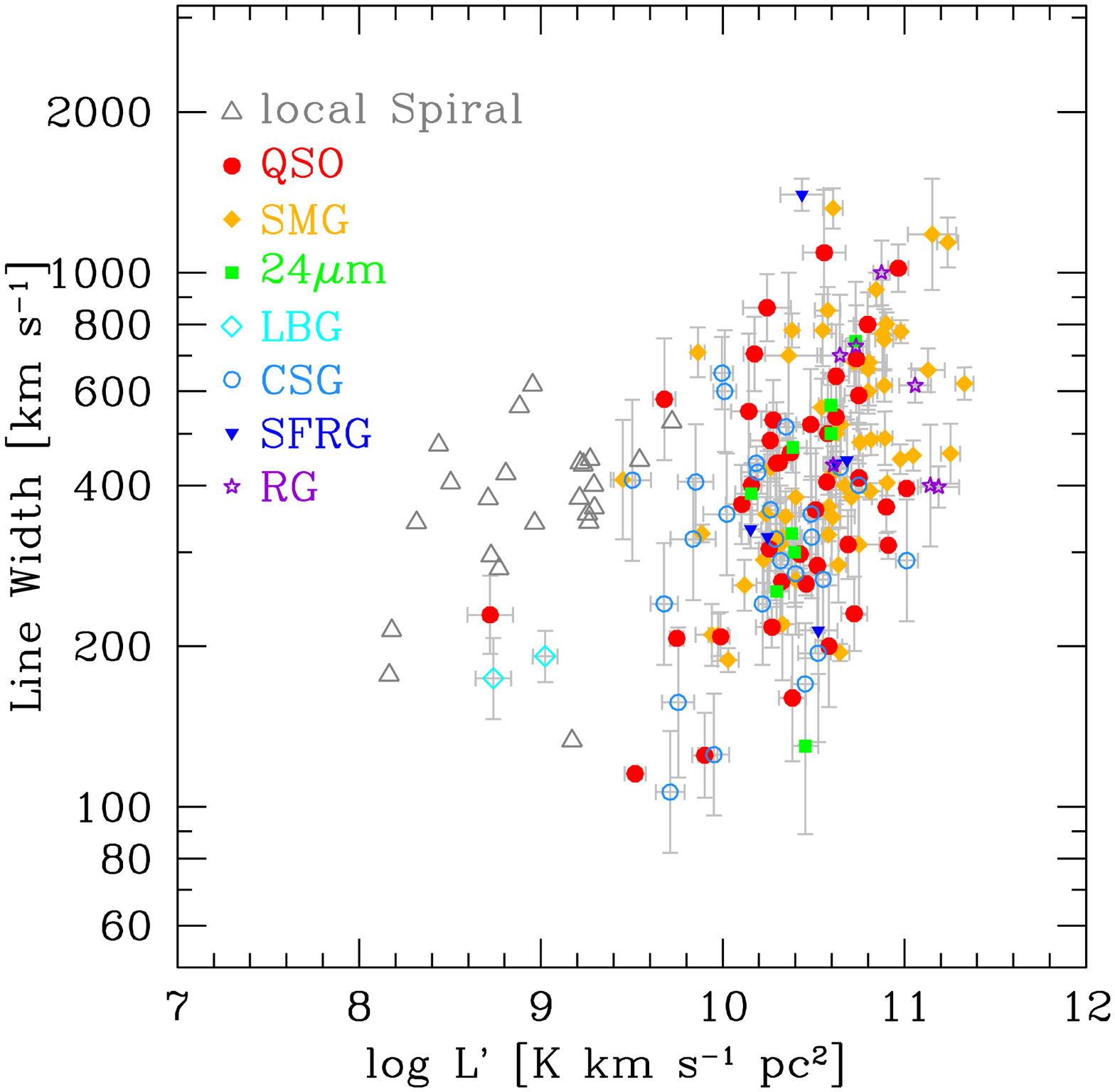,height=7cm}\psfig{figure=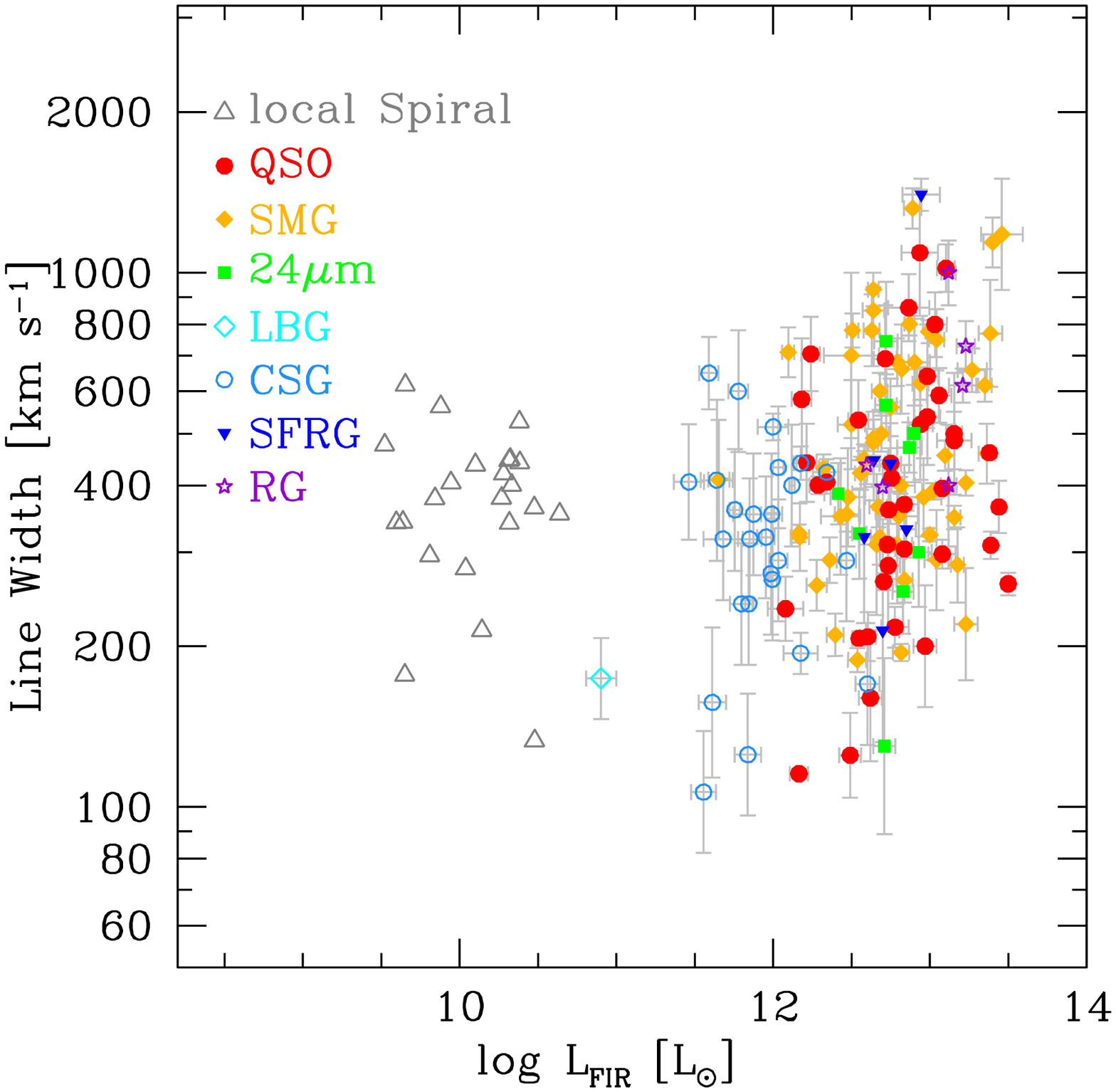,height=7cm}}

\caption{The CO line width (FWHM) versus CO line luminosity (left) and versus 
the FIR luminosity (right panel).  Note that the CSG show systematically 
lower line widths for a
given CO line luminosity than the hyper--starburst quasar hosts and
SMGs (we have corrected the v$_{\rm rot, max}$ values given in Tacconi
et al. (2010) to give accurate FWHM values). The grey datapoints show
local spiral galaxies (with stellar masses $>$10$^{10}$\,M$_\odot$ 
from the HERACLES/THINGS surveys (Leroy et
al. 2009, Walter et al.\ 2008). The local FWHM values are corrected
for inclination, the high--z values are not (in the absence of unknown
inclinations).}  

\label{fig_l_fwhm} \end{figure}

\newpage
\begin{figure}
\centerline{\psfig{figure=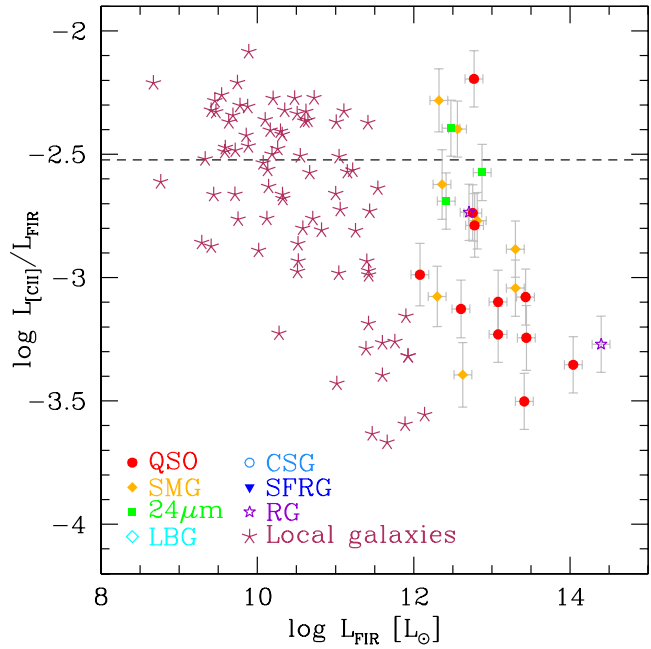,height=9cm}}

\caption{The ratio L$_{\cii}$/L$_{\rm FIR}$ as a function of
L$_{\rm FIR}$ (data from Table 3). The dashed horizontal line indicates a value of
3$\times10^{-3} \sim$ Milky Way value. The L$_{\rm FIR}$ measurements are corrected for lensing (where known).}

\label{fig_cii_fir}
\end{figure}

\newpage
\begin{figure}
\centerline{\psfig{figure=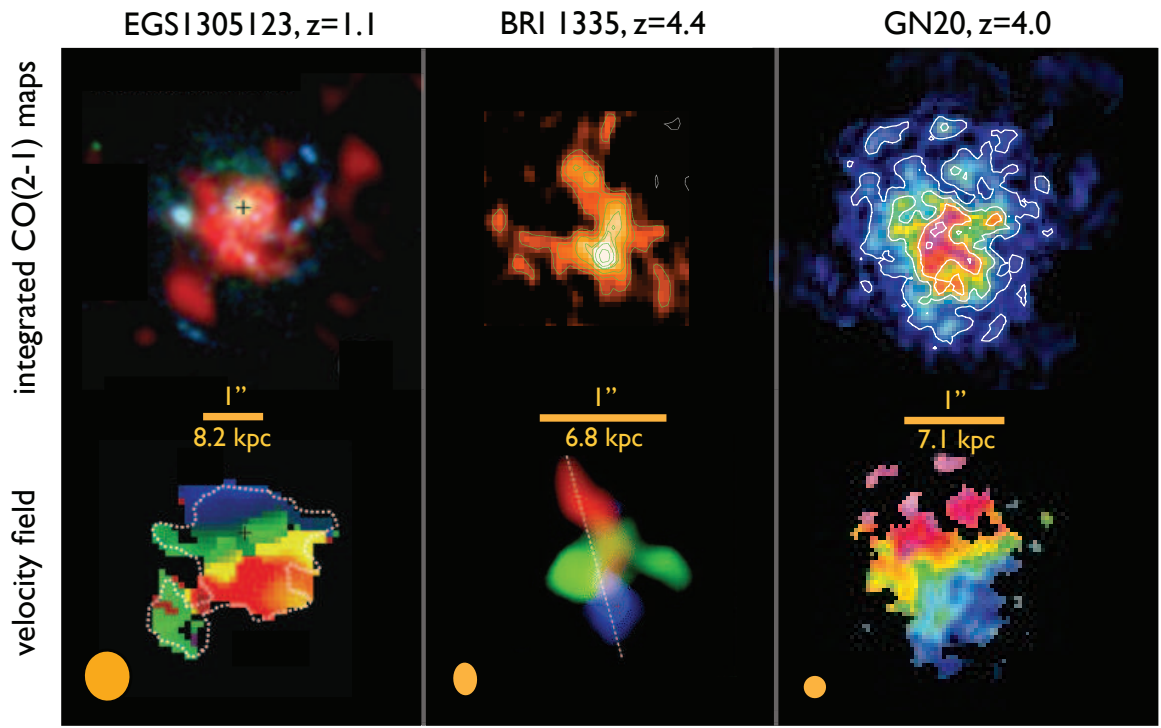,height=10cm}}

\caption{Best examples of resolved molecular line
emission at high redshift from which gas kinematics can be
derived. These are (from left to right): The CSG EGS\,1305123
(Tacconi et al.\ 2010), the QSO BRI\,1335 (Riechers et al.\ 2008b) and
the SMG GN\,20 (Hodge et al.\ 2012). The top row shows the integrated
CO(2--1) maps of the targets. The bottom row shows the velocity fields
of the targets; here the color indicates the velocity at which gas is
moving a given position on the sky. The velocity scale ranges from
(blue to red): --65 to +100 km s$^{-1}$, --154 to +154 km s$^{-1}$, --300
to 300 km s$^{-1}$, respectively.  The beamsizes are given in the
bottom left of each galaxy; the bar indicates 1$"$ on the sky (size in
kpc is also given at the respective redshift).  }

\label{fig_CO_im}
\end{figure}

\newpage
\begin{figure}
\centerline{\psfig{figure=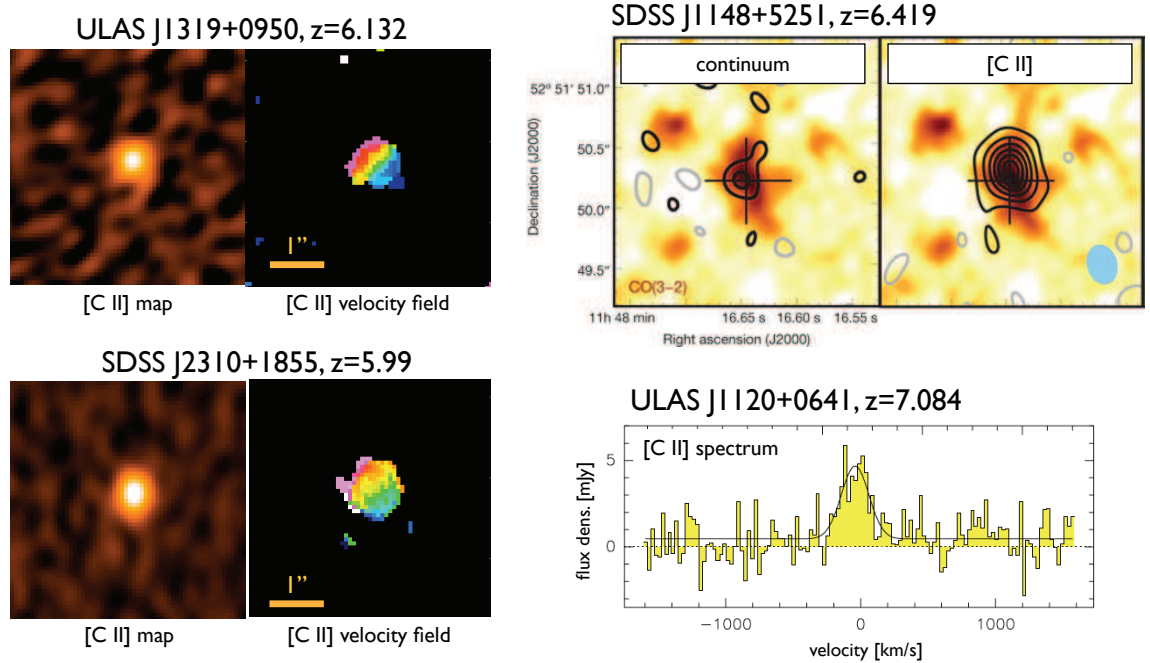,height=9cm}}

\caption{
Results from observations of \cii\ 158$\mu$m emission from $z
\ge$ 6 quasars. Left shows ALMA images of the total intesity plus the
intensity weighted mean gas velocity for two SDSS quasars (Wang et
al.\ 2013). The velocity range in the top figure is $\pm 300$ km
s$^{-1}$ (red to blue), while that in the bottom frame is $\pm 100$ km
s$^{-1}$. Note the clear velocity gradients, consistent with gas
rotation on scales $\sim 7$\,kpc ($\sim$1$"$ at z\,$\sim$\,6).  The upper 
right shows the VLA CO images
of the $z$\,=\,6.42 SDSS quasar, J\,1148+5251, as color scale in both
frames. The contours on the left show the 250\,GHz continuum emission,
and those on the right show the \cii\ 158$\mu$m emission imaged by the
PdBI (from Walter et al.\ 2009a). The lower right shows the PdBI \cii
158$\mu$m spectrum of the most distant spectroscopic redshift quasar
known, a quasar at $z$\,=\,7.08 (Venemans et al.\ 2012). 
}

\label{fig_cii_im}
\end{figure}

\newpage
\begin{figure}
\centerline{\psfig{figure=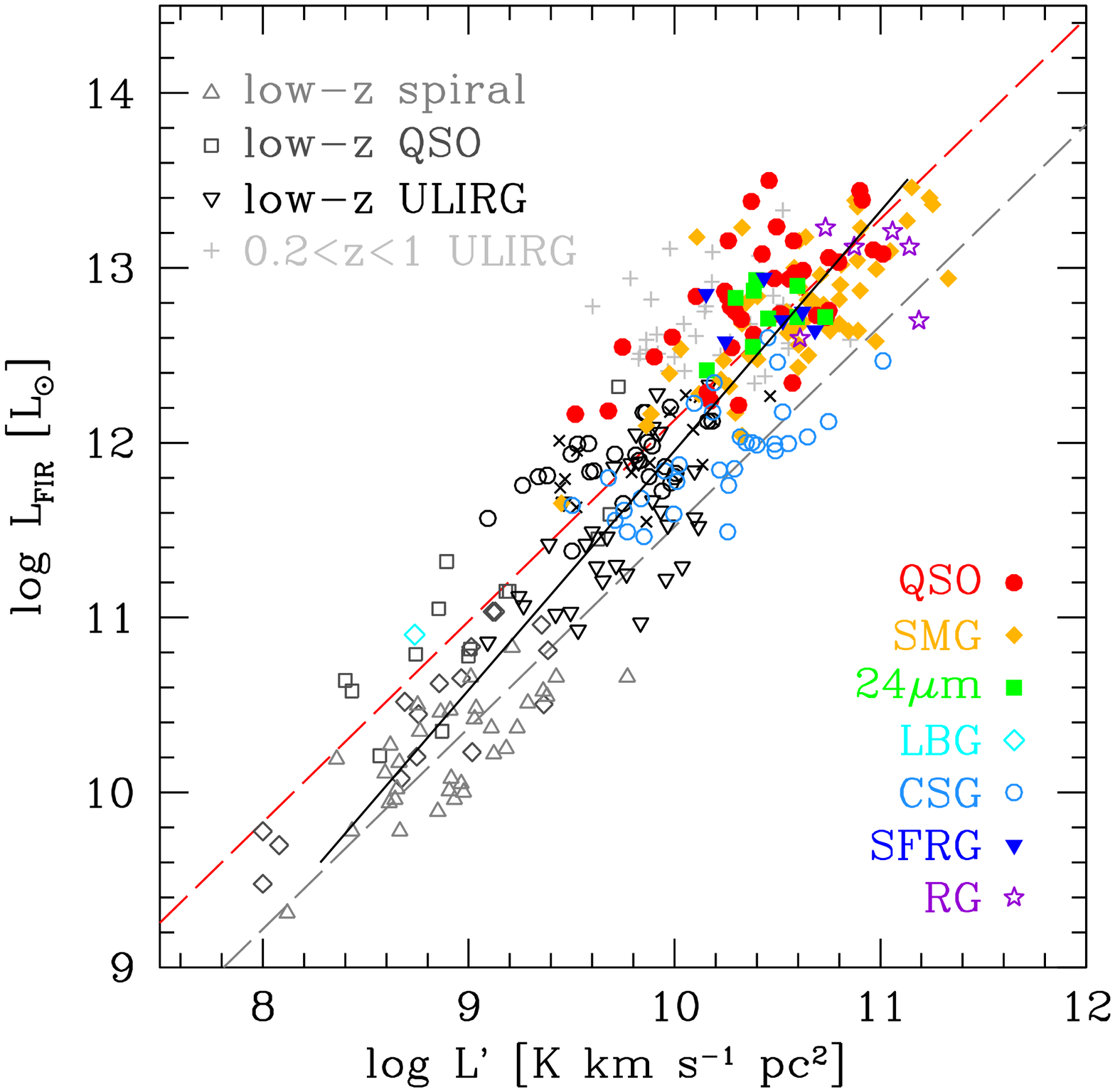,height=7cm}\psfig{figure=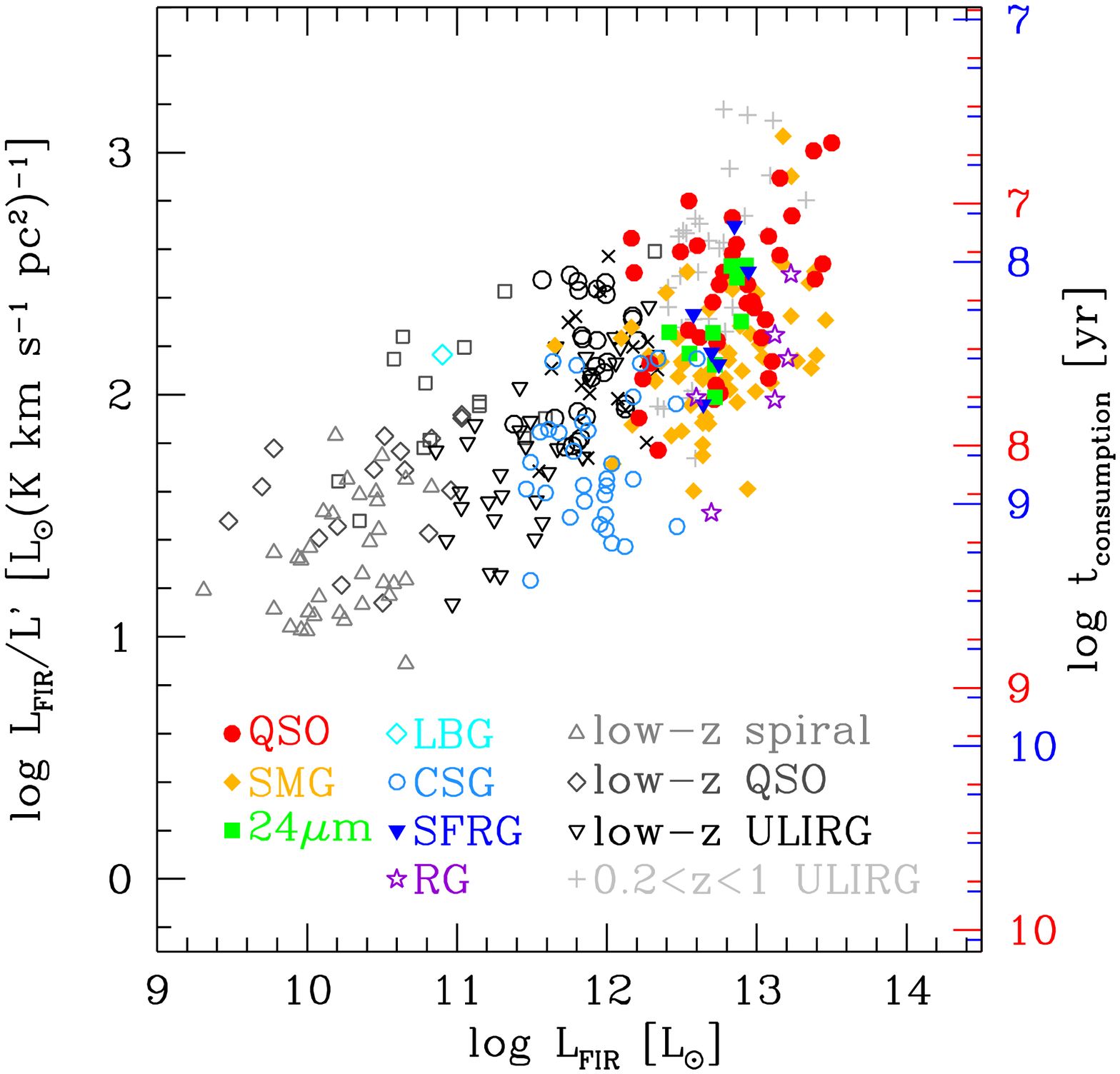,height=7cm}}

\caption {
\small
{\em Left:} L$'_{\rm CO}$ as a function of L$_{\rm FIR}$ for all
systems detected at z$>1$ (colored points). L$'_{\rm CO}$ was calculated
using the lowest available J--measurement and assuming the excitation
correction tabulated in Tab.~\ref{tab_lum_ratios} (see table caption
for details on different source populations). The grey symbols
represent z$<1$ measurements: the 0.2$<$z$<$1 ULIRG sample by Combes
et al.\ 2011 and 2012b (crosses), low--z quasars from the Hamburg--ESO
QSO survey (diamonds, Bertram et al.\ 2007), PG quasars (squares, Evans
et al.\ 2001, Evans 2006, Scoville et al.\ 2003), nearby spiral galaxies
(upward triangles), low--z spirals, starburst galaxies,
and ULIRGs (downward and upward triangles, Gao\& Solomon 2004a,b), the
z$<0.2$ IR QSO sample by Xia et al.\ (2012) and 0.04$<$z$<$0.11 ULIRG
sample by Chung et al.\ (2009, open circles). The full line is a fit to
all data points which gives a slope of 1.35$\pm$0.04. The dashed lines indicate
the best fits for the main sequence galaxies (grey dashed) and
starburst galaxies (red dashed) derived by Genzel et al.\ (2010) and
Daddi et al.\ (2010). {\em Right:} The right hand panel shows the
logarithmic ratio  L$_{\rm FIR}$/L$'_{\rm CO}$ as a function of
L$_{\rm FIR}$. L$_{\rm FIR}$/L$'_{\rm CO}$ is a measure for the star formation
efficiency in an object, and under the assumption of a conversion
factor is the inverse of the consumption time $\tau_{consumption}$. The
consumption time is plotted on the right hand side of the panels for
two different values of $\alpha$, Galactic (red labels) and ULIRG
(blue labels). The consumption time for nearby galaxies and the CSG
are $\sim$1\,Gyr, assuming a Galactic $\alpha$. The consumption times
for the most actively star forming systems are only a few
$\times10^7$\,yr under the typical assumption of a ULIRG $\alpha$.  
}
\label{fig_sf_law}
\end{figure}

\newpage
\begin{figure}
\centerline{\psfig{figure=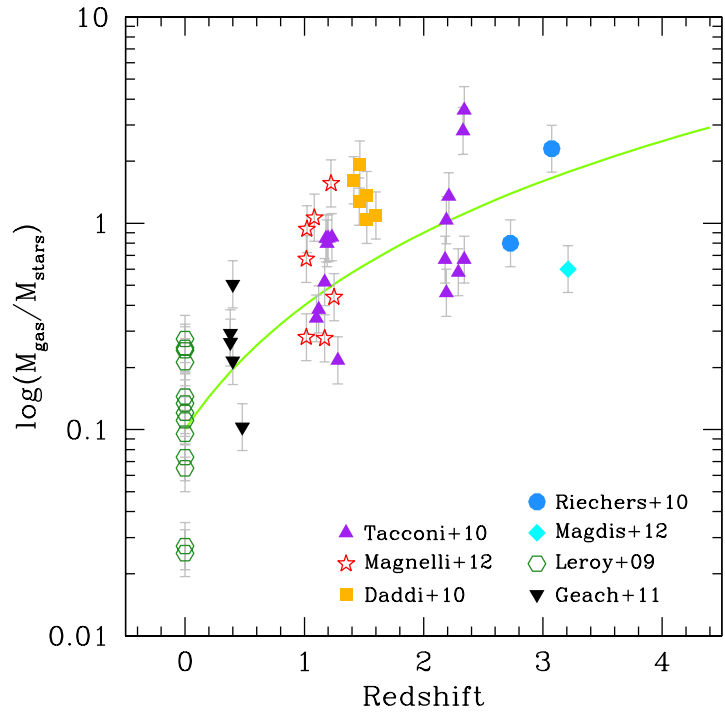,height=10cm}}

\caption{The ratio of gas mass to stellar mass ($\rm
M_{gas}/M_{stars}$), for various galaxy samples. The green circles are
from the z\,=\,0 HERACLES nearby galaxy sample (Leroy et al.\ 2009)
where we only include galaxies with stellar masses
$>10^{10}$\,M$_\odot$, to be consistent with the high--$z$ samples
plotted.  All the points plotted assume $\alpha\!\sim\!4$. The green
curve follows $\rm M_{gas}/M_{stars}$\,=\,0.1$\times$(1\,+\,z)$^2$
(e.g. Geach et al.\ 2011).  } \label{fig_gas_stars} \end{figure}

\newpage
\begin{figure}
\centerline{\psfig{figure=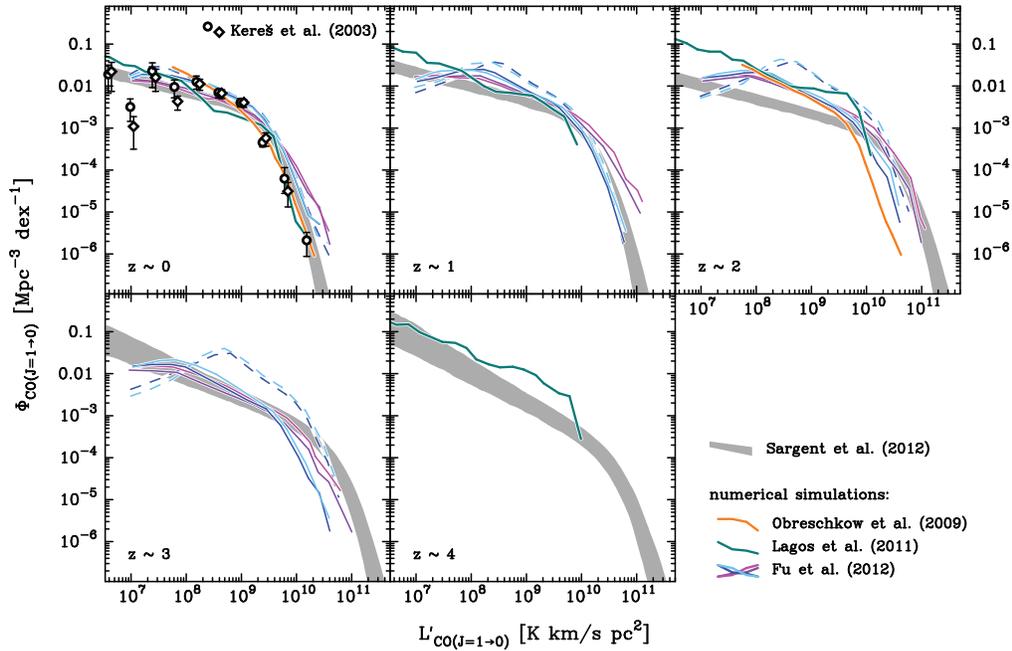,height=9cm}}

\caption {
Models for the evolution of the CO luminosity function based on
semi--analytical cosmological models plus `recipes' to relate gas mass
to CO luminosity (colored lines). The grey shading is from Sargent et
al.\ (2013) and shows the indirectly inferred redshift evolution of
the CO(1--0) luminosity function (grey shading) based on (1) the
evolution of the stellar mass function of star-forming galaxies, (2)
the redshift evolution of the specific SFR of main-sequence galaxies,
(3) the distribution of main-sequence and star-bursting galaxies in
the SFR-$M_{\star}$-plane (Sargent et al. 2012), (4) distinct prescriptions 
of the star formation efficiency of main--sequence and star--bursting galaxies, and (5) a
metallicity-dependent conversion factor $\alpha_{\rm CO}$. The CO
luminosity function includes contributions both from `main--sequence'
and starbursts, where the latter is characterised by a more than
10--fold increase of the star-formation efficiency.  In the upper left
panel the observational constraints on the local CO LF reported in
Kere{\v s} et al.\ (2003) are also shown (see Sargent et al.\ 2013 for
details).
}
\label{fig_lum_functions}
\end{figure}

\newpage
\begin{figure}
\centerline{\psfig{figure=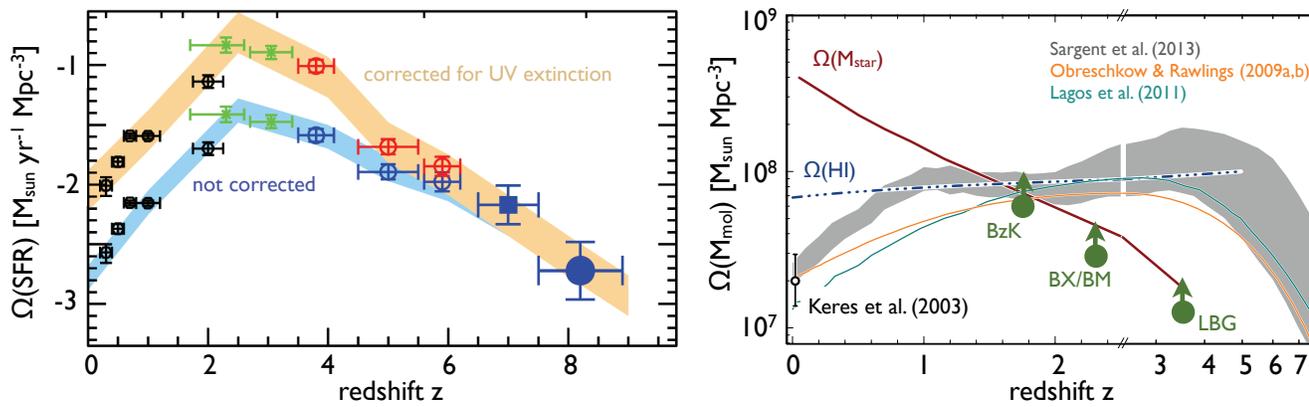,height=5.5cm}}

\caption { {\em Left:} Representation of the evolution of the cosmic
star formation rate density (adopted from the compilation shown in
Bouwens et al.\ 2010). {\em Right:} The evolution of the cosmic cool
gas mass density (from Sargent et al.\ 2013), including predictions
from semi--analytical cosmological models (Obreschkow \& Rawlings
2009a,b, Lagos et al.\ 2011) as well as
the models by Sargent et al.\ (2013). The latter shows the evolution
inferred from the integration of the indirectly inferred molecular gas
mass functions underlying the CO luminosity distributions of
Fig.~\ref{fig_lum_functions}.  Also included are some admittedly
extremely rough limits based on what is know about known galaxy
populations at $z > 1$. (see text for details). These points are shown
to illqustrate the potential impact of molecular deep field (`blind')
surveys.}

\label{omega_h2}
\end{figure}

\end{document}